\documentclass[useAMS,usenatbib]{mnras}
\usepackage{deluxetable}

\title[Variability and pc-scale structure of CSO]{Variability and
  parsec-scale radio structure of candidate compact symmetric objects.}
\author[M. Orienti \& D. Dallacasa]
  {M. Orienti$^{1}$\thanks{E-mail: orienti@ira.inaf.it},
D. Dallacasa$^{1,2}$\\
$^{1}$Istituto di Radioastronomia - INAF, Via P. Gobetti 101, I-40129 Bologna, Italy\\
$^{2}$Dipartimento di Fisica e Astronomia, Universit\`a di Bologna,
Via Gobetti 93/2, I-40129 Bologna, Italy\\
}
\date{Received \today; accepted ?}

\pagerange{\pageref{firstpage}--\pageref{lastpage}} \pubyear{2002}

\def\LaTeX{L\kern-.36em\raise.3ex\hbox{a}\kern-.15em
    T\kern-.1667em\lower.7ex\hbox{E}\kern-.125emX}

\begin{document}

\label{firstpage}

\maketitle

\begin{abstract}

  We report results on multi-epoch Very Large Array (VLA) and
  pc-scale Very Long Baseline Array (VLBA) observations 
  of candidate compact symmetric objects (CSOs) from the faint sample of
    high frequency peakers. New VLBA 
  observations could resolve the radio structure in about 42 per cent of the
  observed sources, showing double components that may be either
  mini-lobes or core-jet structures. Almost all the
  sources monitored by the VLA show 
  some variability on time scale of a decade, and
  only 1 source does not show any significant variation. In 17 sources
  the flux density changes randomly as it is expected in blazars, and
  in 4 sources the spectrum becomes flat in the last observing
  epoch, confirming that samples selected in the GHz regime are highly
  contaminated by beamed objects. In 16 objects,
  the pc-scale and variability properties are consistent with a young
  radio source
  in adiabatic expansion, with a steady decrease of the flux density
  in the optically-thin part of the spectrum, and a flux density
  increase in the optically-thick part. For these sources we estimate
  dynamical ages between a few tens to a few hundreds years.
  The corresponding expansion velocity is generally between
  0.1$c$ and 0.7$c$, similar to values found in CSOs 
with different approaches. 
  The fast evolution that we observe in some CSO 
candidates suggests that not all the objects 
  would become classical Fanaroff-Riley radio sources.

 \end{abstract}

\begin{keywords}
radio continuum: general - galaxies: active - radiation mechanisms: non-thermal 

\end{keywords}

\section{Introduction}

Individual radio source evolution is a field of research where most of
the ingredients are known.
However, many details are still poorly understood, and it is necessary
to investigate how such
ingredients interact and influence each other. In a scenario where
radio sources grow in a self similar way, the evolutionary stage of
each radio source originated by an active galactic nucleus
depends on its linear size. Compact symmetric objects (CSOs), with
  a linear size, LS, $<$ 1 kpc and a two-sided structure that is
  reminiscent of Fanaroff-Riley radio galaxies \citep{fr74}, are
  likely to
  represent radio sources in an early evolutionary stage \citep[see
    e.g.][]{wilkinson94,readhead96}. Following an
evolutionary path, CSOs would become medium-sized objects (MSOs) with
1$<$ LS $<$ 15-20 kpc, which are the progenitors of classical
Fanaroff-Riley radio sources \citep[e.g.,][]{fanti01}. 
This is supported by the estimate of
kinematic and radiative ages: 10$^{2-3}$ yr for CSOs \citep[see
  e.g.,][]{murgia03,polatidis03}, 10$^{4-6}$ yr for MSOs
\citep{murgia99} and 10$^{7-8}$ yr for large radio galaxies
\citep{orru10,harwood17}.\\ 
Several evolution models \citep[e.g.][]{fanti95,readhead96,snellen00,antao12}
have been proposed 
to describe how the physical parameters, such as luminosity, linear
size and velocity, evolve as the
radio emission deploys. However, various aspects concerning the early
stages of the radio evolution
predicted by the models do not match the observations. This indicates
that the initial parameters
considered in evolution models must be improved by a better knowledge
of the physical conditions when the radio emission turns on. 
In particular, the interaction
with the ambient medium may play a crucial role during the early stage
of radio source evolution
\citep[e.g.][]{dd13,morganti13,collier18,keim19,sobolewska19}.\\
To test the physical conditions soon after
  the onset of the radio emission, it is essential to define a fair
  sample of CSOs, large enough to be statistically sound. 
A correct classification requires
  targeted (sub-)parsec scale observations that prove their double
  morphology and identify the core position. 
Another indirect way to search for CSOs is by the analysis of their radio
  spectrum. An empirical anti-correlation was found between
the projected linear size and the peak frequency
\citep[e.g.,][]{odea97}: 
the smaller the
source, the higher the peak frequency is. In this context, the youngest
CSOs should be sought among those sources whose synchrotron spectrum
peaks above a few GHz. \\
High frequency
peakers \citep[HFPs,][]{dd00} are a heterogeneous class of
extragalactic sources, mainly made of blazars and CSOs, and
characterized by a radio spectrum that peaks above $\sim$5
GHz.  
With the aim of searching for CSOs, 
two samples of HFP radio sources in the northern hemisphere have been
constructed and are
currently available: the ``bright''
sample (sources brighter than 300 mJy at 5 GHz) selected by \citet{dd00}
and the ``faint'' sample
(sources with flux density between 50 and 300 mJy at 5 GHz around the
North Galactic Cap) presented in \citet{cstan09}, 
consisting of about 60 sources each. These
samples were built starting from the NRAO VLA Sky Survey (NVSS) and 87
Green Bank (87GB) catalogues and
selecting only those sources with a radio spectrum peaking at 5 GHz or
above, and subsequently
cleaned via quasi-simultaneous multi-frequency
observations with the Very Large Array (VLA). 
Further epochs of quasi-simultaneous multi-frequency spectra
were obtained to distinguish between CSO candidates 
from variable flat spectrum sources that matched the
initial selection criteria owing to their variability \citep{tinti05,
  torniainen05,sadler06,hovatta07,mo07,mingaliev12}. Moreover, all
the objects from the bright sample and $\sim$30 per cent of those from
the faint sample were imaged at
parsec-scale resolution in order to determine their
morphology \citep{mo06,mo12}.
CSOs are not significantly variable \citep{odea98},
possess low polarization, and have
double/triple radio morphology characterized by mini-lobes/hotspots
and relativistic beaming does not play a major role.
On the other hand, blazars do possess strong variability across the whole
electromagnetic spectrum, have high polarization, and have core-jet structures on
pc-scales. However, in the youngest radio sources, substantial
variability is expected as a consequence of the source
expansion, on timescales much longer than in beamed objects, and with
modest amplitude. In newly born radio
sources, a significant evolution of the radio emission can occur on time-scales
of the order of a few decades. Assuming a homogeneous radio source in
adiabatic expansion with the magnetic field frozen in the plasma, the
flux density variation in the optically thick part of the spectrum is
$\Delta S \propto \left(1+ (\Delta t/t_0)
\right)^3$, where $t_0$ is the source age and $\Delta t$ is the time
interval between two observations. If $\Delta t$ is a large
fraction of $t_{0}$, $\Delta S$ can be significant.\\
In this paper we present results on new VLA
observations from 1 to 32 GHz of 35 out of 61 sources from the faint
HFP sample, and Very Long Baseline Array (VLBA) observations at 15 and
24 GHz of a sub-sample of 12 sources. The long time-baseline of multi-epoch
VLA observations (more than a decade) allows us to study the
long-term variability and investigate if some spectral changes are
consistent with a CSO in adiabatic expansion. On the
other hand, dual-frequency observations with milli-arcsecond resolution
provide a deep look into the radio source structure at few parsecs in
size. The combination of VLA and VLBA information is necessary for
determining the nature of each object and removing blazars
that contaminate the sample of CSO candidates. The
final goal is the construction of a sample of genuinely young 
compact symmetric objects. 
The determination of the physical properties in the very early phase of radio
source evolution will provide important constraints on the initial
conditions assumed in the development of evolutionary models.\\
This paper is organized as follows:
in Section 2 we present the observations and data analysis;
results are reported in Section 3 and discussed in Section 4. A brief
summary is presented in Section 5.

Throughout this paper, we assume the following cosmology: $H_{0} =
71\; {\rm km/s\, Mpc^{-1}}$, 
$\Omega_{\rm M} = 0.27$ and $\Omega_{\rm \Lambda} = 0.73$,
in a flat Universe. The spectral index $\alpha$
is defined as 
$S {\rm (\nu)} \propto \nu^{- \alpha}$. \\

\section{Radio observations}

\subsection{VLA observations}

Simultaneous multi-frequency VLA observations of 35 out of the
61 sources from the ‘faint’ HFP sample \citep{cstan09} 
were carried out during two runs in 2012 April and May
(Table \ref{radio-log}). Observations were performed in L (1$-$2 GHz), S
(2$-$4 GHz), C (4.5$-$6.5 GHz), X (8$-$10 GHz), Ku (13$-$15 GHz),
K (20.2$-$22.2 GHz) bands, and for a run also in Ka (31$-$33 GHz) band (project
code: 12A-048).   
Observations had a band width of 2 GHz,
with the exception of L band.
In each frequency band the band-width was
divided into 16 spectral windows. In both runs 3C\,286 was used as
primary calibrator and band pass 
calibrator, with the exception of K and Ka band, were the calibrator
J0927$+$3902 (4C\,39.25) was used as band pass calibrator. Target sources were
observed for about 1.5 min per frequency. Secondary calibrators were
chosen to minimize the antenna slewing.\\
Calibration was performed using the \texttt{CASA} software
\citep{mcmullin07}
following
the standard procedure for the VLA. Parts of L, S and C bands were highly
affected by RFI and we had to flag some spectral windows. 
Errors on the amplitude calibration are conservatively 3 per cent in
L, C, and X bands, 5 per cent in Ku band, and 10 per cent in S, K, and
Ka bands. 
After the a-priori calibration,
imaging was done with the \texttt{CASA} task \texttt{CLEAN} and the
\texttt{AIPS} task \texttt{IMAGR}. Phase-only self-calibration of the
target field was generally performed in L-band, given the presence of
many sources in the field of view, granting substantial amount of flux
density for the model. \\
We produced an image for each spectral window of each band in order to
maximize the spectral coverage (Fig. \ref{radio_spectra}). 
In Table
\ref{vla-flux} we report the flux densities at 1.4, 1.7, 4.5, 5.0, 8.1,
8.4, 15, 22, 32 GHz, in order to have a direct comparison with the values
from the narrow-band receivers of historical VLA observations
\citep{cstan09,mo10}. When RFI affects any of those frequencies,
nothing is reported.\\
The error on the image plane, $\sigma_{\rm rms}$,
is usually around 50 $\mu$Jy/beam, but it may be as high as
0.2$-$0.5 mJy/beam in some spectral windows particularly affected by
RFI. \\

\begin{table}
  \caption{Log of radio observations.}
  \begin{center}
\begin{tabular}{lccc}
\hline
Date & Project & Configuration & Code \\
\hline
2012-04-25 & AO281 & VLA-C & a \\
2012-06-16 & AO281 & VLA-CnB & b \\
2019-01-19 & BO057 & VLBA & c \\
\hline
\end{tabular}
\end{center}
\label{radio-log}
\end{table}

\begin{table*}
\caption{Multi-frequency VLA flux density of faint HFP sources. Column 1:
  source name; column 2: optical counterpart; column 3: redshift: a
  $p$ indicates a photometric redshift; columns 4-12: flux density (in
  mJy) at
  1.4, 1.7, 4.5, 5.0, 8.1, 8.4, 14.5, 21.5, and 32 GHz, respectively;
  columns 13 and 14: spectral index below and above the peak
  frequency, respectively.}
  \begin{center}
\begin{tabular}{cccccccccccccc}
\hline
Source & ID & $z$ & S$_{1.4}$ & S$_{1.7}$ & S$_{4.5}$ & S$_{5.0}$ &
S$_{8.1}$ & S$_{8.4}$ & S$_{14.5}$ & S$_{21.5}$ & S$_{32}$ &
$\alpha_{\rm b}$ & $\alpha_{\rm a}$ \\
(1) & (2) & (3) & (4) & (5) & (6) & (7) & (8) & (9) & (10)&(11)&(12)&(13)&(14)\\
\hline
J0754$+$3033 & Q &0.769 & 66$\pm$2 & 75$\pm$2 & 142$\pm$4 & 145$\pm$4
& 143$\pm$4 & 141$\pm$4 & 108$\pm$5 & 96$\pm$10 & - & -0.6 & 0.4\\
J0819$+$3823 & Q & - & 19$\pm$1 & 25$\pm$1 & 115$\pm$3 & 118$\pm$4 & -
& - & 41$\pm$2 & 25$\pm$3 & - & -1.5 & 1.1\\
J0821$+$3107 & Q & 2.625& 96$\pm$5 &
106$\pm$5&71$\pm$2&68$\pm$2&52$\pm$2&50$\pm$2&34$\pm$4&33$\pm$5& - & -
& 0.5\\
J0951$+$3451 & G & 0.358 & 27$\pm$1 & 36$\pm$1 & 63$\pm$2 & 64$\pm$2 &
57$\pm$2 & 56$\pm$2 & 35$\pm$2 & 27$\pm$3 & 21$\pm$2 & -1.0 & 0.6\\
J0955$\pm$3355& Q & 2.491& 46$\pm$2 & 61$\pm$3& 68$\pm$2 & 66$\pm$2&
47$\pm$1 & 45$\pm$1 & 26$\pm$2 & 20$\pm$3 & - & -0.9 & 0.8\\
J1008$+$2533 & Q & 1.96 & 54$\pm$3 & 69$\pm$3 & 102$\pm$3 & 105$\pm$3
& 103$\pm$3 & 104$\pm$3 & 115$\pm$6 & 130$\pm$13 &- & -1.1 & 0.0\\
J1020$+$4320 & Q & 1.964 & 117$\pm$4 & 158$\pm$5 & 306$\pm$9 &
303$\pm$9 & 243$\pm$7 & 239$\pm$7 & 161$\pm$8 & 124$\pm$12& - & -1.0 &
0.5\\
J1025$+$2541 & G & 0.457 & 26$\pm$1 & 31$\pm$1 & 42$\pm$1 & 38$\pm$1 &
23$\pm$1 & 21$\pm$1 & 11$\pm$1 & 5$\pm$1 & - & -1.2 & 1.2\\
J1035$+$4230 & Q & 2.44 & 23$\pm$1 & 28$\pm$1 & 77$\pm$2 & 88$\pm$3 &
84$\pm$3 & 83$\pm$3 & 59$\pm$3 & 40$\pm$4 & 29$\pm$3 & -1.1 & 0.7\\
J1044$+$2959 & Q & 2.983 & 52$\pm$2 & 66$\pm$2 & 132$\pm$4 & 134$\pm$4
& 124$\pm$4 & 123$\pm$4 & 95$\pm$5 & 75$\pm$8 & - & -0.8 & 0.5\\
J1046$+$2600 & - & - & 16$\pm$1 & 20$\pm$1 & 35$\pm$1 & 33$\pm$1 &
23$\pm$1 & 22$\pm$1 & 11$\pm$1 & 6$\pm$1 & - &-1.1 & 1.1\\
J1052$+$3355 & Q & 1.407& - & 23$\pm$5 & 24$\pm$1 & 22$\pm$1 &
13$\pm$1 & 11$\pm$1 & 7$\pm$1 & 6$\pm$1 & - & - &0.9\\
J1054$+$5058 & Q & - & 12$\pm$1 & 13$\pm$1 & 21$\pm$1 & 21$\pm$1 &
28$\pm$1 & 30$\pm$1 & 36$\pm$2 & - & - & -0.5 & -\\
J1058$+$3353 & G & 0.265 & - & 21$\pm$3 & - & - & 78$\pm$2 & 80$\pm$2
& 116$\pm$6 & 118$\pm$12 & - & -0.7 & -\\
J1107$+$3421 & - & - & 29$\pm$1 & 42$\pm$2 & 64$\pm$2 & 60$\pm$2 &
37$\pm$1 & 35$\pm$1 & 18$\pm$1 & 9$\pm$1 & - &-1.4 & 1.1\\
J1137$+$3441& Q &0.835 & 13$\pm$2 & 30$\pm$3 & 59$\pm$2 & 62$\pm$2 &
63$\pm$2 & 63$\pm$2 & 61$\pm$3 & 59$\pm$6 & - & -0.9 & 0.1\\
J1218$+$2828 & G & 0.18p& - & - & 55$+$2 & 56$\pm$2 & 41$\pm$1 &
39$\pm$1 & 29$\pm$2 & 42$\pm$4 & - & -0.5 & 0.6\\
J1240$+$2323 & G & 0.38p & 18$\pm$5 & 24$\pm$3 & - & - & 50$\pm$2 &
50$\pm$2 & 45$\pm$1 & 42$\pm$1 & - & -0.6 & 0.2\\
J1240$+$2425& Q & 0.831 & 63$\pm$3 & 54$\pm$3 & 30$\pm$1 & 28$\pm$1 &
18$\pm$1 & 17$\pm$1 & 10$\pm$1 & 7$\pm$1 & - & - & 0.9\\
J1258$\pm$2820 & Q & - & -& - & 45$\pm$2 & 49$\pm$2 & 51$\pm$2 &
51$\pm$2 & 45$\pm$2 & 34$\pm$3 & - & -0.3 & 0.5\\
J1309$+$4047 & Q & 2.91 & - & 62$\pm$2 & 128$\pm$4 & 127$\pm$4 &
103$\pm$3 & 101$\pm$3 & 64$\pm$3 & 37$\pm$4 & - & -0.9 & 0.7\\
J1420$+$2704 & Q & - & - & 19$\pm$1 &69$\pm$3 & 70$\pm$2 & 65$\pm$2 &
64$\pm$2 & 45$\pm$2 & 29$\pm$3 & 22$\pm$2 & -1.0 & 0.7\\
J1421$+$4645 & Q & 1.668 & - & 126$\pm$4 & 237$\pm$7 & 244$\pm$7 &
244$\pm$7 & 244$\pm$4 & 204$\pm$10 & 184$\pm$18 &136$\pm$14 & -0.7 & 0.4\\
J1459$+$3337 & Q & 0.645 & 50$\pm$5 & 71$\pm$5 & 216$\pm$7 & 223$\pm$7
& 188$\pm$6 & 183$\pm$6 & 108$\pm$6 & 68$\pm$7 & 43$\pm$4 & -1.0 & 1.0\\
J1512$+$2219& G & 0.40p& 14$\pm$2 & 24$\pm$3 & 24$\pm$1 & 21$\pm$1 &
10$\pm$1 & 9$\pm$1 & 3$\pm$1 & 2$\pm$1 & - & -1.7 & 1.3\\
J1528$+$3816 & Q & 0.749 & 27$\pm$1 & 32$\pm$1 & 52$\pm$2 & 53$\pm$2 &
56$\pm$2 & 57$\pm$2 & 58$\pm$6 & 57$\pm$6 & 59$\pm$6 & -0.6 & 0.0\\
J1530$+$2705 & G & 0.033 & 24$\pm$3 & 29$\pm$3 & 50$\pm$2 & 50$\pm$2 &
41$\pm$1 & 40$\pm$1 & 28$\pm$2 & 26$\pm$3 & - & -0.8 & 0.4\\
J1530$+$5137 & G & 0.632p & 53$\pm$2 & 60$\pm$2 & 98$\pm$3 & 100$\pm$3
& 111$\pm$3 & 112$\pm$3 & 111$\pm$6 & 115$\pm$12 & 127$\pm$13 & -0.5 &
0.0\\
J1547$+$3518& Q & - & - & 23$\pm$2 & 41$\pm$1 & 44$\pm$1 & 51$\pm$1 &
53$\pm$2 & 53$\pm$3 & 57$\pm$6 & - & -0.5 & -\\
J1602$+$2646 & G & 0.372 & - & 44$\pm$2 & 162$\pm$5 & 189$\pm$6 &
297$\pm$9 & 310$\pm$9 & 345$\pm$17 & 303$\pm$30 &266$\pm$27 & -1.0 & 0.3\\
J1613$+$4223 & Q & - & 43$\pm$1 & 70$\pm$2 & 197$\pm$6 & 188$\pm$6 &
107$\pm$3 & 102$\pm$3 & 33$\pm$3 & 15$\pm$2 & - & -1.9 & 1.6\\
J1616$+$4632& Q & 0.950 & 82$\pm$4 & 83$\pm$4 & 77$\pm$2 & 79$\pm$2 &
78$\pm$2 & 78$\pm$2 & 62$\pm$3 & 58$\pm$6 & - & - & 0.2\\
J1617$+$3801 & Q & 1.607 & 26$\pm$1 & 34$\pm$1 & 72$\pm$2 & 77$\pm$2 &
115$\pm$3 & 121$\pm$4 & 128$\pm$7 & 107$\pm$11 &83$\pm$8 & -1.0 & 0.4\\
J1624$+$2748 & G & 0.541p & 19$\pm$1 & 21$\pm$1 & 97$\pm$3 & 105$\pm$3
& 155$\pm$5 & 162$\pm$5 & 175$\pm$9 & 172$\pm$17 & 153$\pm$15 & -1.2 &
-\\
J1719$+$4804 & Q & 1.084 & 70$\pm$2 & 83$\pm$3 & 83$\pm$3 & 77$\pm$2 &
47$\pm$1 & 44$\pm$1 & 18$\pm$2 & 11$\pm$1 & - & -0.8 & 0.7\\
\hline
\end{tabular}
\end{center}
\label{vla-flux}
\end{table*}

\begin{table}
\caption{Pc-scale radio morphology and accurate source position of
  faint HFP sources with VLBA observations. Column 1: source name;
  column 2: morphology: CD = compact double; MR = marginally resolved;
  Un = unresolved. Columns
  3 and 4: right ascension and declination of the main component; 
Column 5: VLBA calibrator observed for
  phase-referencing. The uncertainty on the position is about 0.16 mas.}
\begin{center}
\begin{tabular}{ccccc}
  \hline
  Source & M. &RA & Dec & Cal.\\
         &        &(J2000)   &(J2000)     & (B1950)\\
  \hline
  J0754$+$3033& CD &07:54:48.8514 & 30:33:55.020 & 0738$+$313 \\
  J0819$+$3823& CD &08:19:00.9562 & 38:23:59.810 & 0821$+$394 \\
  J1002$+$5701& Un &10:02:41.6661 & 57:01:11.484 & 1014$+$615 \\
  J1025$+$2541& CD &10:25:23.7918 & 25:41:58.362 & 1012$+$232 \\
  J1035$+$4230& MR &10:35:32.5776 & 42:30:18.959 & 1020$+$400 \\
  J1044$+$2959& Un &10:44:06.3428 & 29:59:01.004 & 1059$+$282 \\
  J1046$+$2600& MR &10:46:57.2508 & 26:00:45.104 & 1040$+$244 \\
  J1107$+$3421& CD &11:07:34.3382 & 34:21:18.596 & 1101$+$384 \\
  J1420$+$2704& Un &14:20:51.4879 & 27:04:27.045 & 1417$+$273 \\
  J1459$+$3337& CD &14:59:58.4359 & 33:37:01.776 & 1504$+$377 \\
  J1613$+$4223& Un &16:13:04.8033 & 42:23:18.893 & 1638$+$398 \\
  J1719$+$4804& Un &17:19:38.2496 & 48:04:12.248 & 1726$+$455 \\
  \hline
  \end{tabular}
\end{center}
\label{phase-vlba}
\end{table}

\begin{table*}
\caption{VLBA flux density of HFP sources. Column 1: source name;
  column 2: source component; columns 3 and 4: flux density (in mJy) at 15 and
  24 GHz, respectively; column 5: VLBA spectral index between 15 and
  24 GHz; column 6: VLA spectral index between 15 and 22 GHz; columns
  7 and 8: fractional flux density between VLBA and VLA, S$_{\rm
    VLBA}/$S$_{\rm VLA}$ at 15 and 24 GHz, respectively. For the
  source J1002$+$5701 the VLA spectral index and flux density 
  refers to data from \citet{cstan09}.}
\begin{center}
\begin{tabular}{cccccccc}
\hline
Source & Comp & S$_{\rm 15}$ & S$_{24}$ &$\alpha_{15}^{24}$ &
$\alpha_{\rm VLA}$& F$_{15}$ & F$_{24}$\\
& & mJy & mJy &  & & \% & \%\\
\hline
J0754$+$3033 & E & 50.7$\pm$3.5 & 38.0$\pm$2.7&0.6$\pm$0.2 & - & - & -\\
& W & 26.0$\pm$1.8 & 14.2$\pm$1.0 & 1.3$\pm$0.2 & - & - & -\\
& Tot & 79.4$\pm$5.5 & 54.7$\pm$3.8 &0.8$\pm$0.2 & 0.3$\pm$0.3&0.73&0.57\\
J0819$+$3823& E & 4.7$\pm$0.3 & - & - & -& - & -\\
& W & 21.7$\pm$1.5 & - & - & - & - & -\\
& Tot& 26.4$\pm$1.8 & 10.4$\pm$0.7 & 2.0$\pm$0.2& 1.0$\pm$0.3&0.64 & 0.42\\
J1002$+$5701& Tot & 8.0$\pm$0.6 & 3.1$\pm$0.2 & 2.0$\pm$0.2 &
1.0$\pm$0.3&0.40 & 0.24\\
J1025$+$2541& E & 7.8$\pm$0.5 & - & - & - & - & -\\
& W & 2.0$\pm$0.1 & - & - & - & - & -\\
& Tot & 9.8$\pm$0.7 & 6.9$\pm$0.5 & 0.7$\pm$0.2 & 1.6$\pm$0.5&0.89&1.4\\
J1035$+$4230& E & - & 13.0$\pm$0.9 & - & - & - & -\\
            & W & - & 5.2$\pm$0.4 & - & - & - & -\\
& Tot & 28.5$\pm$2.0 & 18.2$\pm$1.3 & 1.0$\pm$0.2& 0.8$\pm$0.3&0.48&0.45\\
J1044$+$2959& Tot & 73.8$\pm$5.2 & 45.1$\pm$3.1 & 1.0$\pm$0.2& 0.5$\pm$0.3&0.78&0.60\\
J1046$+$2600& E & - & 5.3$\pm$0.4 & - & -& - & -\\
& W & - & 4.1$\pm$0.3 & - & -& - & -\\
& Tot & 11.2$\pm$0.8 & 9.4$\pm$0.7 & 0.4$\pm$0.2& 1.2$\pm$0.3&1.00&0.64\\
J1107$+$3421& E & 4.2$\pm$0.5 & 3.2$\pm$0.2 & 0.6$\pm$0.3 & -& - &-\\
& W & 3.1$\pm$0.5 & 4.0$\pm$0.5 & -0.5$\pm$0.4 & - & - & -\\
& Tot & 7.3$\pm$0.7 & 7.2$\pm$0.7 & 0.0$\pm$0.3& 1.4$\pm$0.3& 0.40 & 0.80\\
J1420$+$2704& Tot & 28.0$\pm$2.0 & 18.4$\pm$1.3 &0.9$\pm$0.2 & 0.9$\pm$0.3&0.62&0.63\\
J1459$+$3337& E & 5.3$\pm$0.4 & 3.3$\pm$0.3 & 1.0$\pm$0.3 & -& - & -\\
& W & 20.1$\pm$1.4 & 11.5$\pm$0.8 & 1.2$\pm$0.2 & - & - & -\\
& Tot & 25.4$\pm$1.8 & 14.8$\pm$1.0 & 1.1$\pm$0.2 & 0.9$\pm$0.3& 0.23&0.22\\
J1613$+$4223&Tot& 28.9$\pm$2.0 & 5.8$\pm$0.4 & 3.4$\pm$0.2 & 1.6$\pm$0.4&0.87&0.39\\
J1719$+$4804&Tot& 4.2$\pm$0.3 & 3.5$\pm$0.3 & 0.4$\pm$0.2 & 1.0$\pm$0.3&0.23&0.32\\
\hline
\end{tabular}
\end{center}
\label{vlba-flux}
\end{table*}

\begin{table*}
\caption{Variability and peak frequency of faint HFP sources. Column 1:
  source name (sources which are
  still considered CSO candidates are marked with a diamond); columns 2-5: variability computed between two
  consecutive epochs, V$_{\rm ep}$ ($a$=1998, $b$=1999, $c$=2000, $d$=2003,
  $e$=2004, $f$=2006, $g$=2007, $h$=2012. For the precise dates see
  \citet{cstan09} and \citet{mo10}). Column 6: 
variability computed between the first
  epoch \citep[1998-2000,][]{cstan09} and last epoch (2012) of VLA
  observations, V$_{\rm tot}$; columns 7-11: peak frequency during the first
  (1998-1999), second (2003), third (2004), fourth (2006-2007) and
  last (2012) observing epochs, respectively; column 12: variability
  classification - NV = non variable, SV = slightly variable, V =
  variable (see Section \ref{sec_variability}). } 
  \begin{center}
\begin{tabular}{cccccccccccc}
\hline
Source &V$_{\rm ep (1-2)}$&V$_{\rm ep (2-3)}$&V$_{\rm
    ep (3-4)}$&V$_{\rm
  ep (4-5)}$&V$_{\rm tot (1-5)}$&$\nu_{\rm p,1}$&$\nu_{\rm p,2}$&$\nu_{\rm
  p,3}$&$\nu_{\rm p,4}$&$\nu_{\rm p,5}$& Class.\\
(1)&(2)&(3)&(4)&(5)&(6)&(7)&(8)&(9)&(10)&(11)&(12)\\
\hline
J0754$+$3033$^{\diamond}$&  26.5$^{b,d}$& 1.5$^{d,e}$&  1.9$^{e,f}$&  6.1$^{f,h}$&
57.4&8.8$\pm$0.6&7.8$\pm$0.4&8.3$\pm$0.5&7.4$\pm$0.2&6.5$\pm$1.0& V\\
J0819$+$3823$^{\diamond}$&  6.1$^{a,d}$&  2.7$^{d,e}$&  8.5$^{e,f}$& 13.5$^{f,h}$ &   7.4&6.1$\pm$0.9&5.7$\pm$1.0&6.0$\pm$0.9&6.2$\pm$0.9&5.6$\pm$1.9&SV\\
J0821$+$3107$^{\diamond}$& 52.5$^{b,f}$ & 41.0$^{f,h}$ & - & - & 130.0 & 3.4$\pm$0.3 & - & -
&2.7$\pm$0.3&$<$2.0& V \\
J0951$+$3451$^{\diamond}$&  0.2$^{a,d}$&  4.0$^{d,f}$ &  2.7$^{f,h}$ &  - &  12.6&6.1$\pm$0.5&6.0$\pm$0.6&
- & 5.6$\pm$0.5&5.6$\pm$1.0 & SV\\
J0955$+$3335& 6.2$^{b,d}$ & 9.7$^{d,f}$ & 9.8$^{f,h}$ & - & 124.1 & 5.7$\pm$0.6&5.2$\pm$0.5& -&4.6$\pm$0.4&2.4$\pm$0.4 & V\\
J1008$+$2533&  17.8$^{a,e}$ & 1.4$^{e,f}$ & 3.4$^{f,h}$ & - &   19.7&5.9$\pm$0.5& -
&4.9$\pm$0.4&5.4$\pm$0.4&4.1$\pm$0.7 &SV\\
J1020$+$4320&  1.7$^{b,e}$&  17.7$^{e,h}$ &   - &  - &  23.3&4.6$\pm$0.4& -
&4.6$\pm$0.5& - &5.2$\pm$1.2&SV\\
J1025$+$2541&  15.2$^{a,e}$& 19.7$^{e,h}$&   -  &  - &  36.9&4.2$\pm$0.7& -
&3.7$\pm$0.5& - &3.3$\pm$1.2 &V\\
J1035$+$4230$^{\diamond}$&  7.8$^{b,e}$&   8.4$^{e,h}$&   -  &  - &  20.8&7.1$\pm$0.6& -
&6.7$\pm$0.8& - &7.0$\pm$1.9 &SV\\ 
J1044$+$2959$^{\diamond}$&  4.1$^{b,e}$&  22.6$^{e,h}$&   -  &  - &  35.9&7.1$\pm$0.7& -
&4.8$\pm$0.3& - &6.1$\pm$1.2 &V\\
J1046$+$2600&  4.9$^{a,e}$&  22.4$^{e,h}$&   -  &  - &  13.4&4.7$\pm$0.7& -
&4.8$\pm$0.6& - &4.0$\pm$1.3 &SV\\
J1052$+$3355$^{\diamond}$& 42.0$^{b,e}$ & 4.8$^{e,f}$ & 73.3$^{f,h}$ & - & 206.0 &5.2$\pm$0.6& -
&1.6$\pm$0.1 &4.2$\pm$0.5&$<$2 & V\\
J1054$+$5058$^{\diamond}$&  7.6$^{c,f}$&   9.1$^{f,h}$&   -  &  - &   7.1&$>$22 & - & - &$>$22 &
$>$15 & SV\\
J1058$+$3353& 41.4$^{b,d}$ & 105.4$^{d,h}$ & - & - & 72.2 &6.7$\pm$0.6&26.5$\pm$0.3& -
& - &31.7$\pm$0.6 & V\\
J1107$+$3421$^{\diamond}$& 14.5$^{b,d}$&   8.5$^{d,e}$&  2.8$^{e,f}$ &22.7$^{f,h}$&
49.9&4.6$\pm$0.6&4.4$\pm$0.8&4.6$\pm$0.6&4.2$\pm$0.6&3.9$\pm$1.4 & V\\
J1137$+$3441& 32.5$^{b,d}$ & 105.4$^{d,h}$ & - & - & 19.3&14.3$\pm$0.4& $>$34& - & -
&9.7$\pm$0.7 & V\\
J1218$+$2828 & 19.5$^{a,d}$ & 176.2$^{d,h}$ & - & - & 131.2 &6.9$\pm$0.7&7.5$\pm$0.6&
- & - &3.9$\pm$0.4 & V\\
J1240$+$2323 & 6.2$^{a,d}$ & 4.2$^{d,f}$ & 20.6$^{f,h}$ & - & 9.4 &7.8$\pm$0.6&9.3$\pm$0.4& -
&9.8$\pm$0.4&9.8$\pm$0.6 & SV\\
J1240$+$2425& 14.8$^{a,d}$ & 2.6$^{d,f}$ & 193.0$^{f,h}$ & - & 156.5
&3.8$\pm$0.5&$<$1.4 & - &2.7$\pm$0.3&$<$ 1 & V\\
J1258$+$2820 & 21.3$^{b,d}$ & 4.6$^{d,f}$ & 3.2$^{f,h}$ & - & 22.9 &
4.7$\pm$0.3&7.0$\pm$0.4& - &14.6$\pm$0.3& 8.1$\pm$0.5 &SV\\
J1309$+$4047$^{\diamond}$&  7.0$^{b,d}$&   1.6$^{d,f}$&  5.3$^{f,h}$ &  -  &
2.4&5.4$\pm$0.6&5.7$\pm$0.6& - &5.4$\pm$0.7&4.8$\pm$1.0 &SV\\
J1420$+$2704$^{\diamond}$&  8.6$^{b,d}$&   2.8$^{d,g}$&  6.0$^{g,h}$ &  - &
24.6&7.2$\pm$0.8&6.5$\pm$0.6&6.6$\pm$0.6& - &6.9$\pm$1.3 & SV\\
J1421$+$4645&  -  & -    & -    &  - &  44.1&5.5$\pm$0.3& - & -& -
&7.8$\pm$1.0& V\\ 
J1459$+$3337& 37.3$^{b,d}$& 182.2$^{d,h}$&   -  &  - &
250.8&21.2$\pm$0.9&15.8$\pm$0.8& - & - &2.8$\pm$0.9 & V\\
J1512$+$2219$^{\diamond}$& -  & - & - & - & 92.3 &2.8$\pm$0.3& - & - & - &
1.6$\pm$0.5 &V\\
J1528$+$3816&  -  &   -  &   -  &  - &  19.4&17.7$\pm$0.4& - & - & -
&15.2$\pm$0.7 & SV\\
J1530$+$2705& 58.1$^{b,d}$ & 4.7$^{d,g}$ & 20.6$^{g,h}$ & - & 22.9 &10.2$\pm$0.6&5.6$\pm$0.7&
- &7.2$\pm$0.7&4.6$\pm$0.6 &V\\
J1530$+$5137&  -  &   -  &   -  &  - &  52.7&15.9$\pm$0.2& - & - & -
&19.9$\pm$0.7 & V\\
J1547$+$3518$^{\diamond}$& 5.9$^{b,d}$ & 7.2$^{d,f}$ & 8.4$f,h$ & - & 23.9
&17.5$\pm$0.5&16.3$\pm$0.1& - &16.4$\pm$0.4&14.9$\pm$0.6 & SV\\
J1602$+$2646&  3.1$^{b,d}$&  22.5$^{d,h}$&   -  &  - &
28.5&15.9$\pm$0.5&16.2$\pm$0.5& - & - &15.0$\pm$1.5 & V\\
J1613$+$4223$^{\diamond}$&  6.7$^{b,e}$&   0.8$^{e,g}$&  1.2$^{g,h}$ &  - &
12.6&4.7$\pm$0.8&4.5$\pm$0.8&4.5$\pm$0.8& - &4.4$\pm$2.0 &SV\\
J1616$+$4632& 26.1$^{b,f}$ & 126.9$^{f,h}$ & - & - & 62.2 & $>$22 & - & - &
8.6$\pm$0.3&2.6$\pm$0.2 & V\\
J1617$+$3801&  1.3$^{b,d}$&   9.7$d,h$&   -  &  - &  14.4&12.0$\pm$0.4&9.6$\pm$0.5&
- & - &12.5$\pm$1.7 & SV\\
J1624$+$2748$^{\diamond}$&  3.8$^{b,d}$&   4.7$^{d,h}$&   -  &  - &
2.7&13.0$\pm$0.7&14.2$\pm$0.6& - & - &16.9$\pm$2.3 & NV\\
J1719$+$4804$^{\diamond}$& 27.6$^{b,d}$&   2.3$^{d,e}$& 34.3$^{e,g}$&120.7$^{g,h}$&
254.3&10.8$\pm$0.4&6.3$\pm$0.4&6.2$\pm$0.4&4.9$\pm$0.4&2.8$\pm$0.9 & V\\
\hline
\end{tabular}
\end{center}
\label{vla-variability}
\end{table*}

\subsection{VLBA observations and data reduction}

VLBA observations at 15 and 24 GHz of a sub-sample of 12 faint HFP
sources were carried out on 2019 January 19 in dual polarization and
an aggregate bit rate of 2Gbps (project code BO057). The target sources were
selected on the basis of their peak frequency below 7 GHz, in order to
have observations in the optically-thin part of the spectrum. \\
Each source was observed for about 25 min at 15 GHz and for 30 min at 24
GHz,  spread into twelve to fifteen  short  scans of about 2 min each,
switching between frequencies and sources in order to improve the
coverage of the ({\it u,v}) plane.
Target sources are too faint for fringe fitting, and the
observations were performed in phase-referencing mode. Phase
calibrators are reported in Table \ref{phase-vlba}. \\
Calibration and data reduction were performed following the standard
procedures described in the Astronomical Image Processing System
({\texttt AIPS}) cookbook. J0927+3902 was used to generate the band pass
correction. The amplitudes were calibrated using antenna system
temperatures and antenna gains and applying an atmospheric opacity
correction. The uncertainties on the amplitude calibration were found
to be approximately 7 per cent at both frequencies.\\
Images were produced in {\texttt AIPS} with the task {\texttt
  IMAGR}. Phase-only self-calibration was performed for those sources
stronger than 10 mJy. The rms noise level on the image plane is
between 0.07 and 0.3 mJy beam$^{-1}$. Flux densities are
reported in Table \ref{vlba-flux}. When the source is resolved we
define E and W the eastern and western components, respectively
  (in Fig. \ref{vlba-images} North is up and East is left). \\
The uncertainty on the source
position was estimated by comparing the positions at 15 and 24 GHz,
and it is about 0.16 mas. \\

\begin{figure*}
\begin{center}
\includegraphics{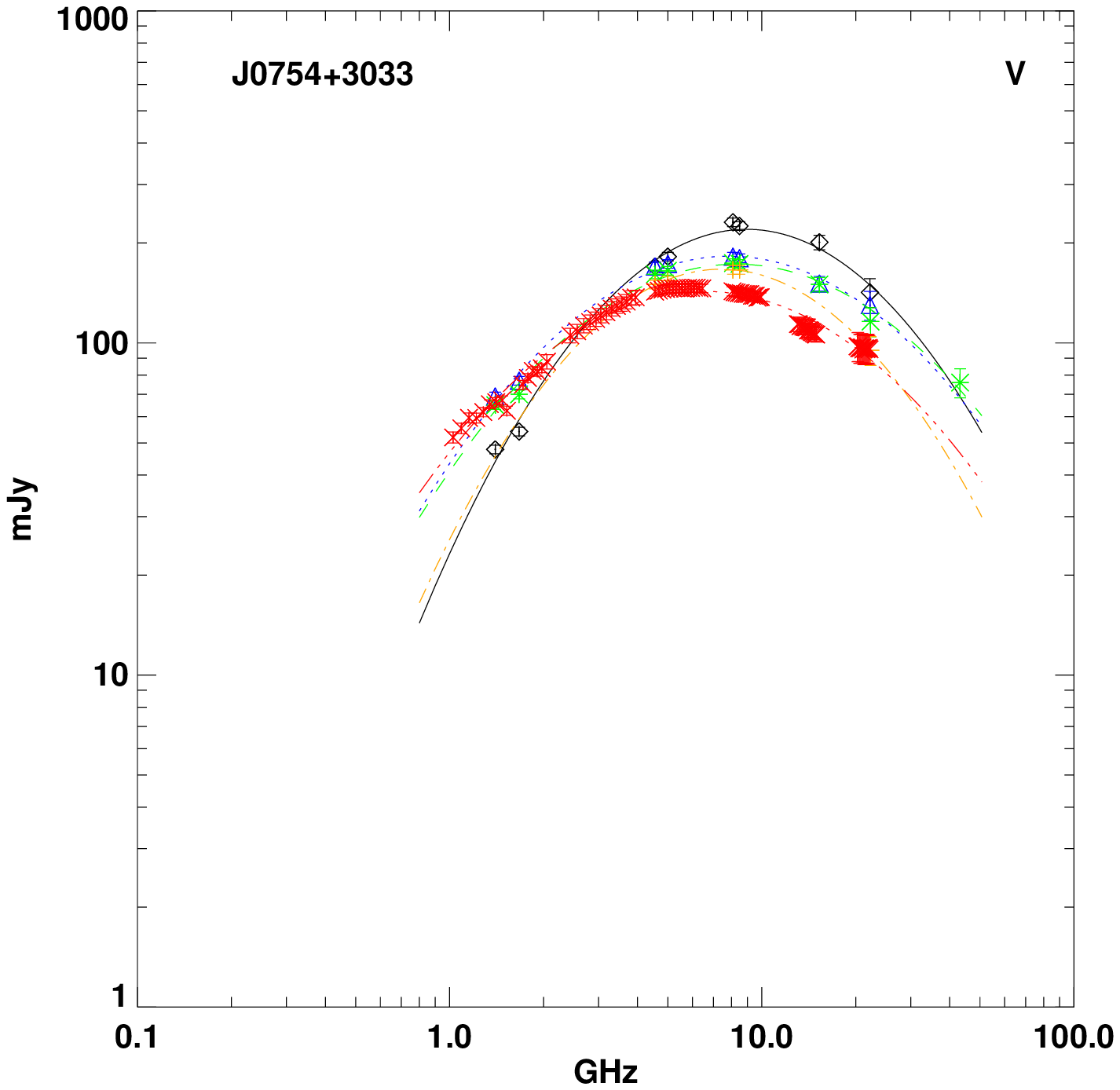}
\includegraphics{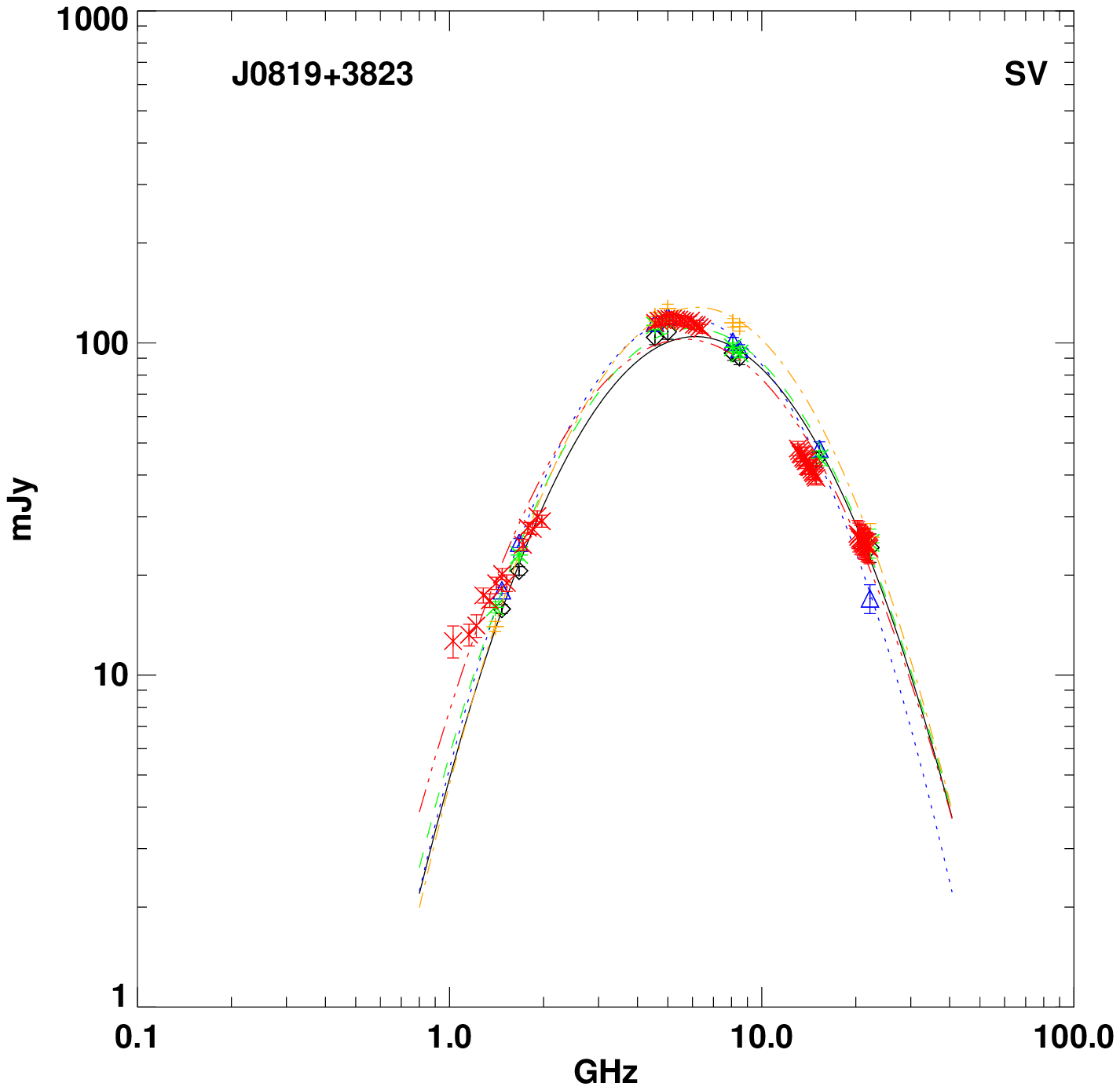}
\includegraphics{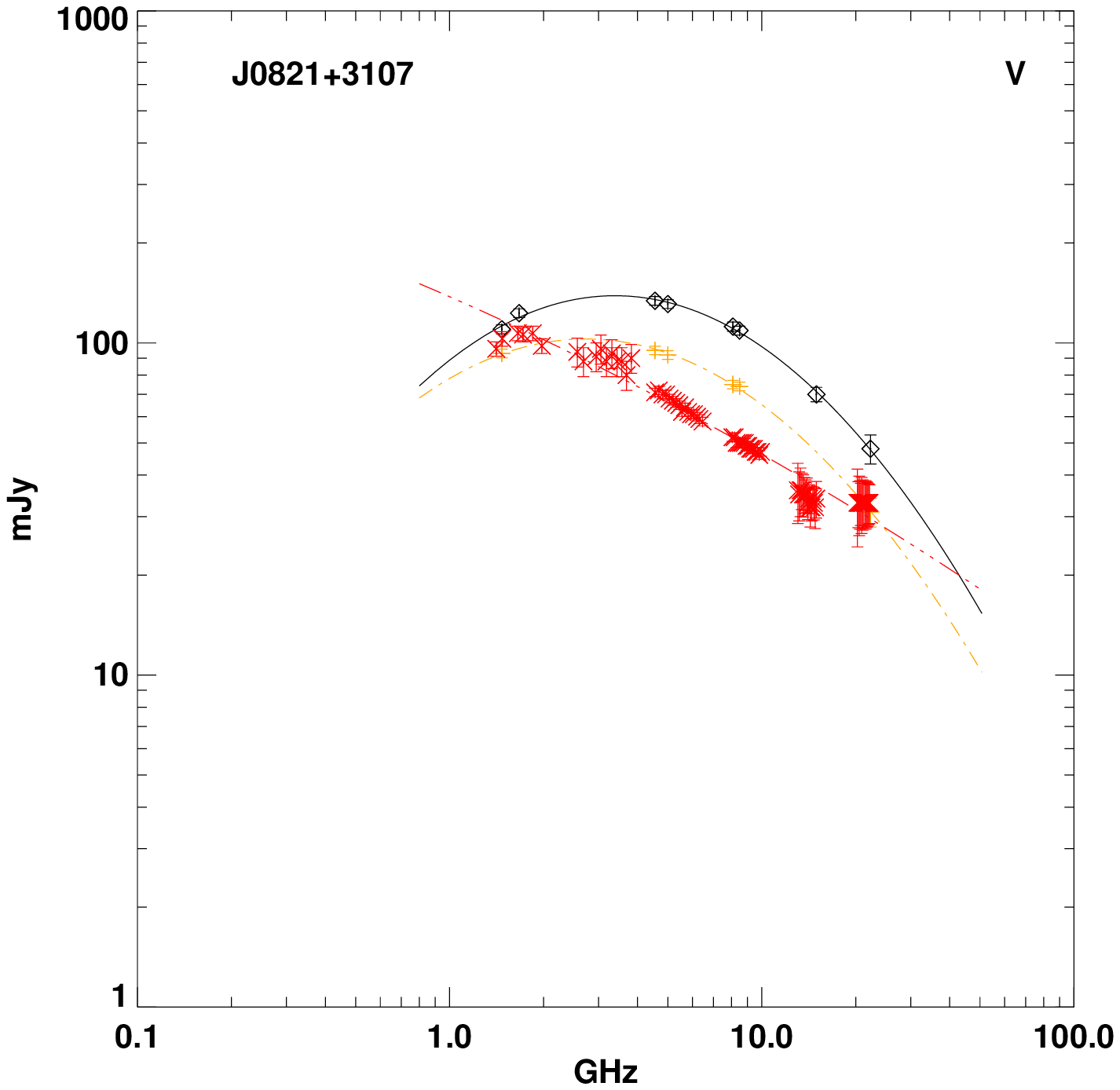}
\includegraphics{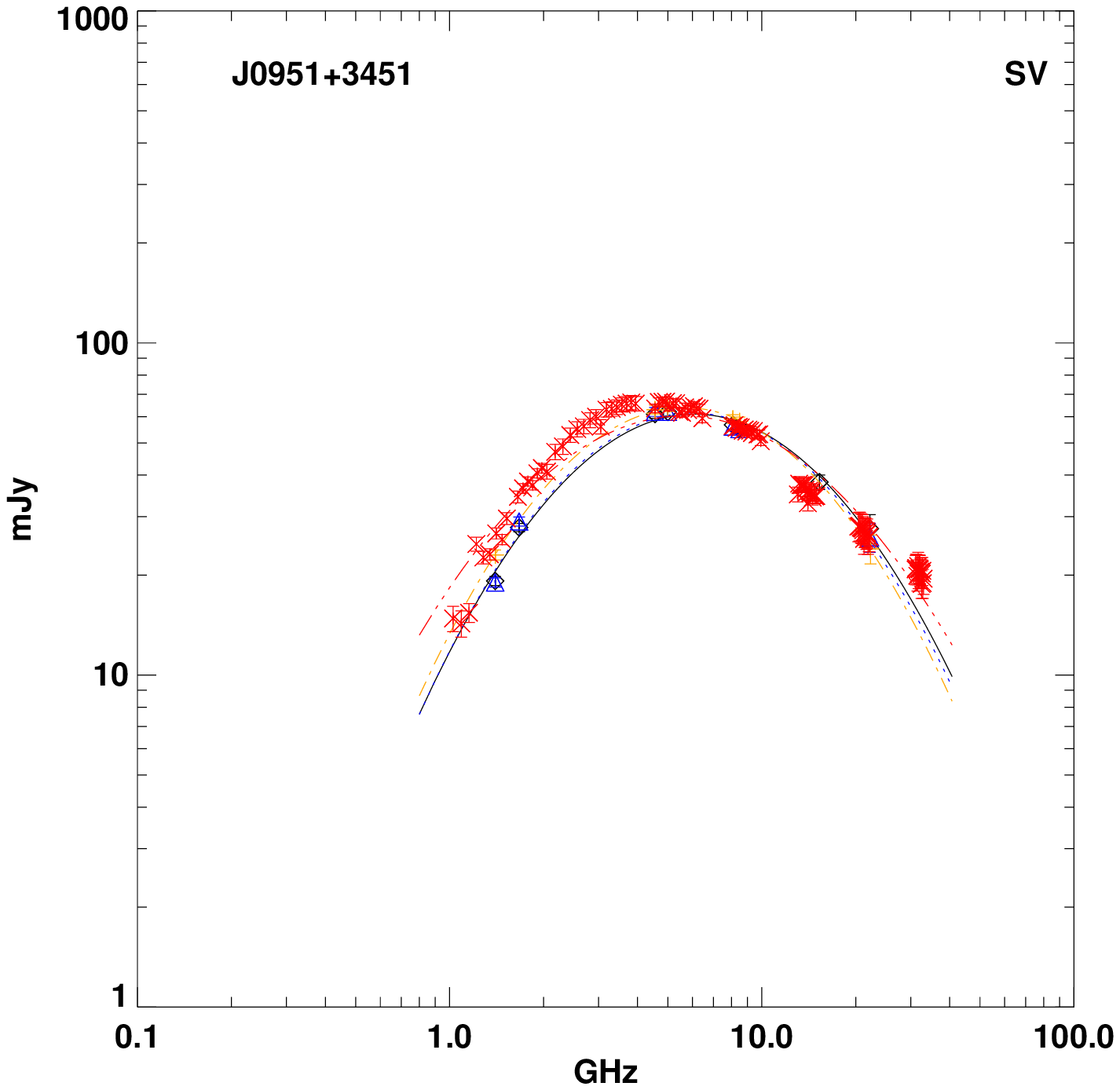}
\includegraphics{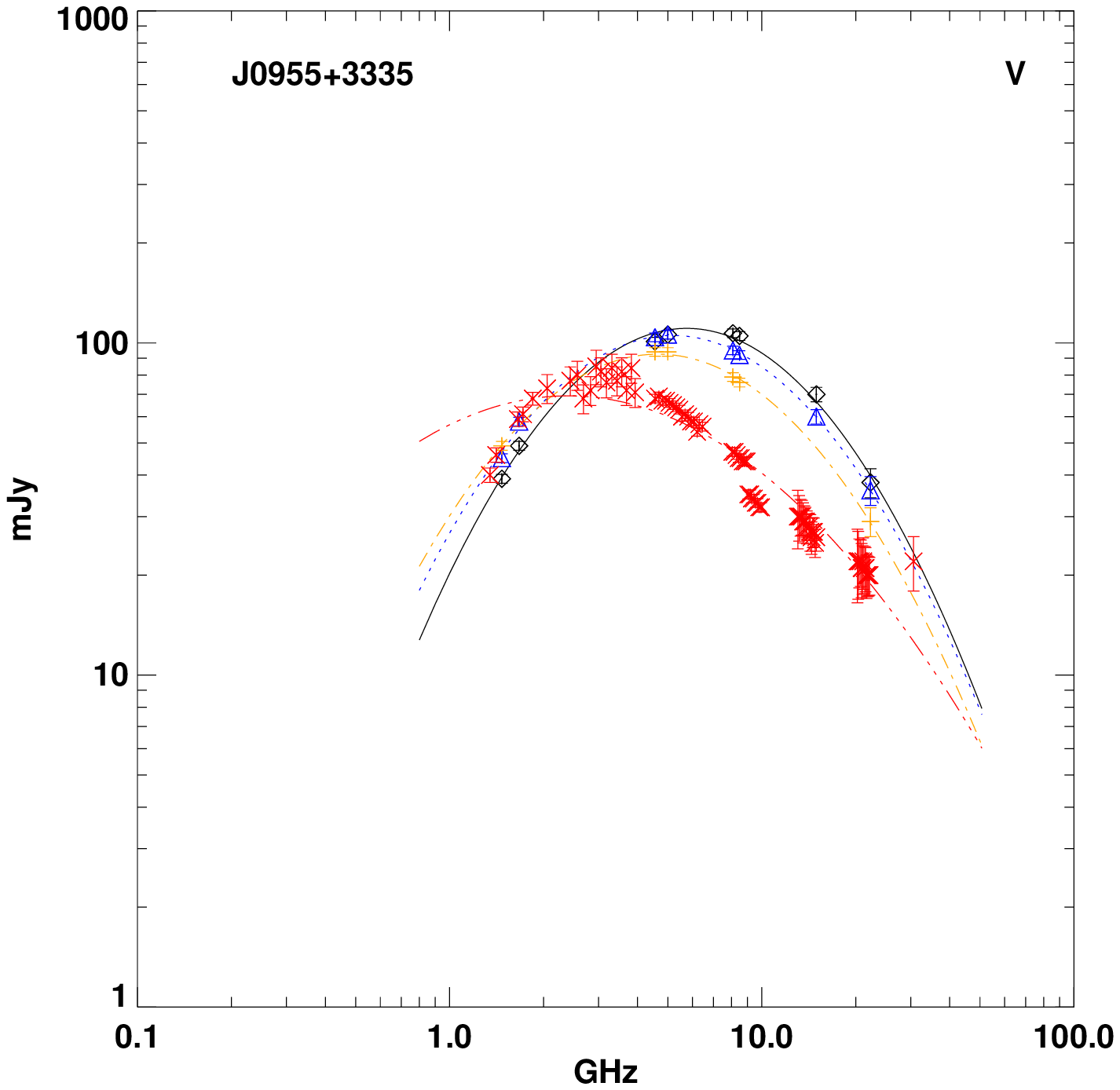}
\includegraphics{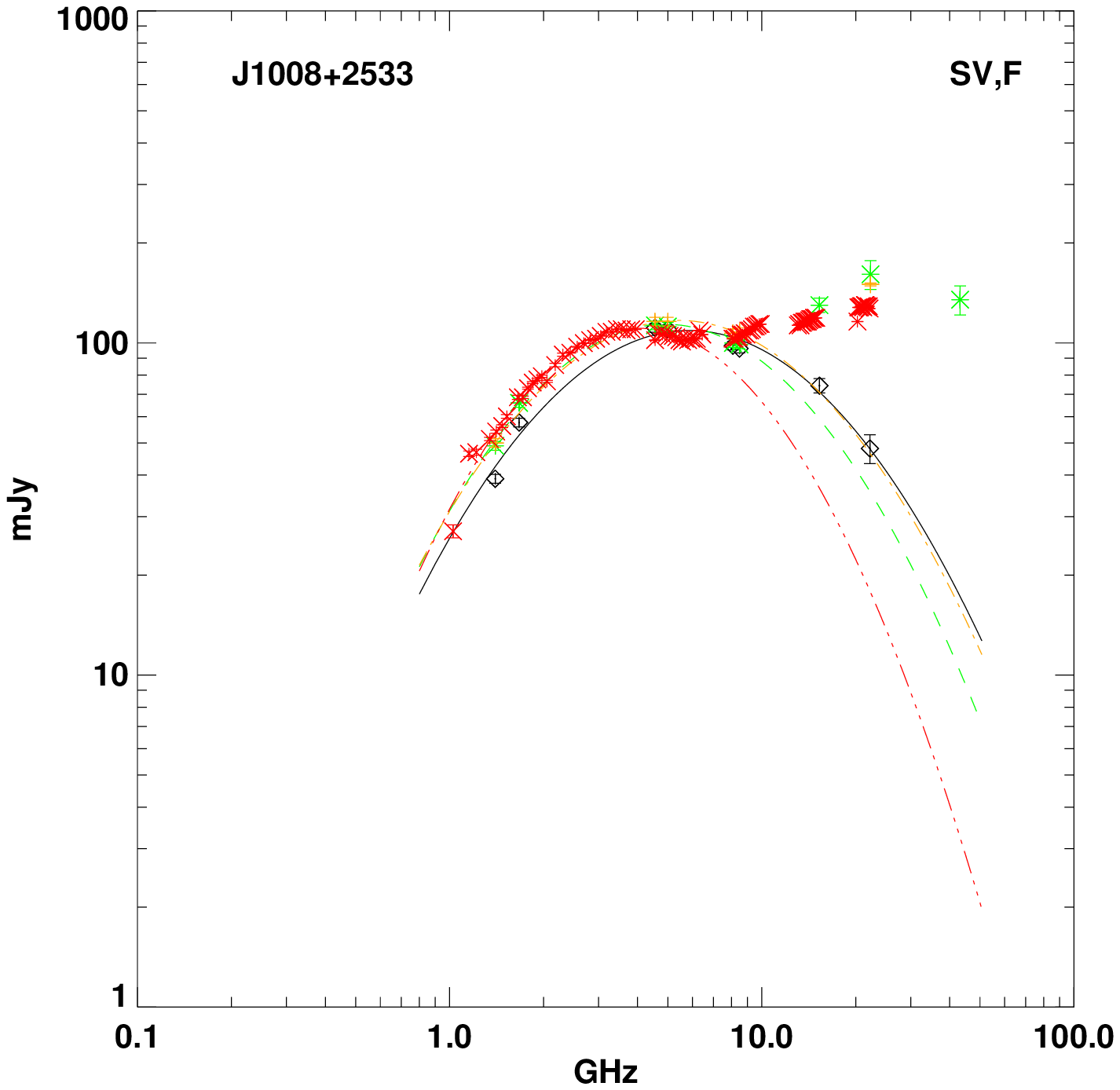}
\includegraphics{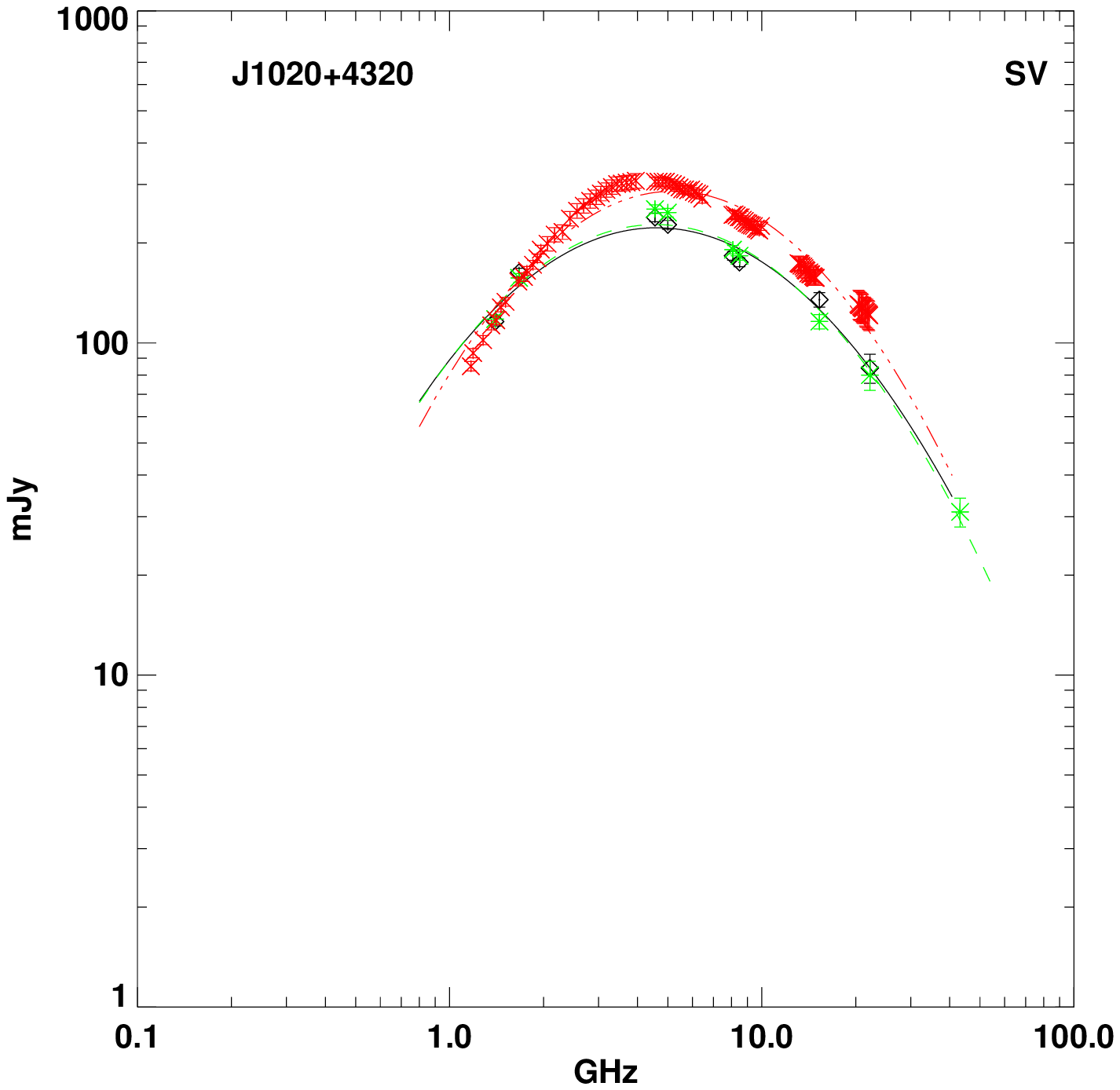}
\includegraphics{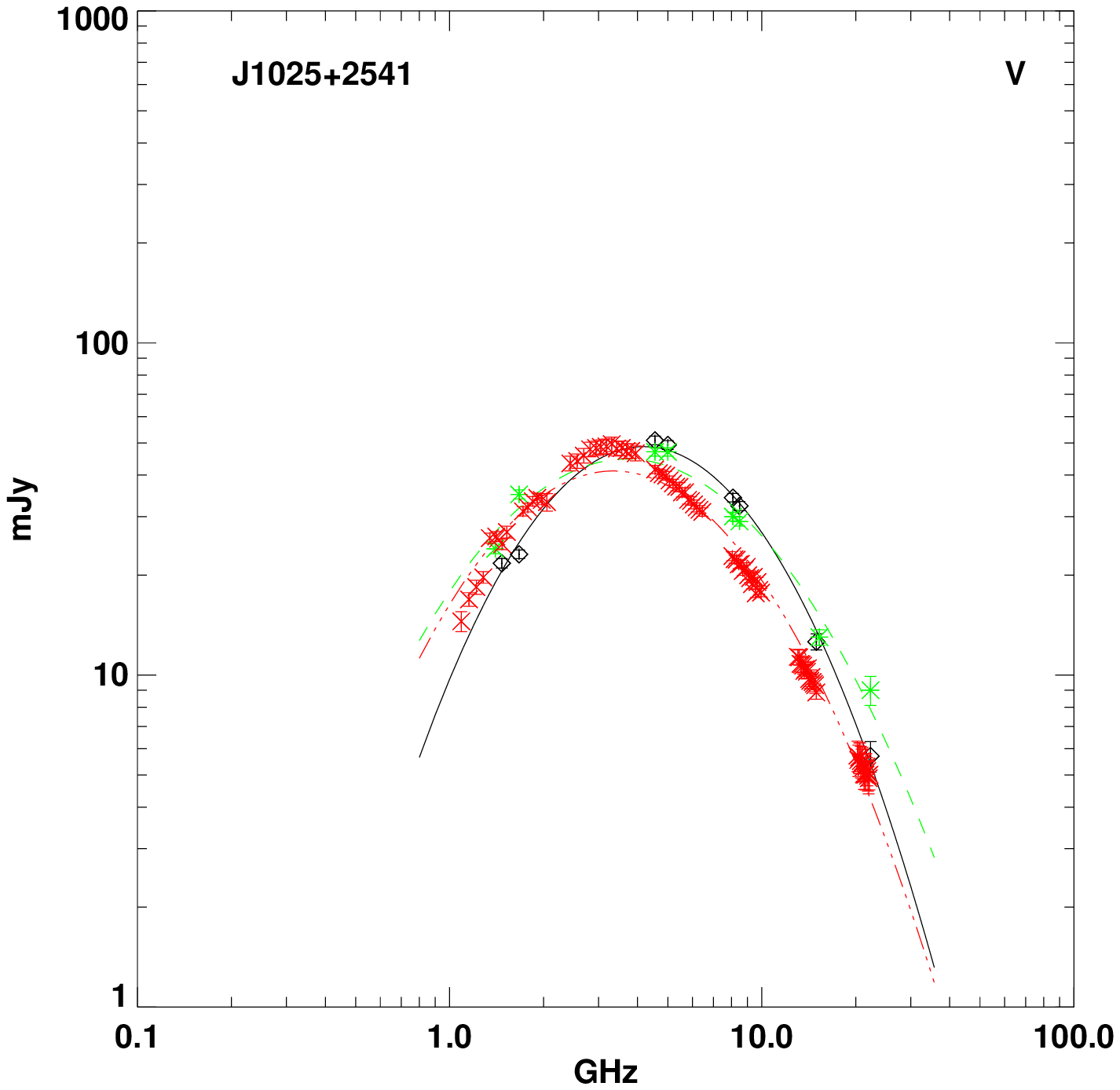}
\includegraphics{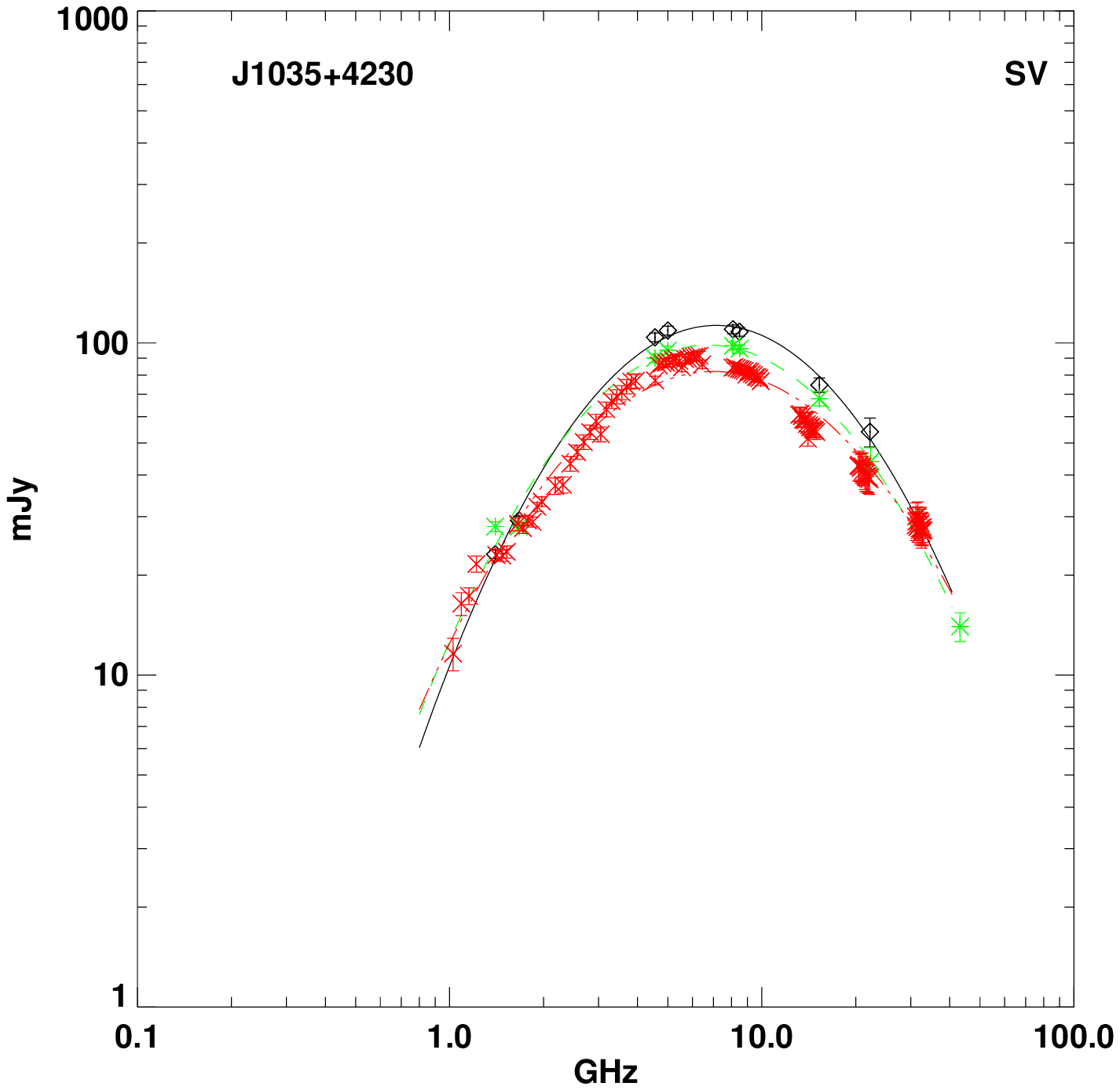}
\vspace{20cm}
\caption{Radio spectra of the 35 HFPs from the ``faint'' HFP
  sample observed with the VLA during the observing runs presented in
  this paper. Black diamonds and black solid line refer to the first
  epoch observations \citep[1998-2000,][]{cstan09}; blue triangles and
  a blue dotted
  line refer to observations in 2003; green asterisks and a green dashed line
  refer to observations in 2004; orange $+$ signs and orange dash-dot line refer to
  observations in 2006/2007; red crosses and red dash-dot-dot line refer to
  the last
 epoch (2012). Flux densities and precise observing dates for the
 epoch 2003, 2004, and 2006-2007 can be found in \citet{mo10}. 
The designations V, SV, and NV, mean that the source is
   classified as variable, slightly variable, and non variable,
   respectively, while F indicates a flat radio spectrum during at
   least one epoch (see Section 3.2).} 
\label{radio_spectra}
\end{center}
\end{figure*}

\addtocounter{figure}{-1}
\begin{figure*}
\begin{center}
\includegraphics{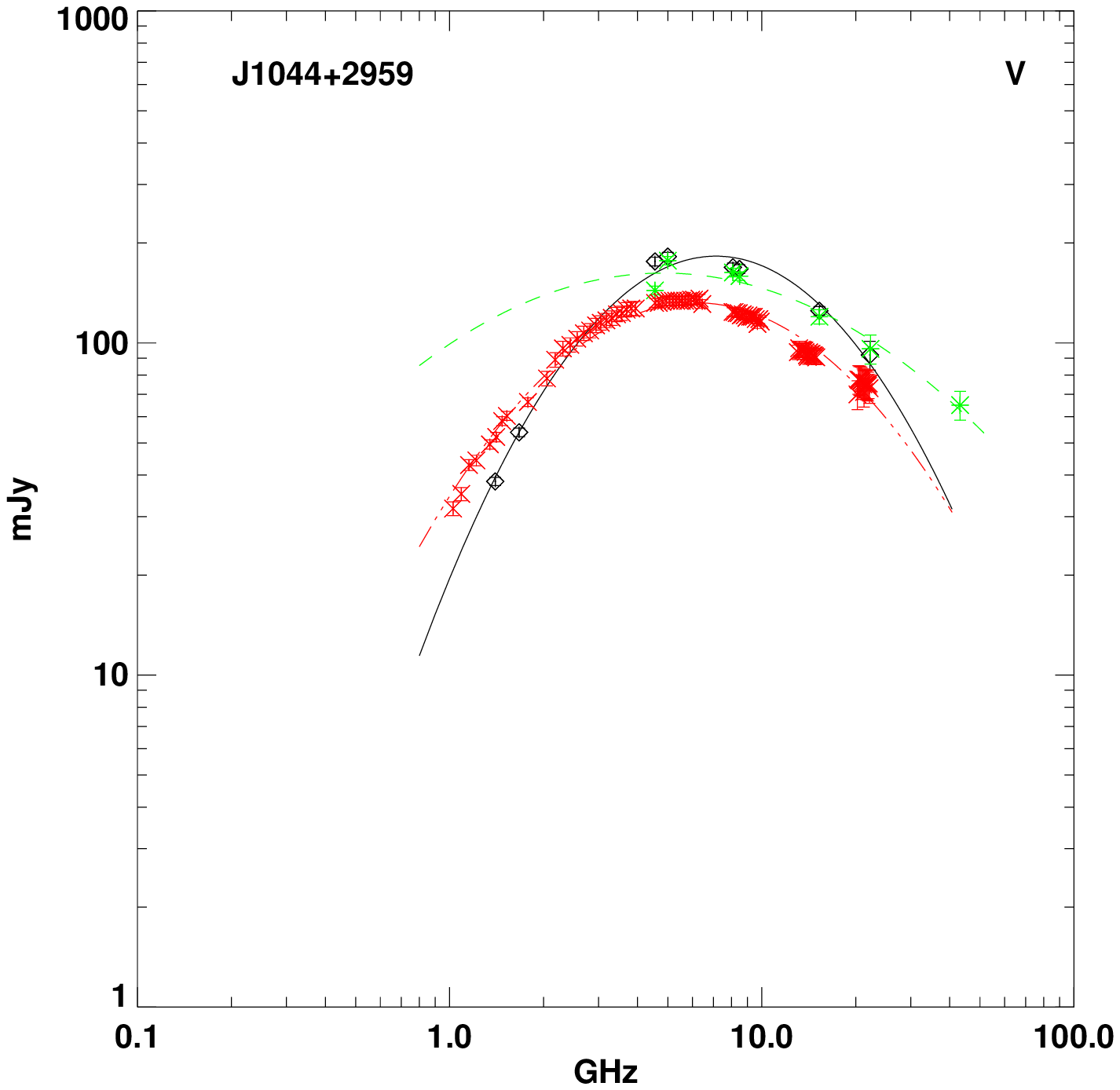}
\includegraphics{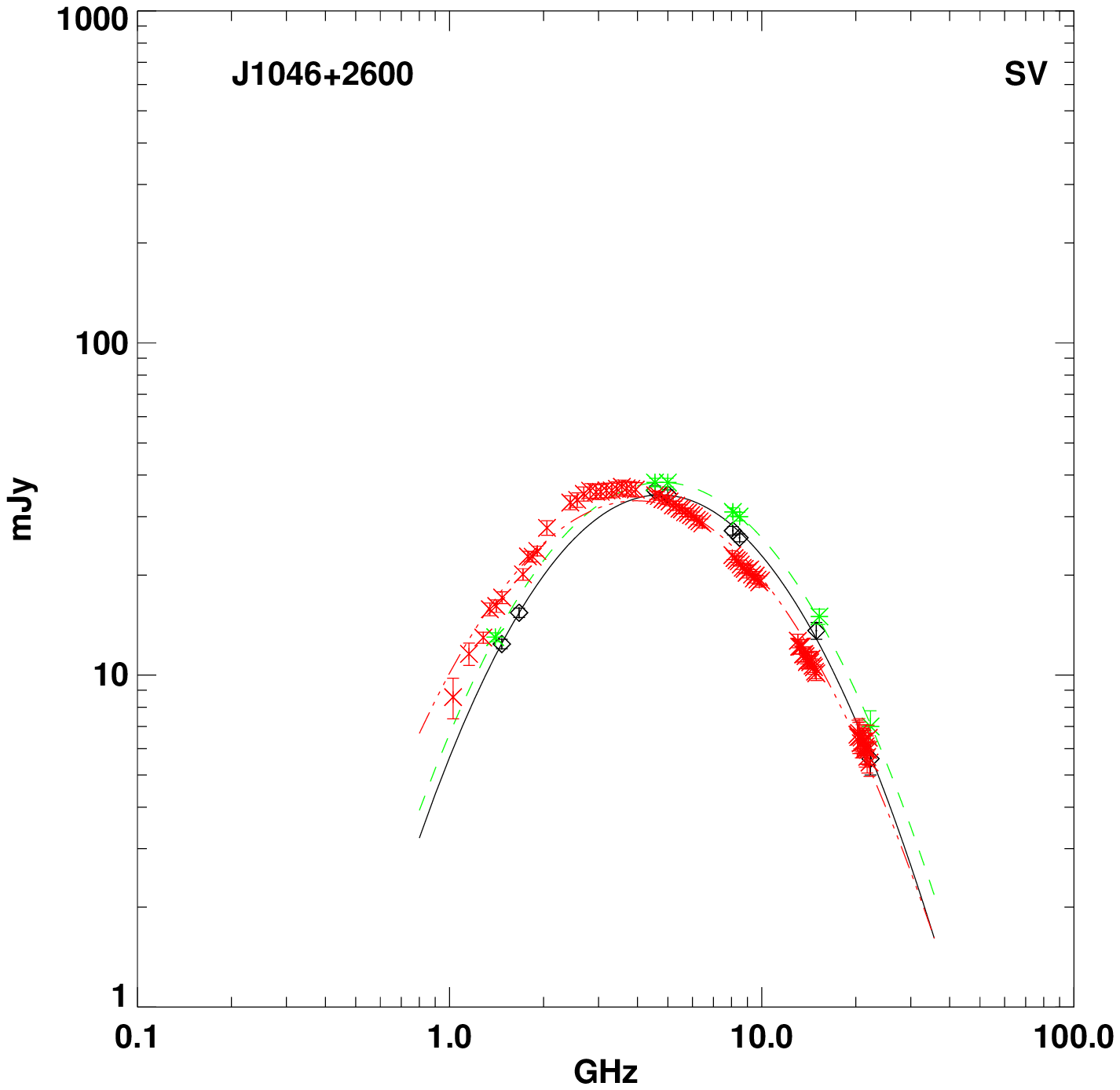}
\includegraphics{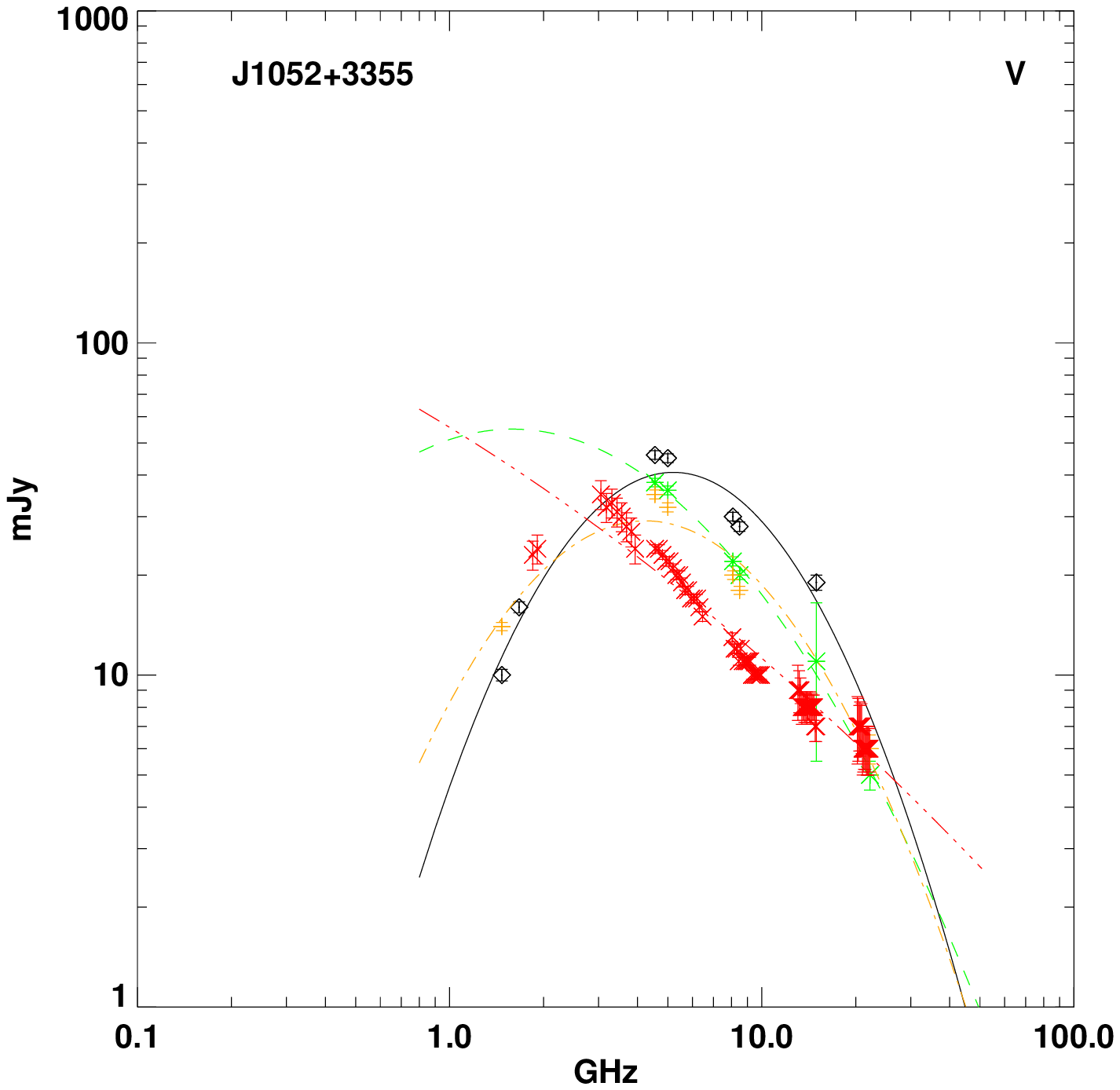}
\includegraphics{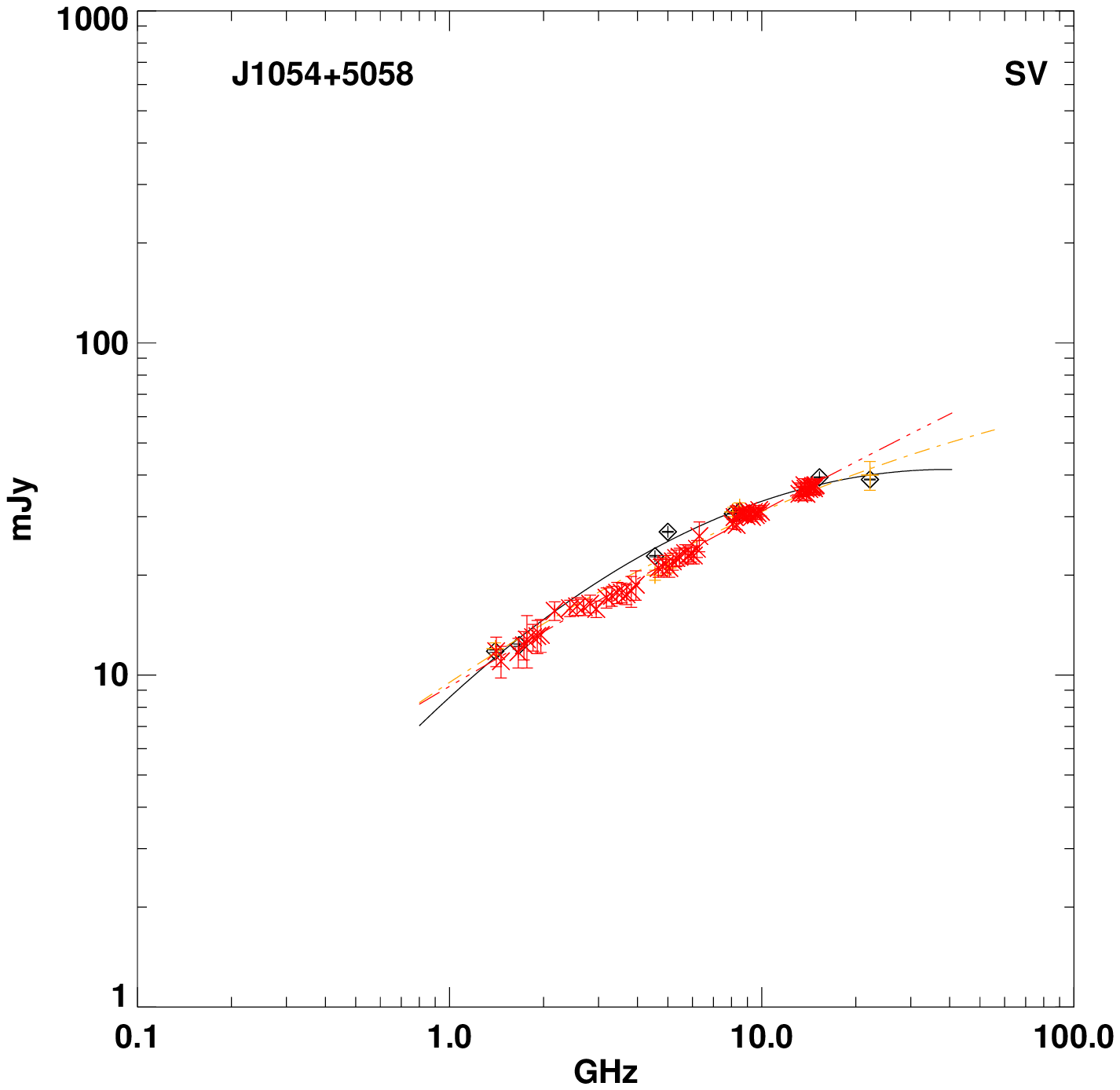}
\includegraphics{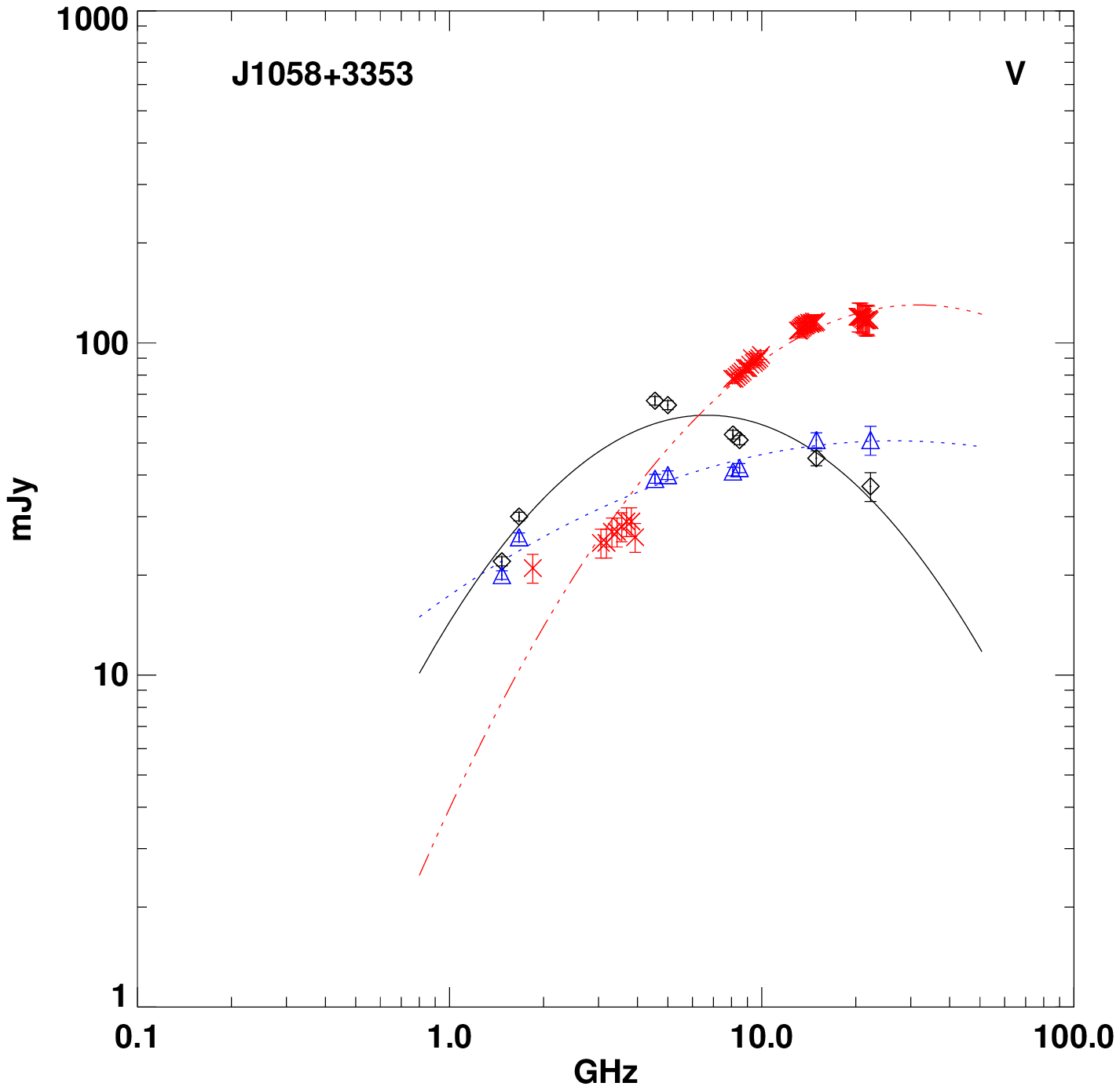}
\includegraphics{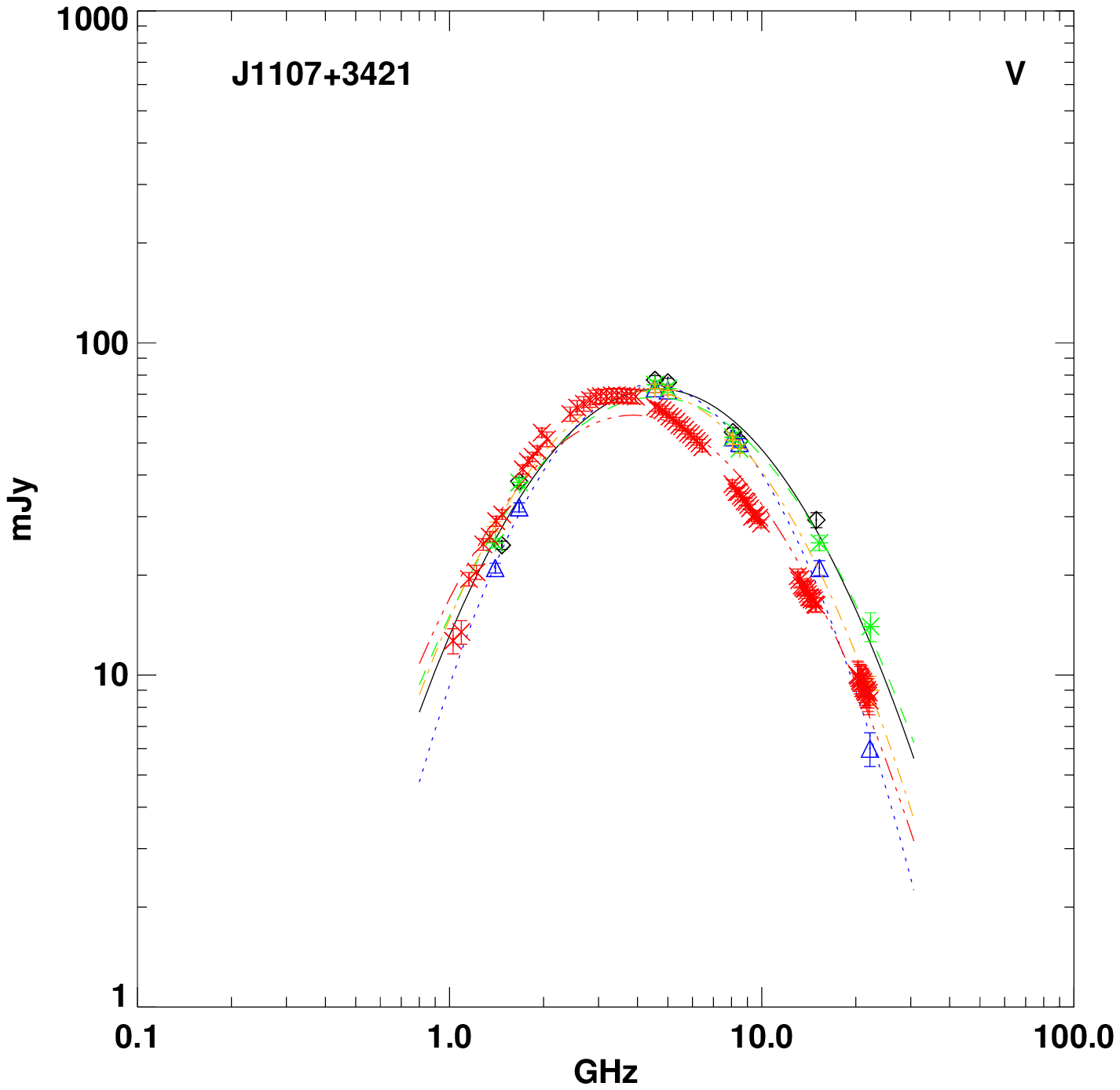}
\includegraphics{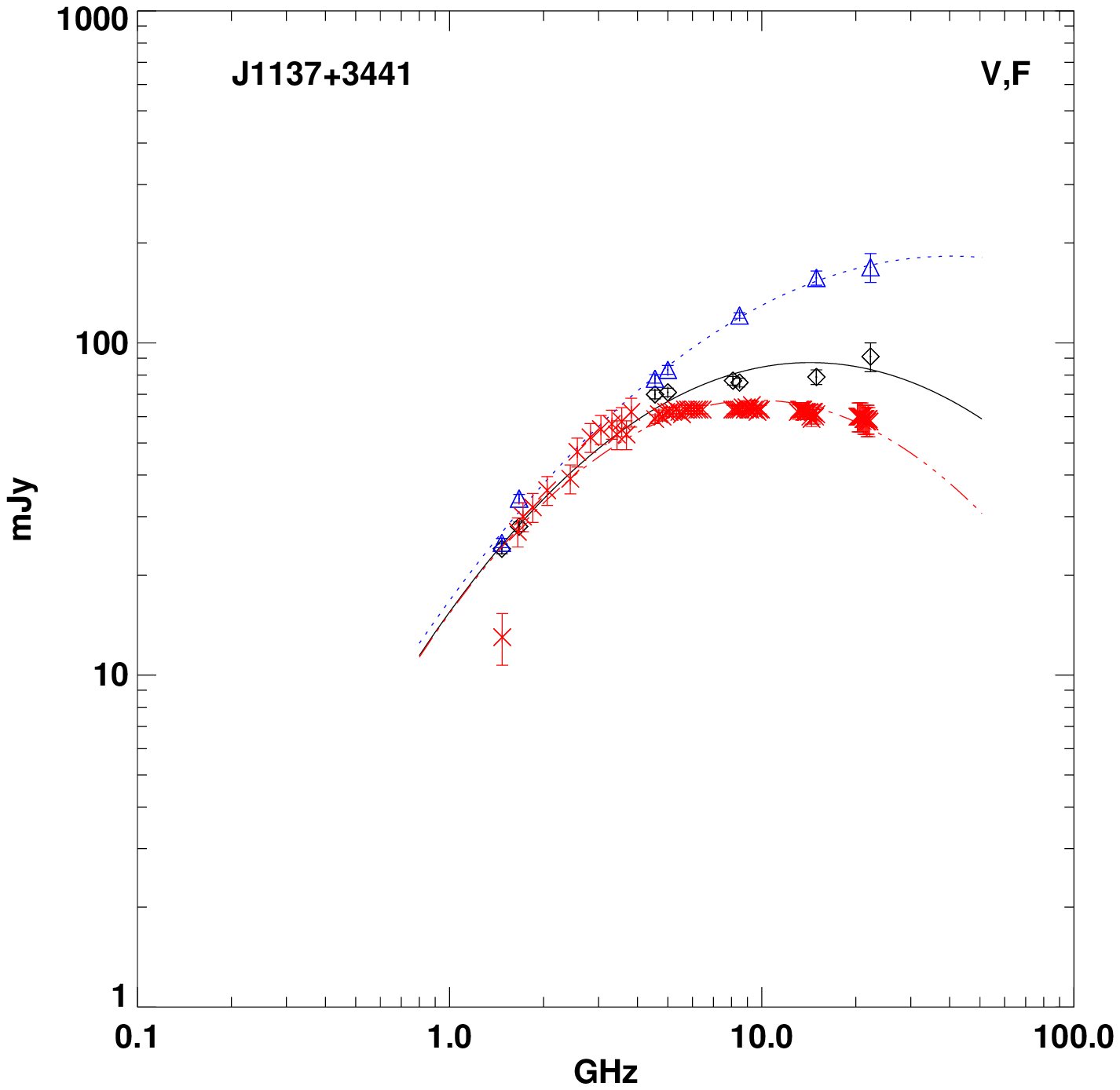}
\includegraphics{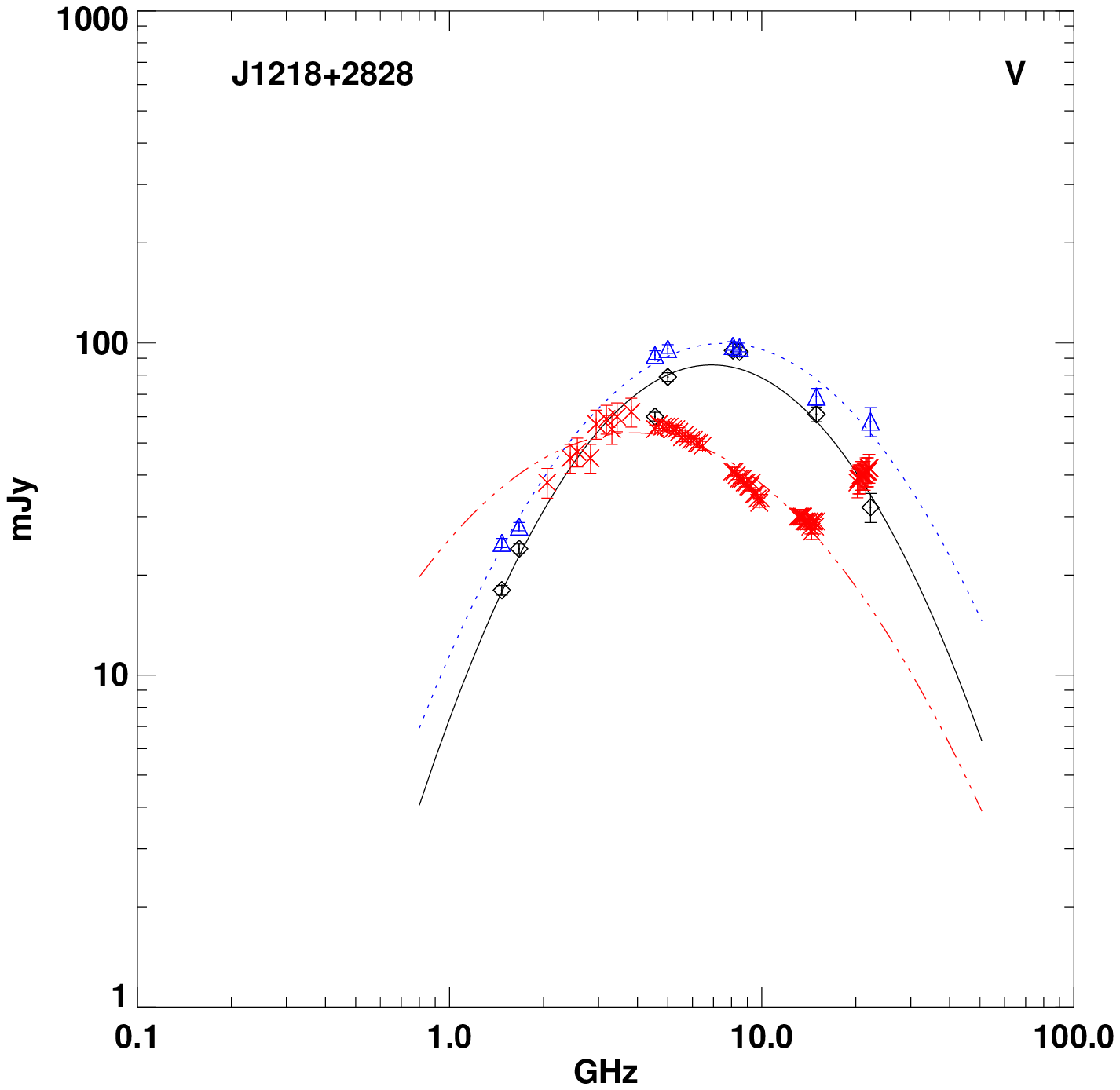}
\includegraphics{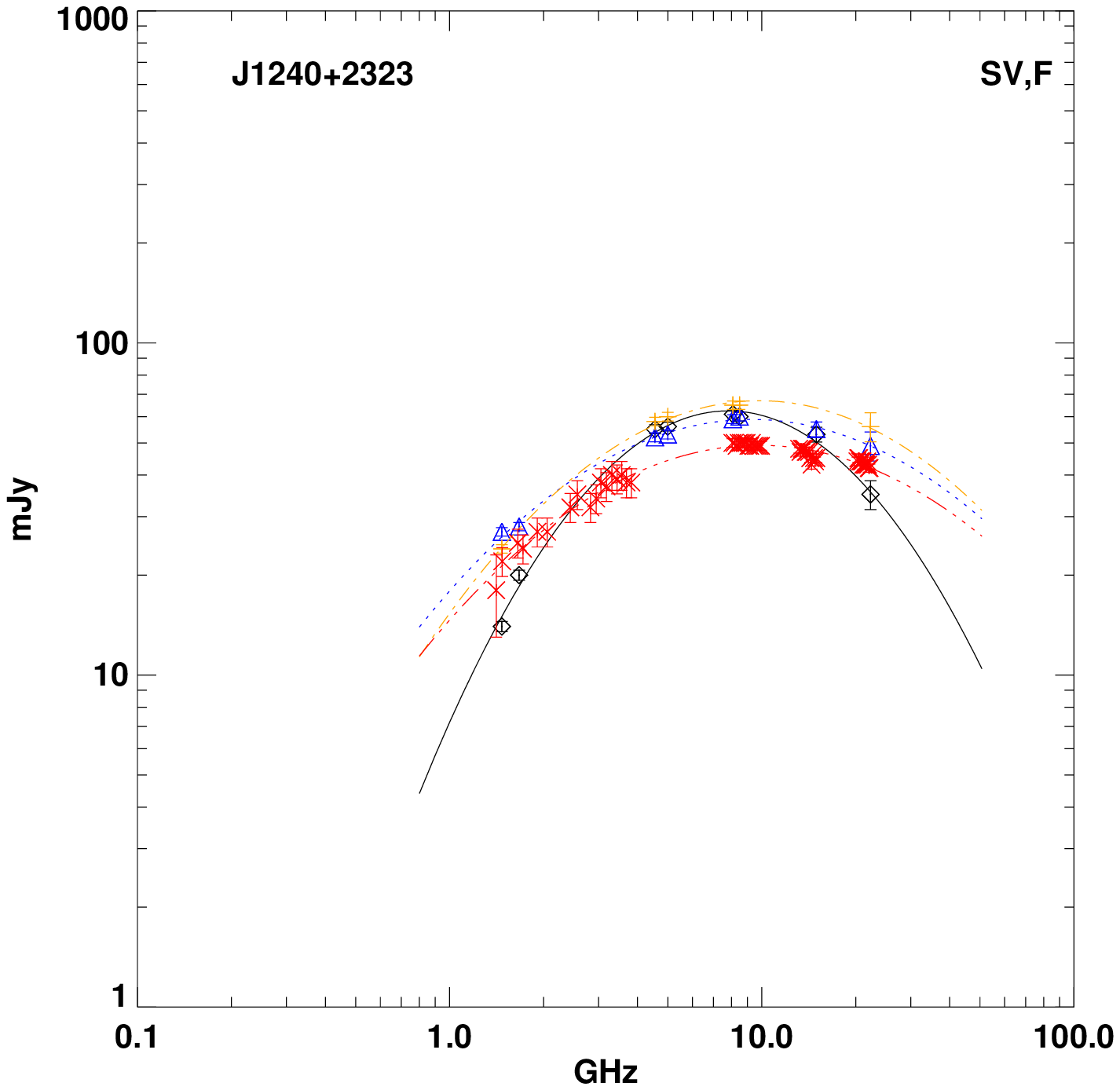}
\vspace{20cm}
\caption{Continued.}
\end{center}
\end{figure*}

\addtocounter{figure}{-1}
\begin{figure*}
\begin{center}
\includegraphics{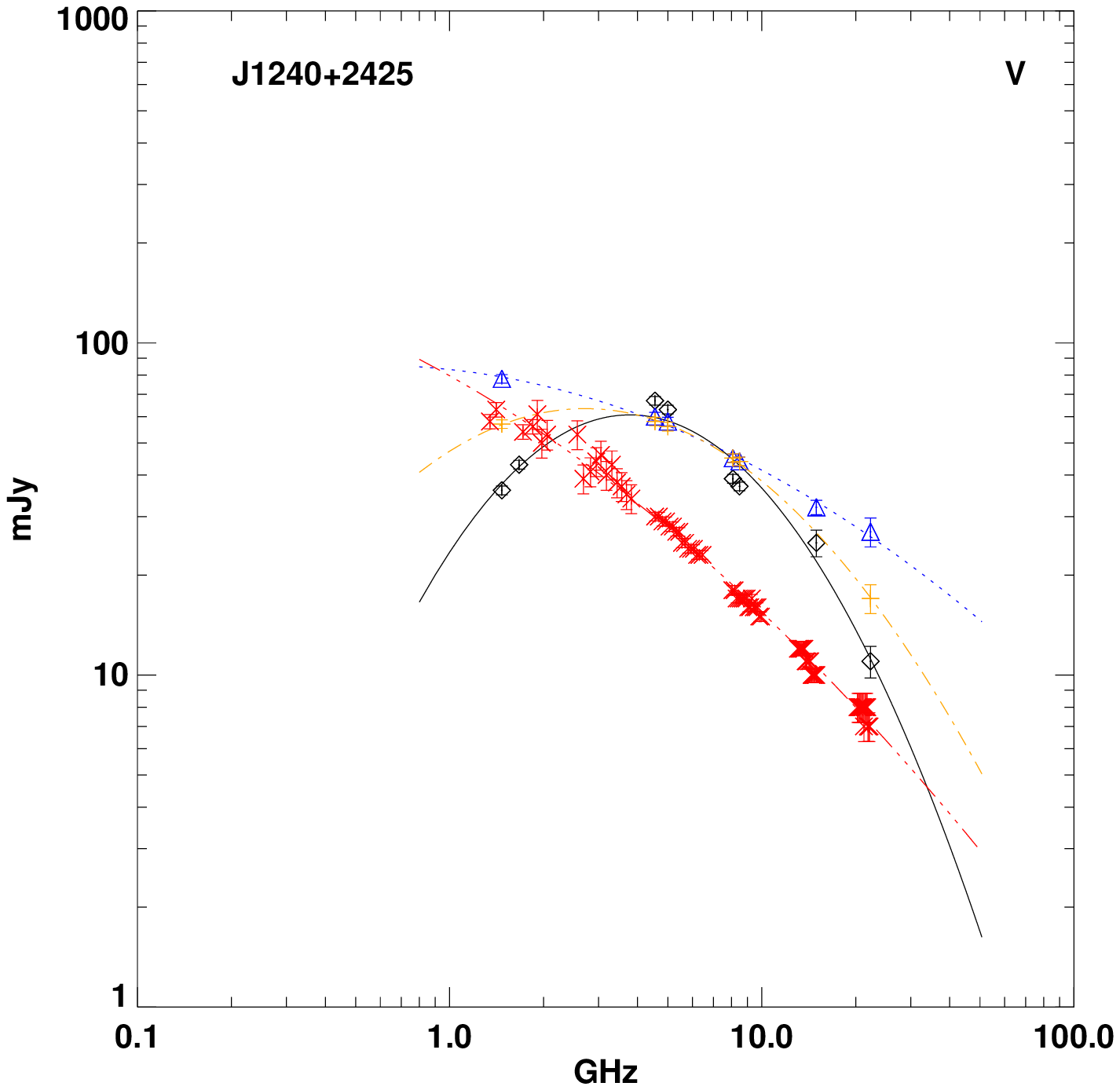}
\includegraphics{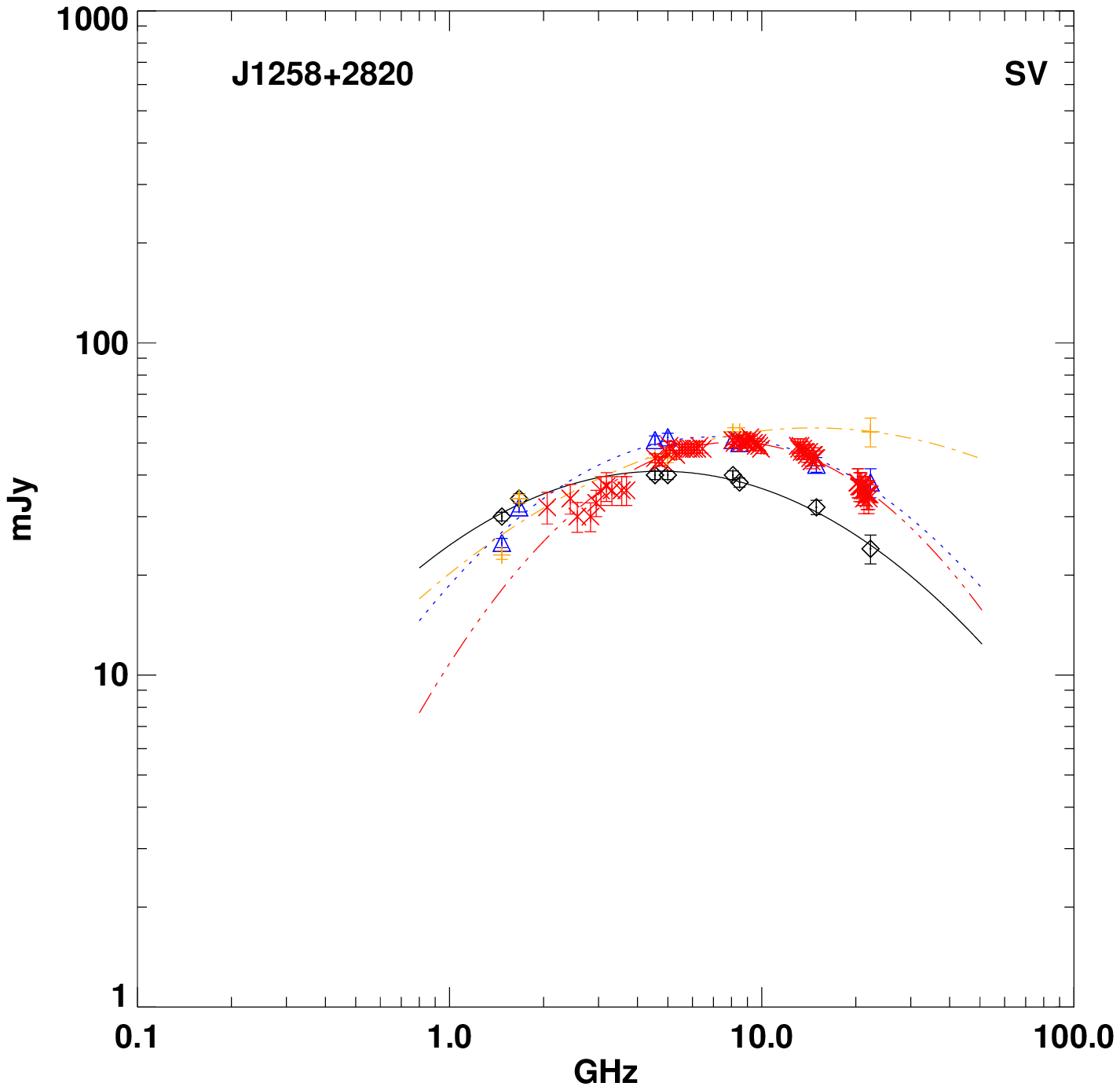}
\includegraphics{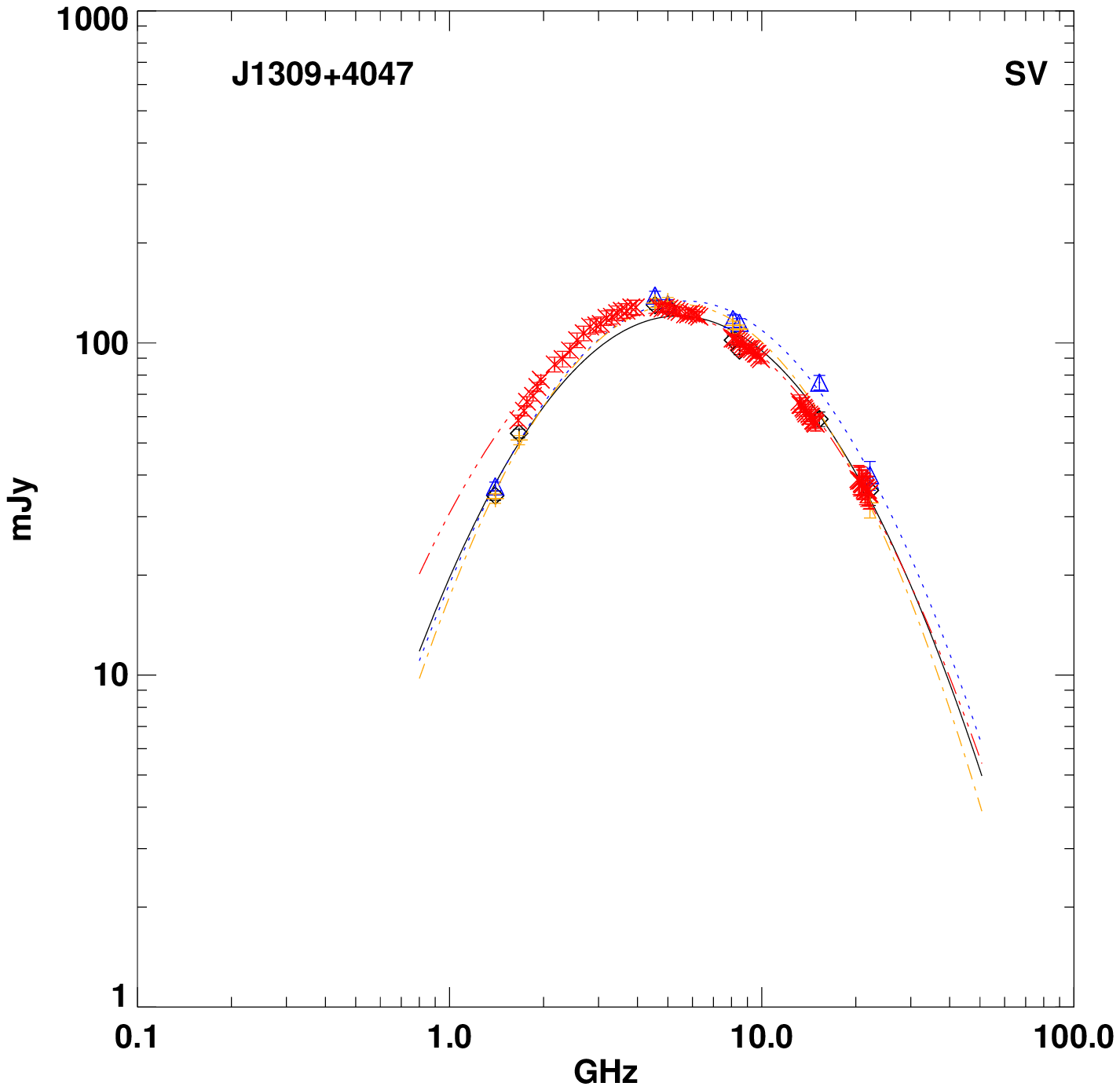}
\includegraphics{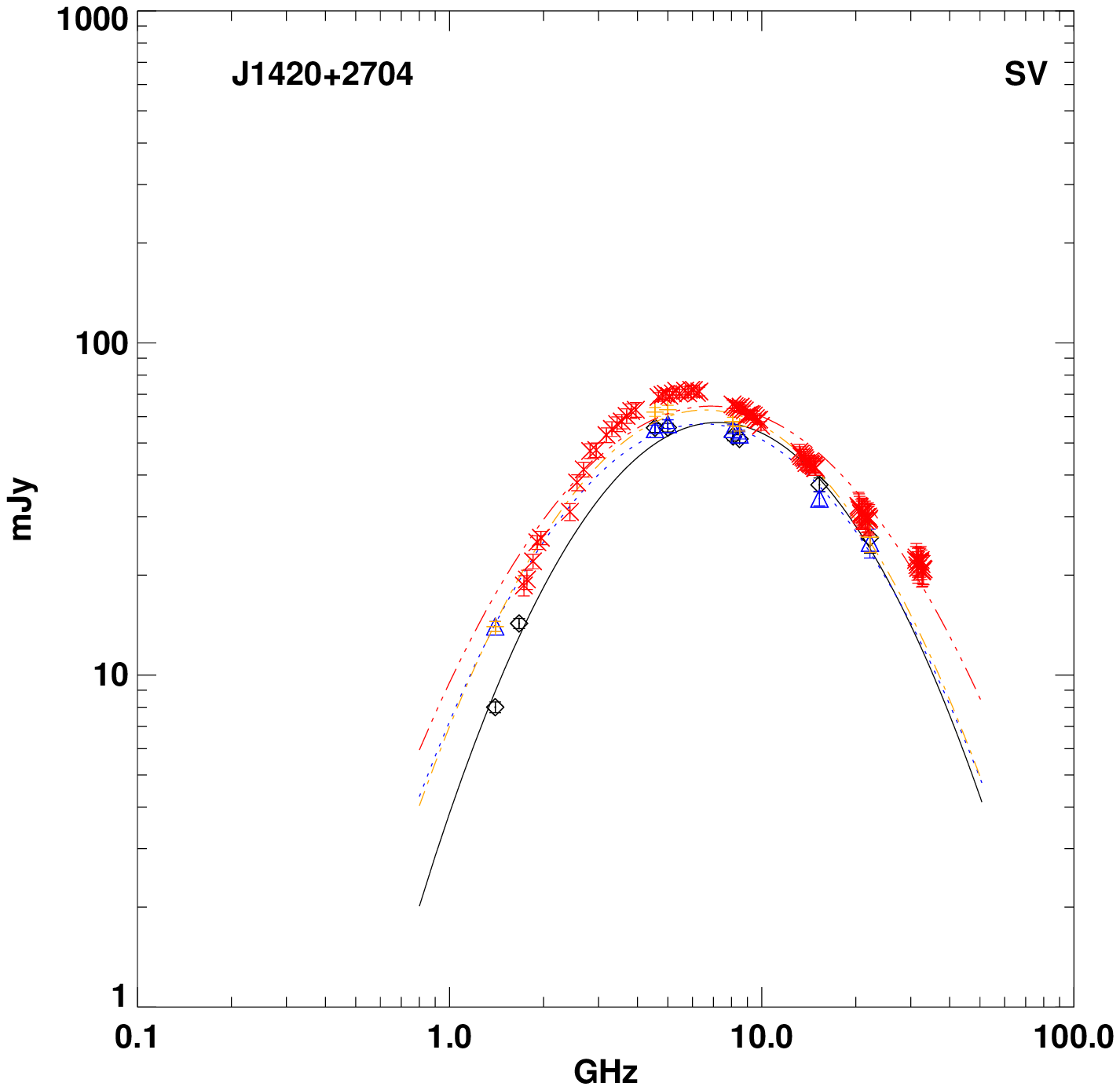}
\includegraphics{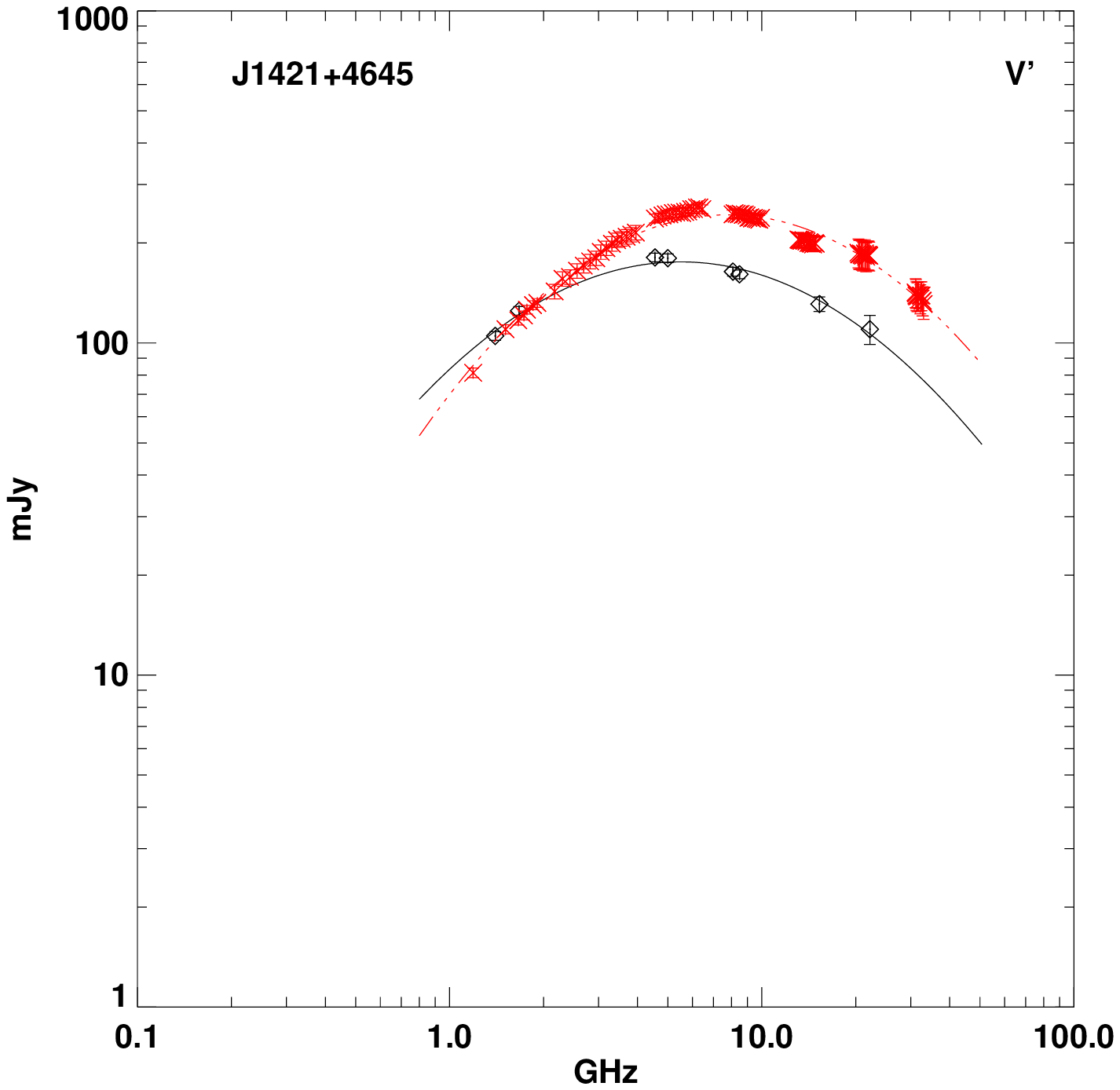}
\includegraphics{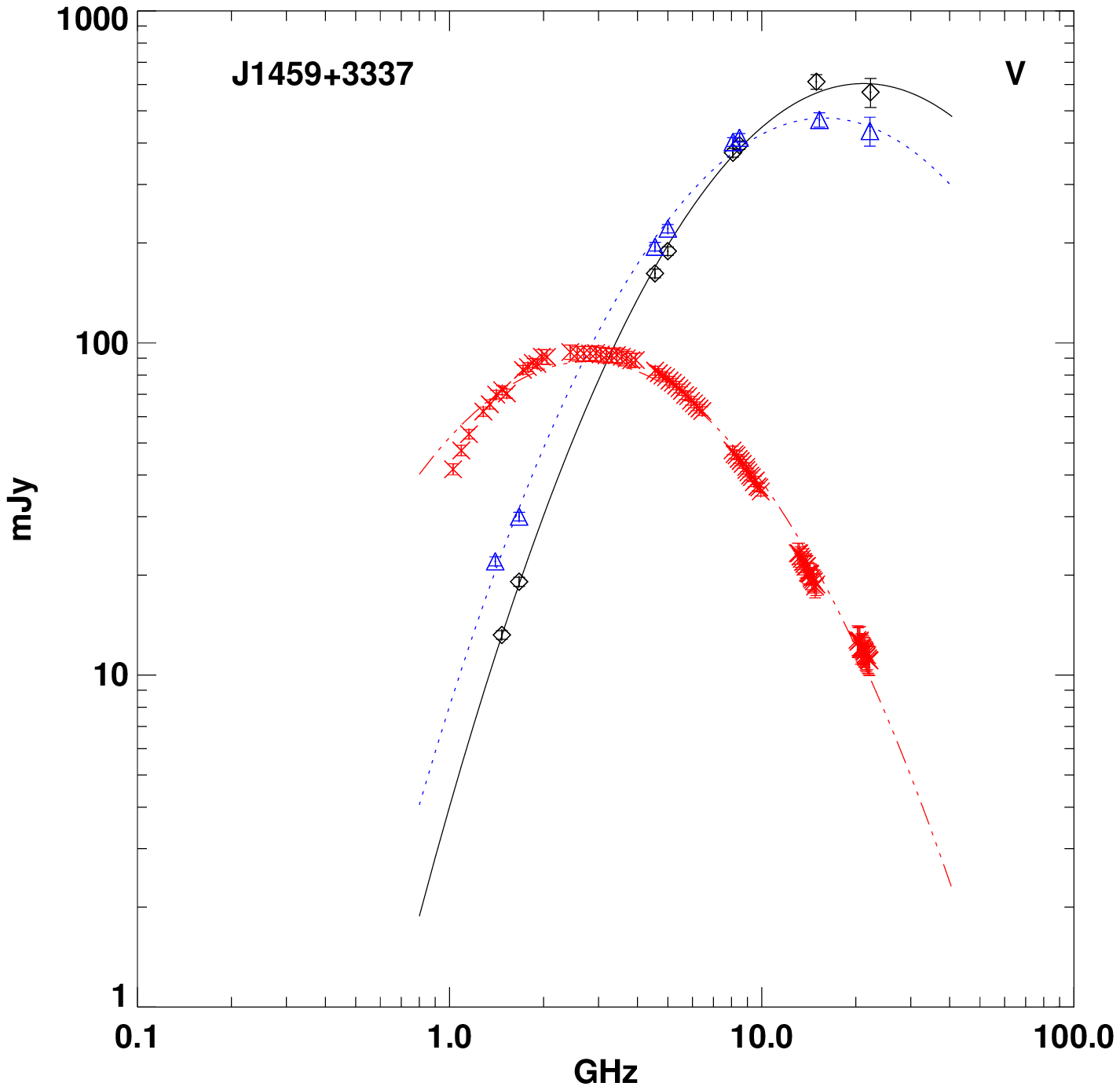}
\includegraphics{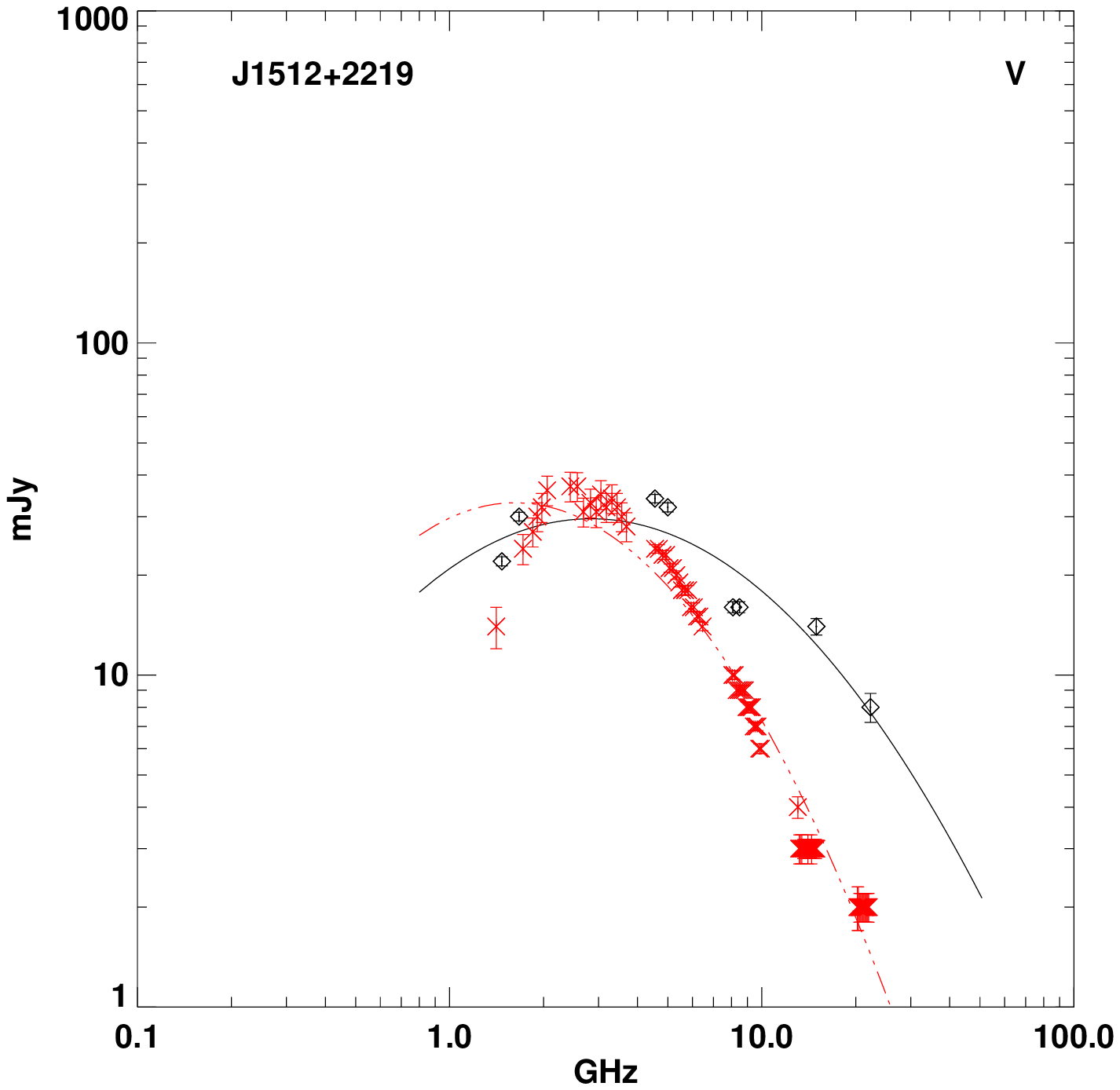}
\includegraphics{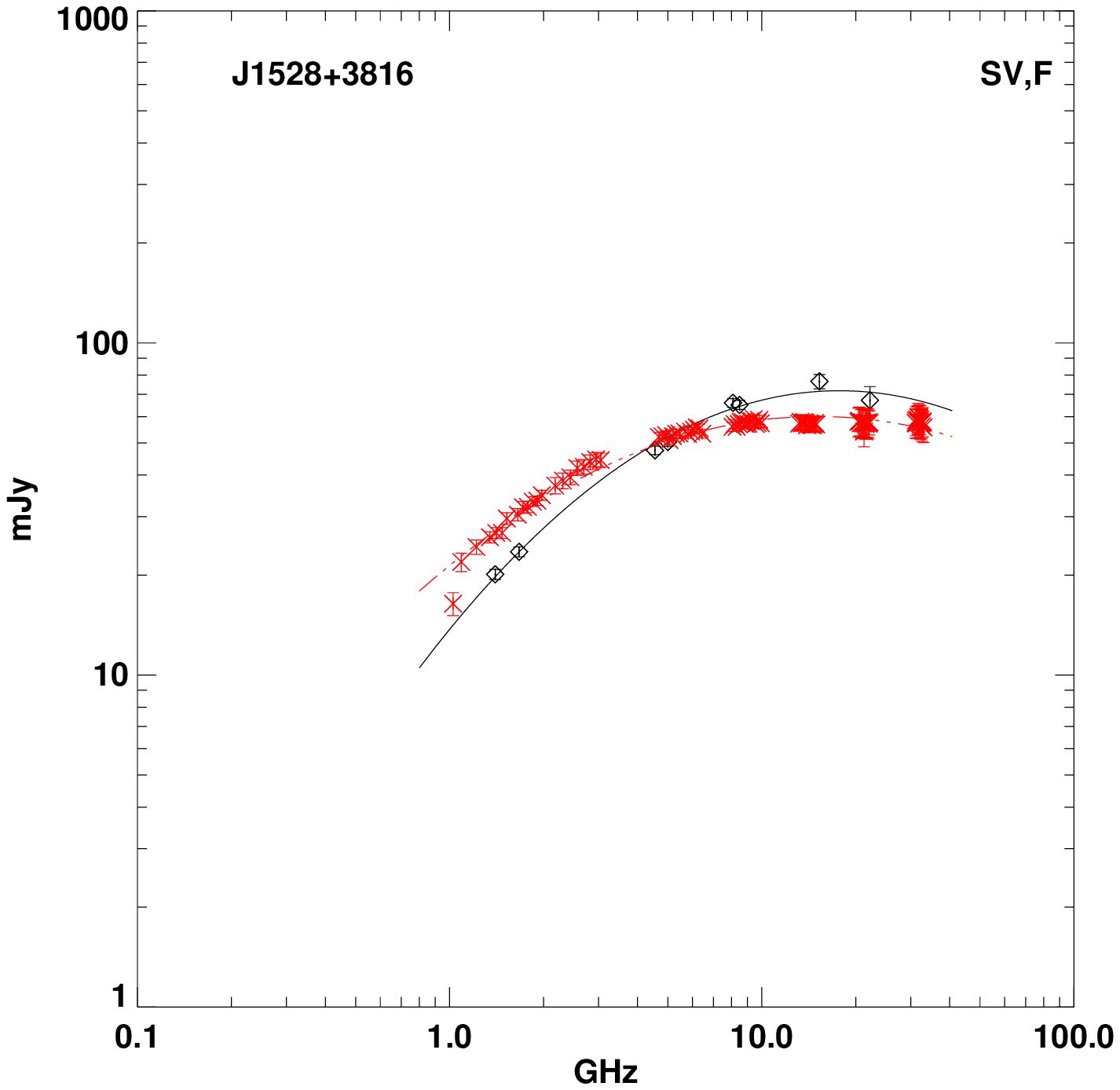}
\includegraphics{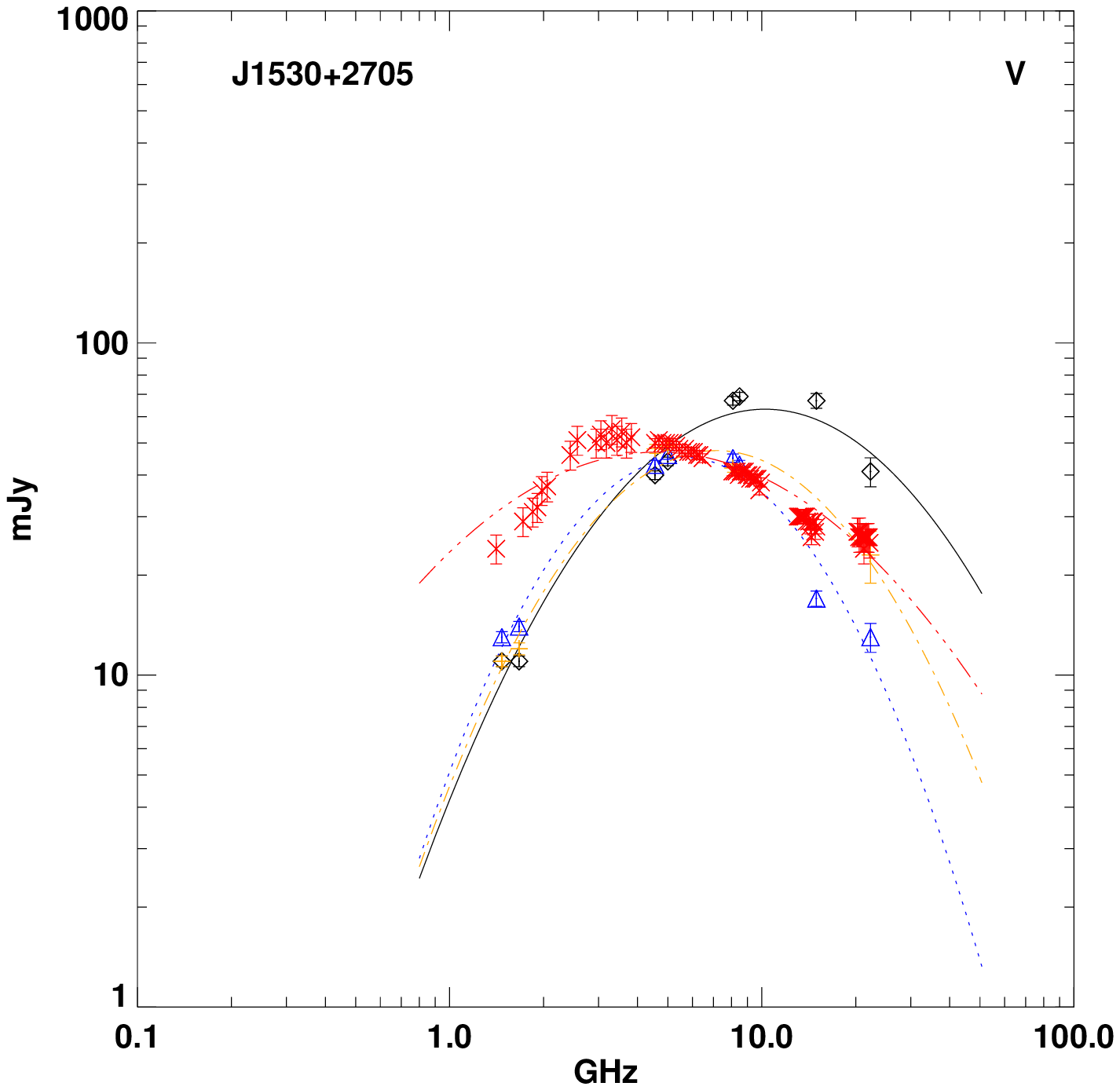}
\vspace{20cm}
\caption{Continued.}
\end{center}
\end{figure*}

\addtocounter{figure}{-1}
\begin{figure*}
\begin{center}
\includegraphics{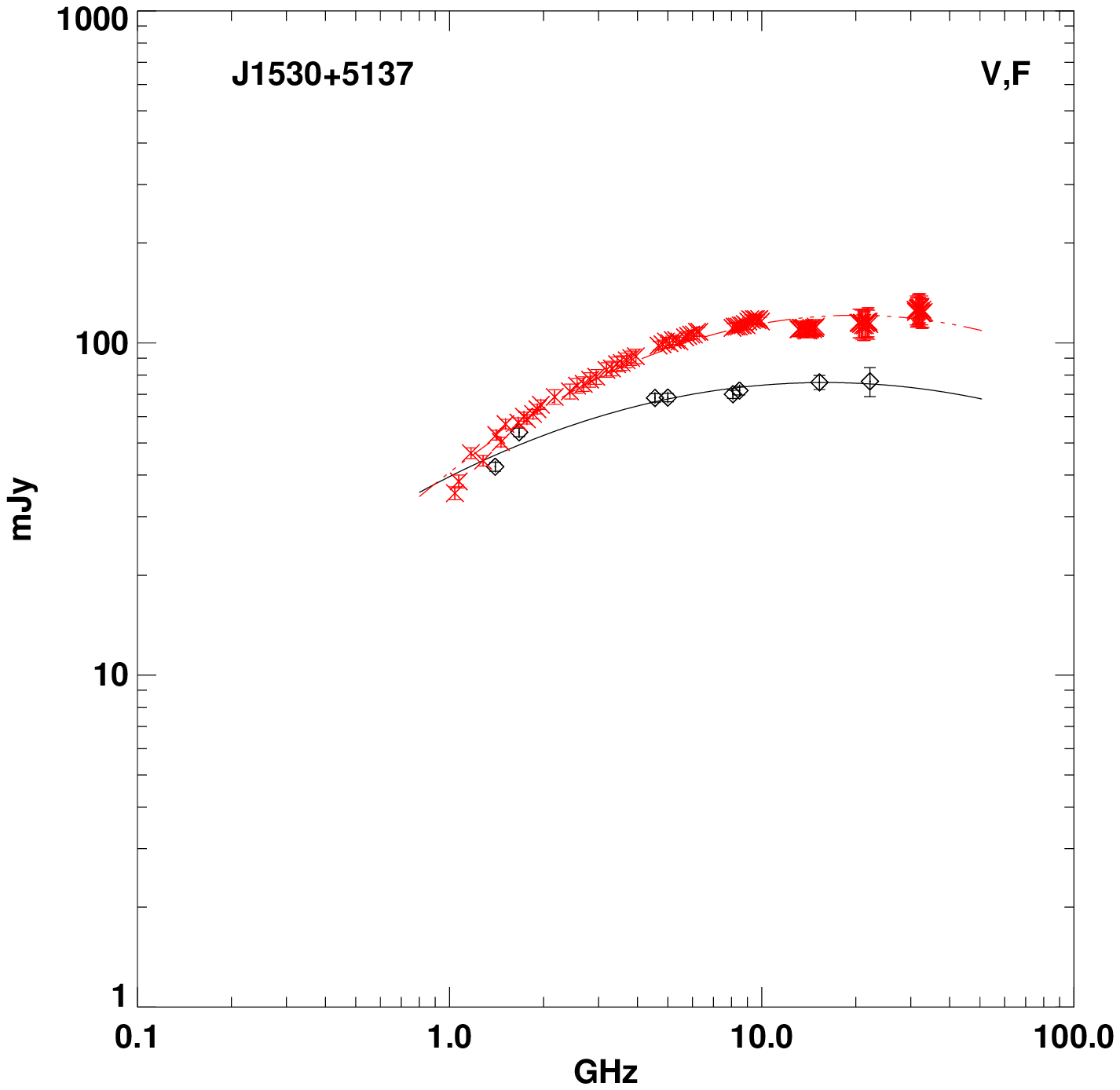}
\includegraphics{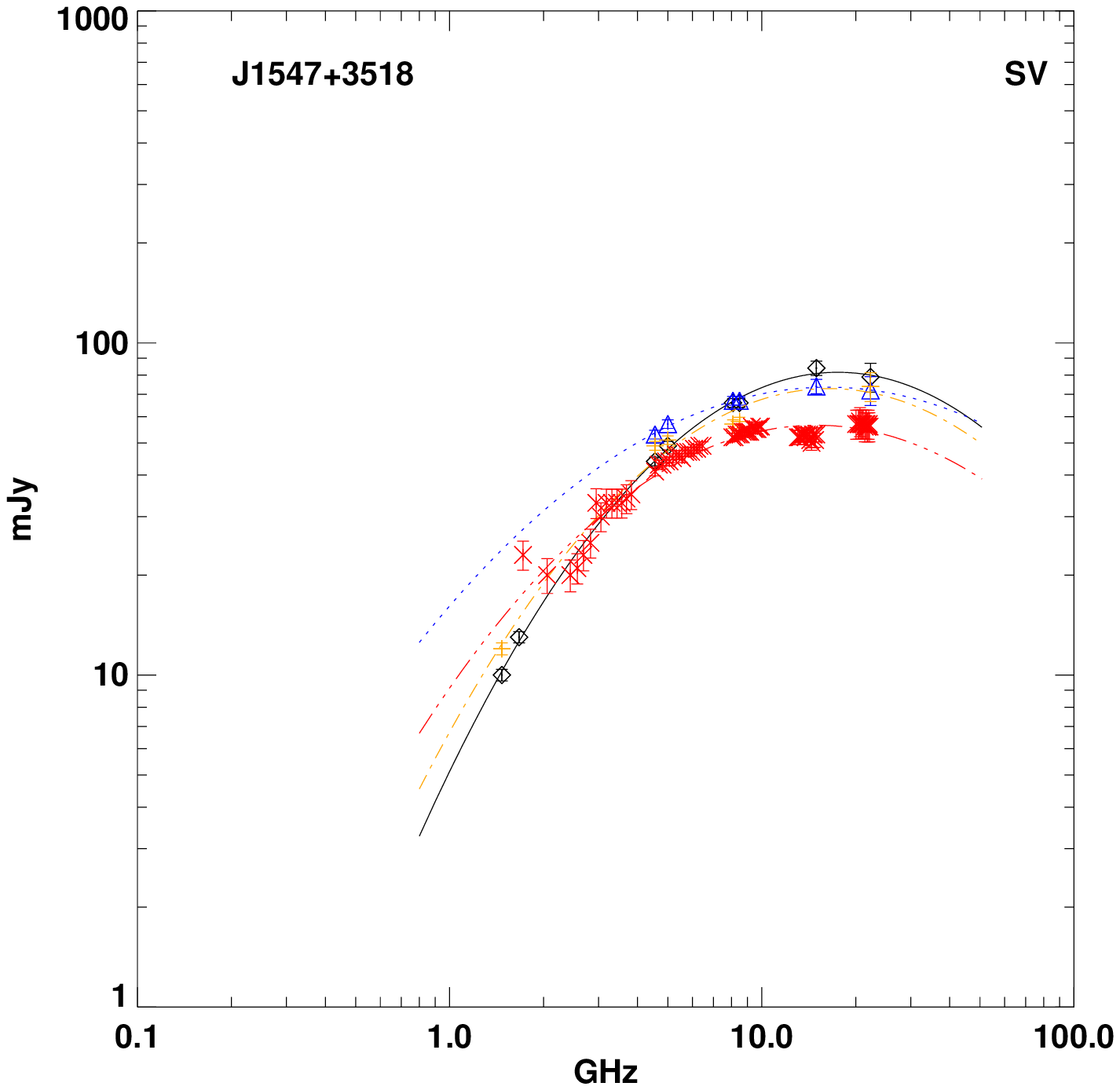}
\includegraphics{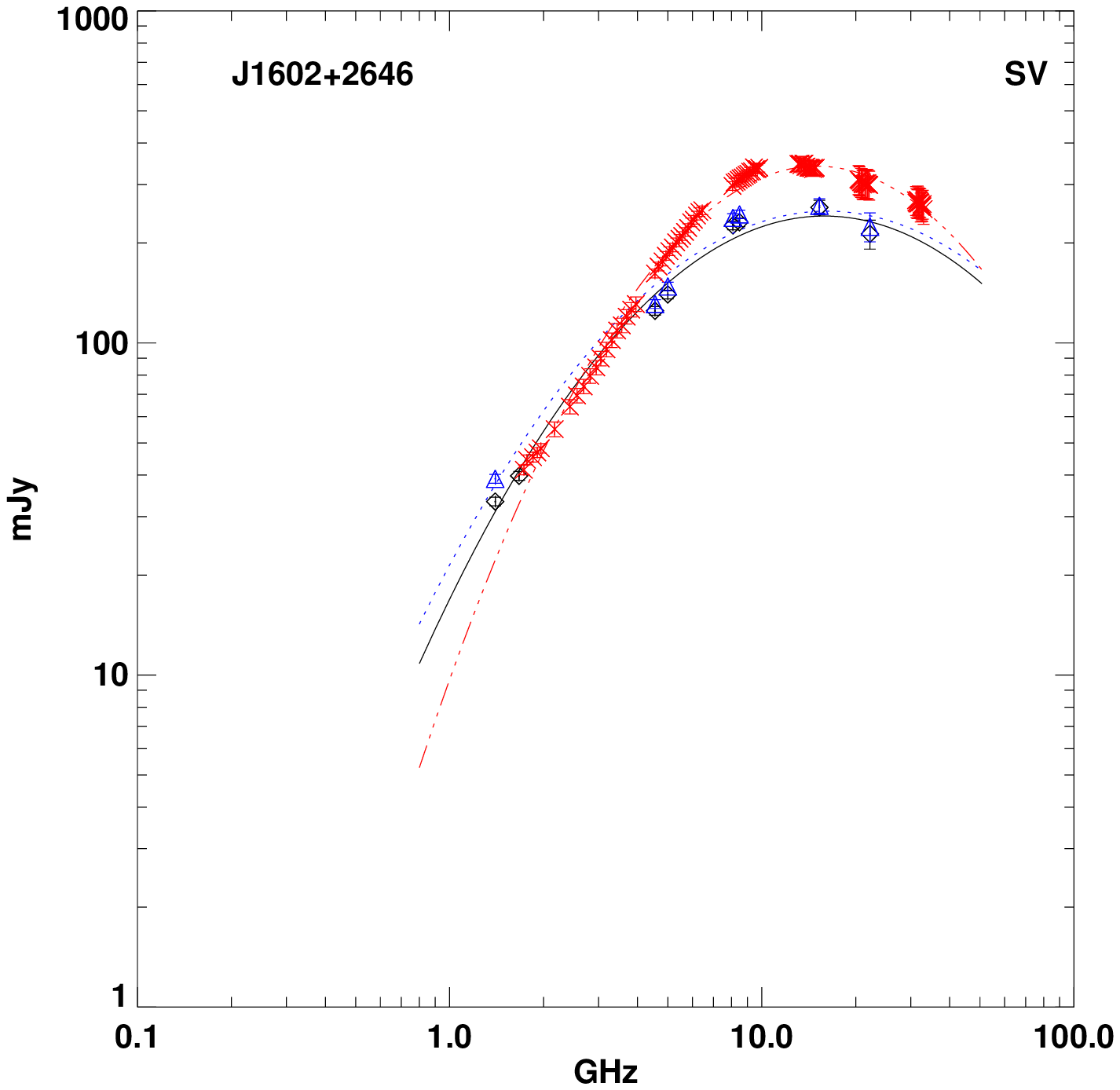}
\includegraphics{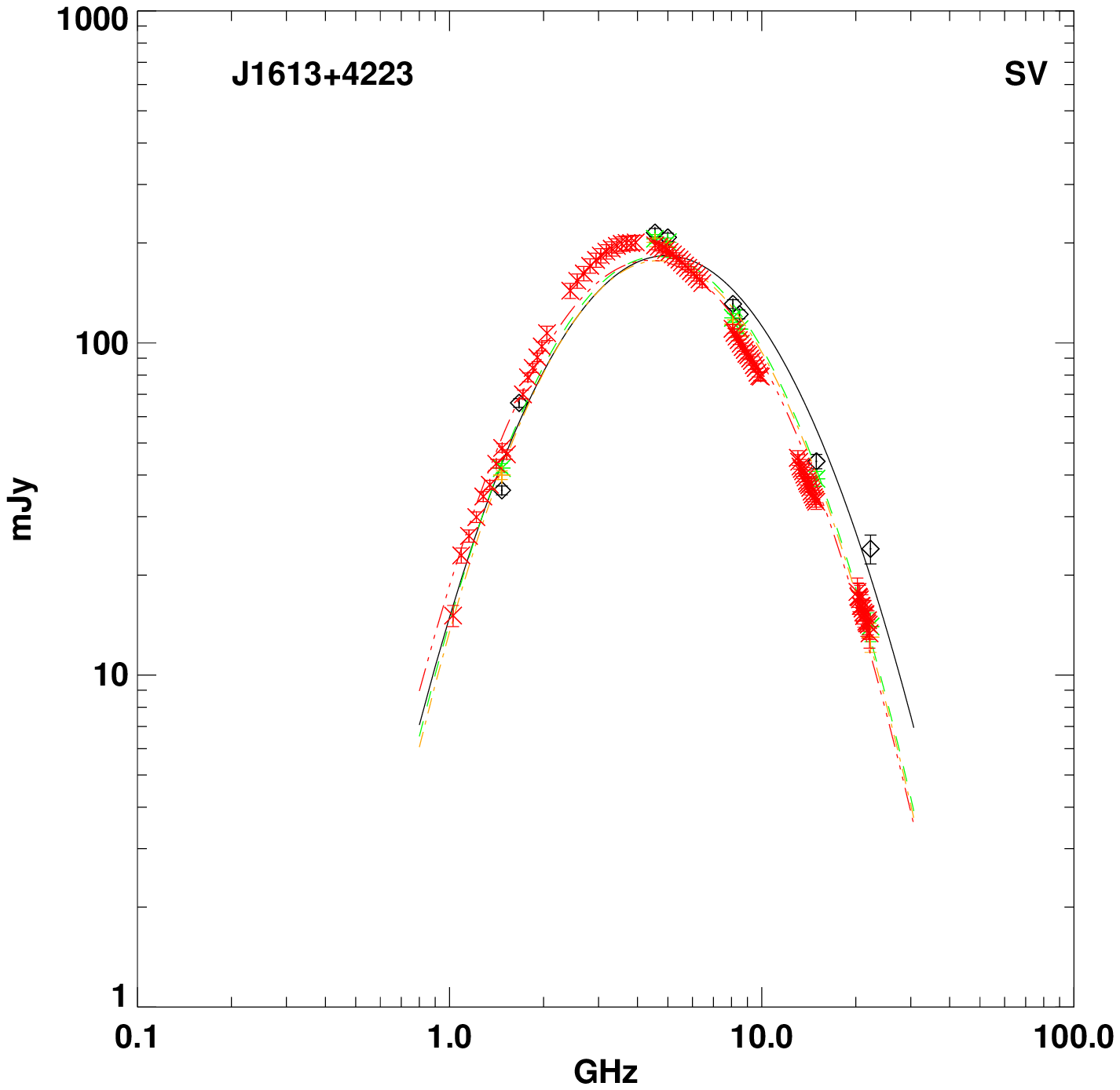}
\includegraphics{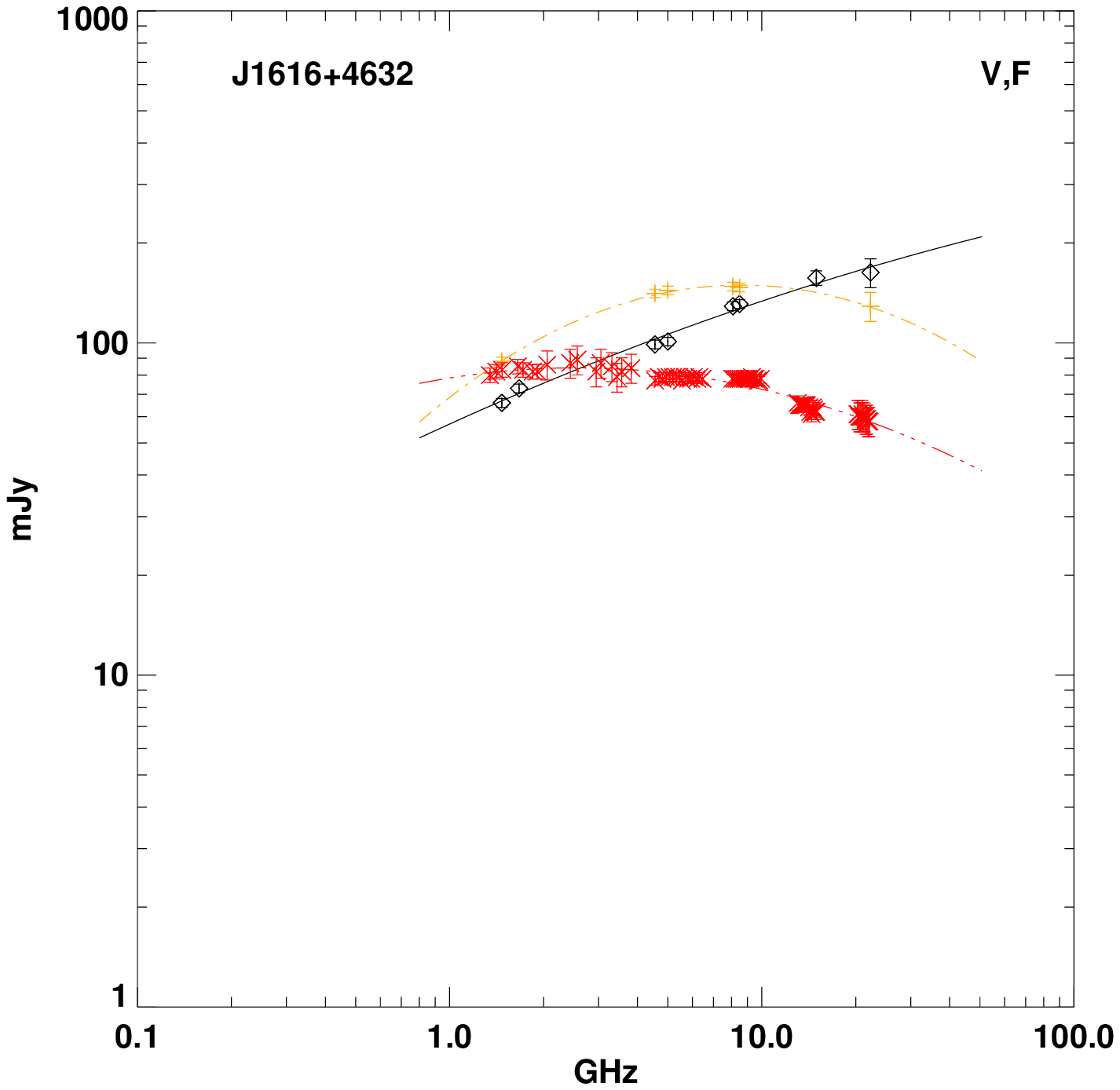}
\includegraphics{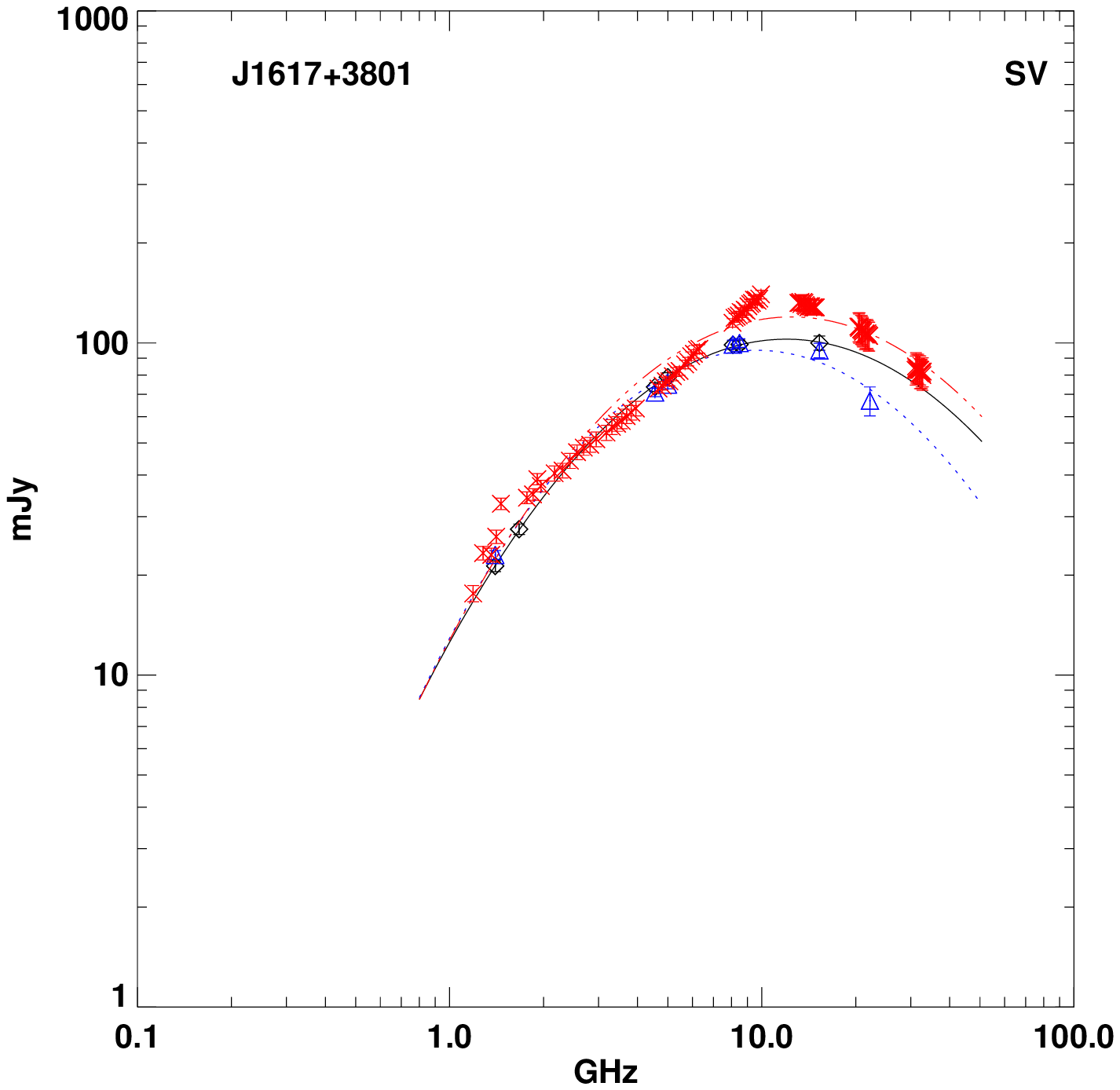}
\includegraphics{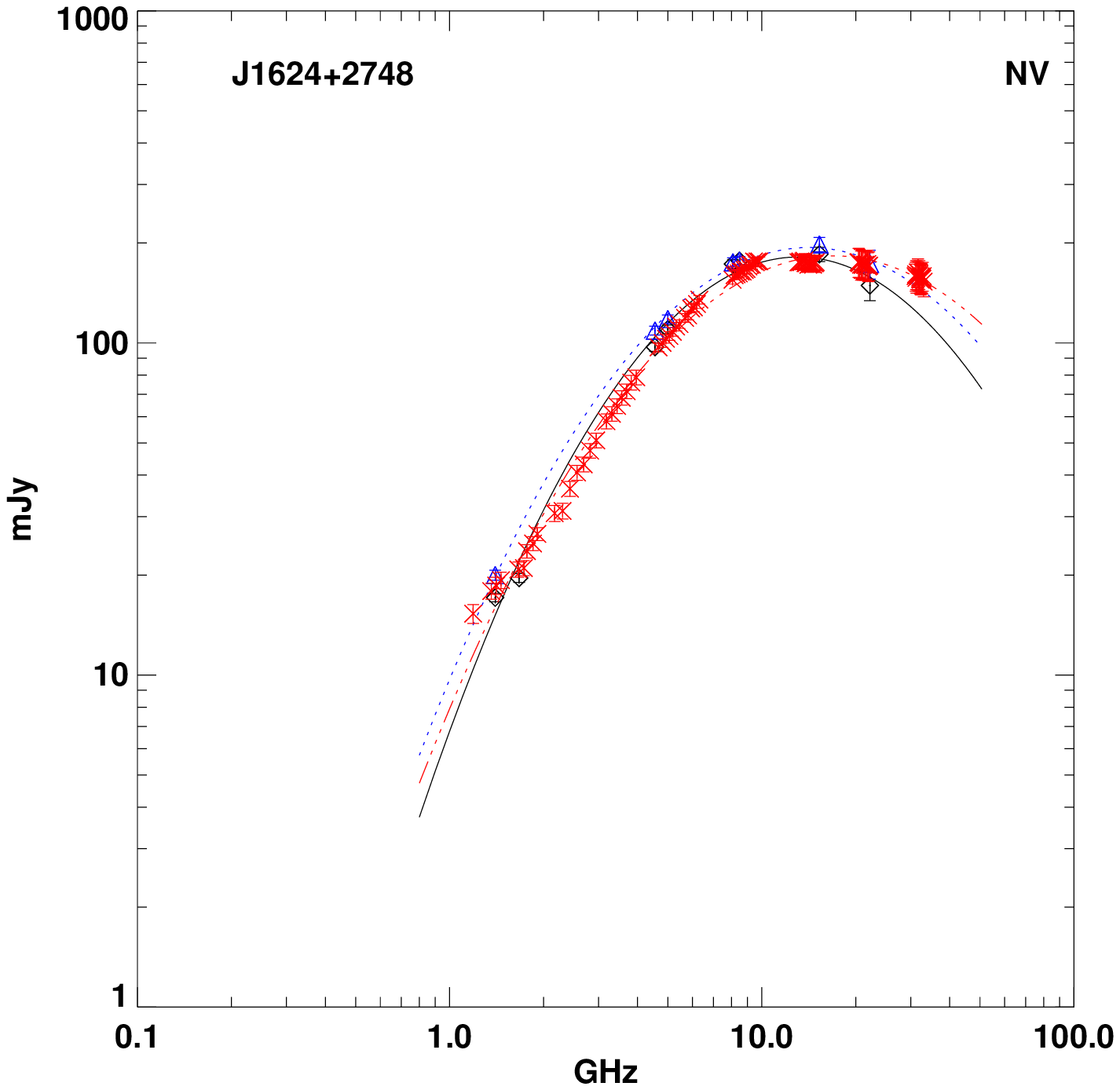}
\includegraphics{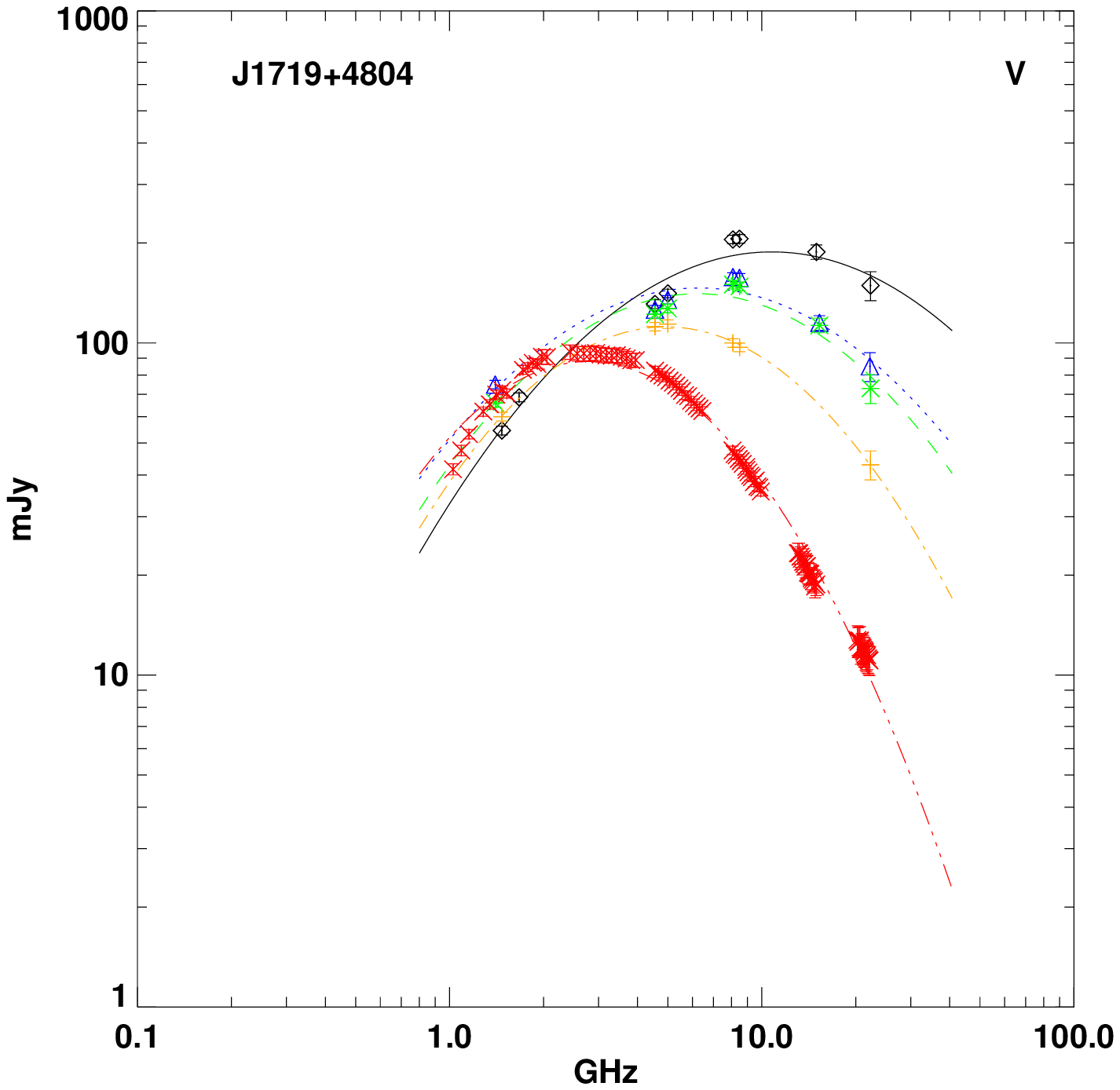}
\vspace{20cm}
\end{center}
\caption{Continued.}
\end{figure*}

\begin{figure*}
  \begin{center}
\includegraphics{0754_U_WINDOW.PS}
\includegraphics{0754_K_WINDOW.PS}
\includegraphics{0819_U_WINDOW.PS}
\includegraphics{0819_K_WINDOW.PS}
\includegraphics{1025_U_WINDOW.PS}
\includegraphics{1025_K_WINDOW.PS}    
\vspace{22cm}
\caption{VLBA images at 15 GHz ({\it left}) and at 24 GHz ({\it
    right}). On each image we provide the source name, the observing frequency, the peak brightness (peak br.) and the first contour (f.c.), which is 3 times the off-source noise level on the image plane. Contours increase by a factor of 2. The beam is plotted in the bottom left-hand corner of each image.}
\label{vlba-images}
  \end{center}
\end{figure*}

\addtocounter{figure}{-1}
\begin{figure*}
  \begin{center}
\includegraphics{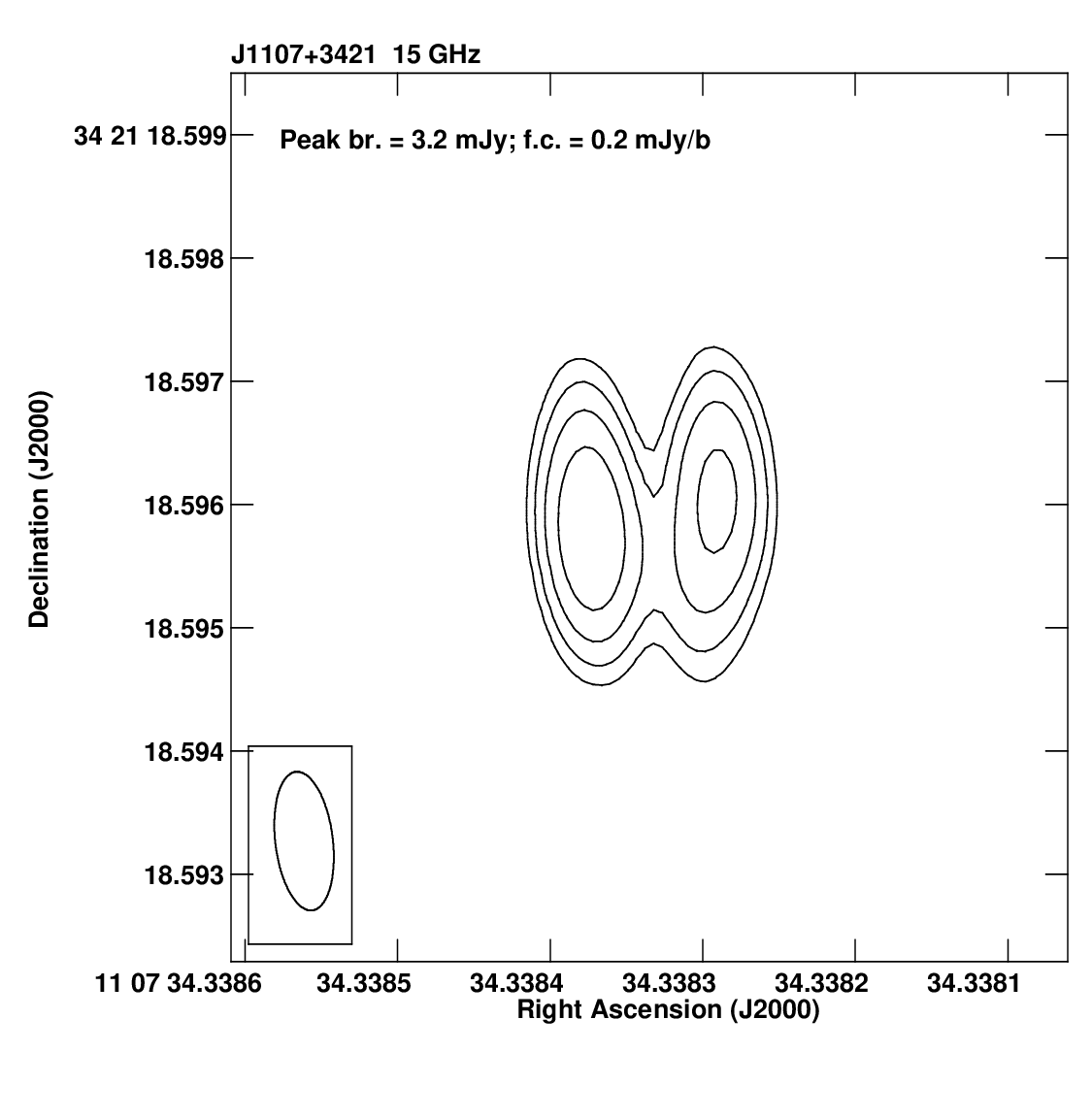}
\includegraphics{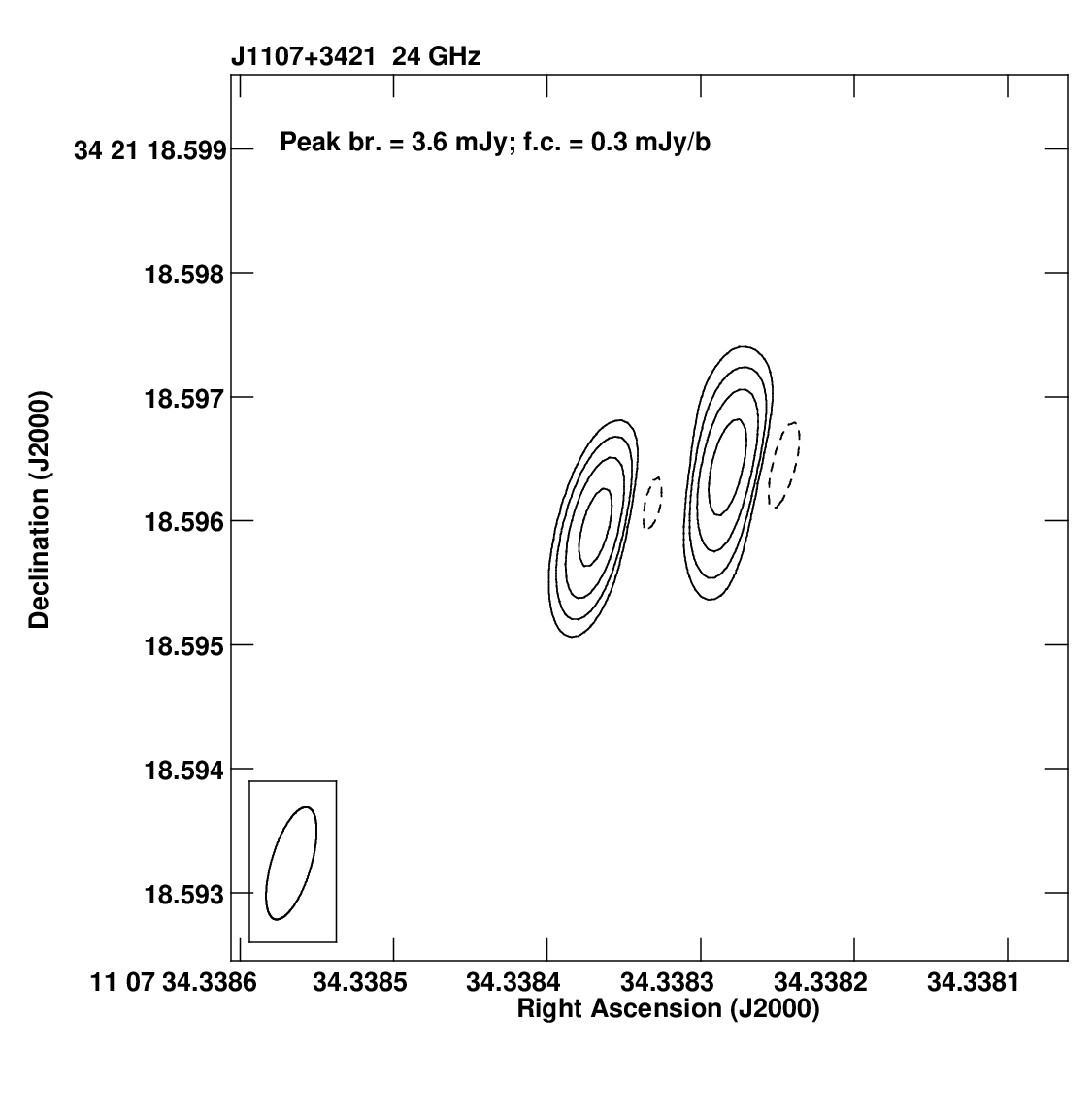}
\includegraphics{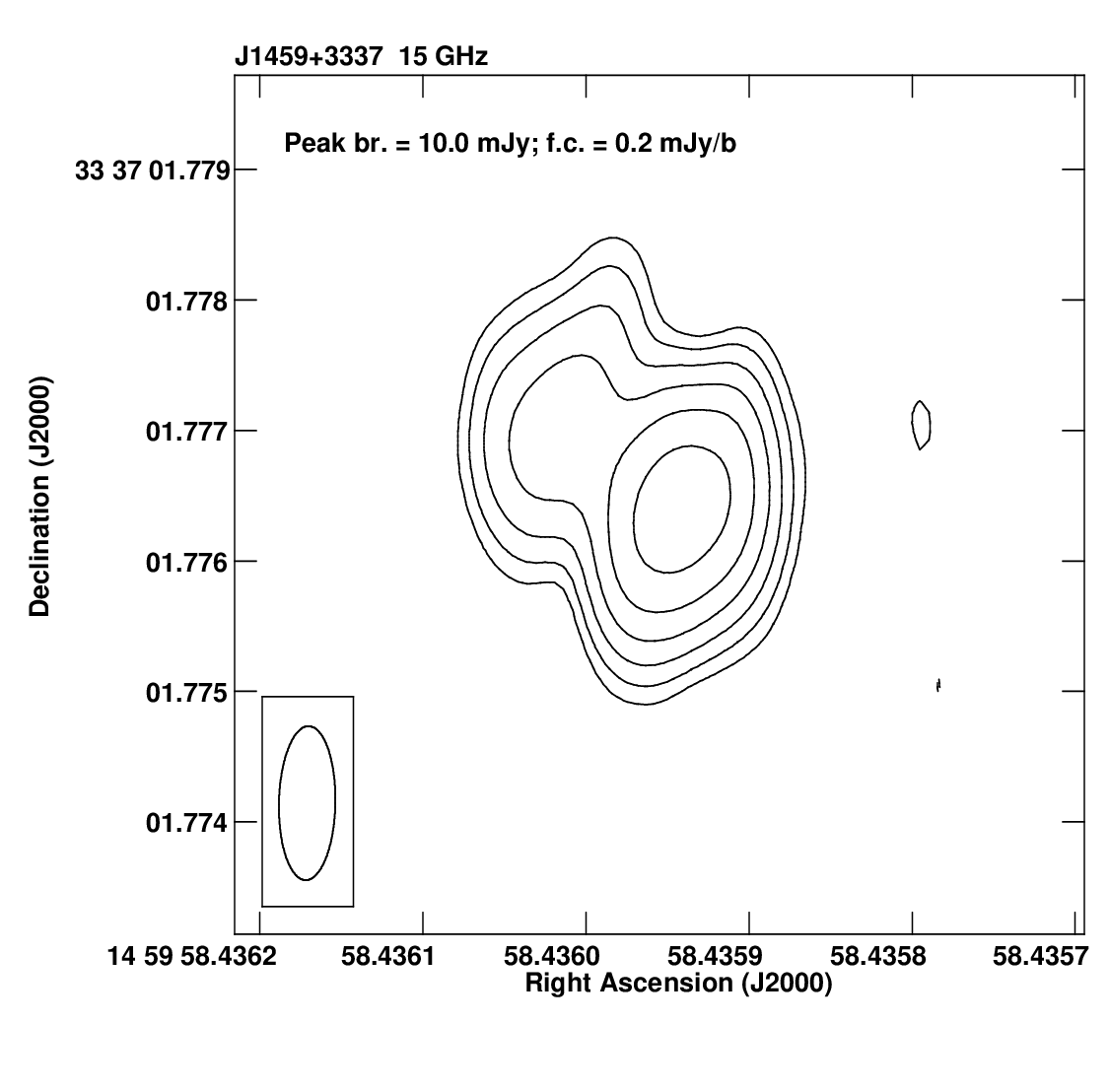}
\includegraphics{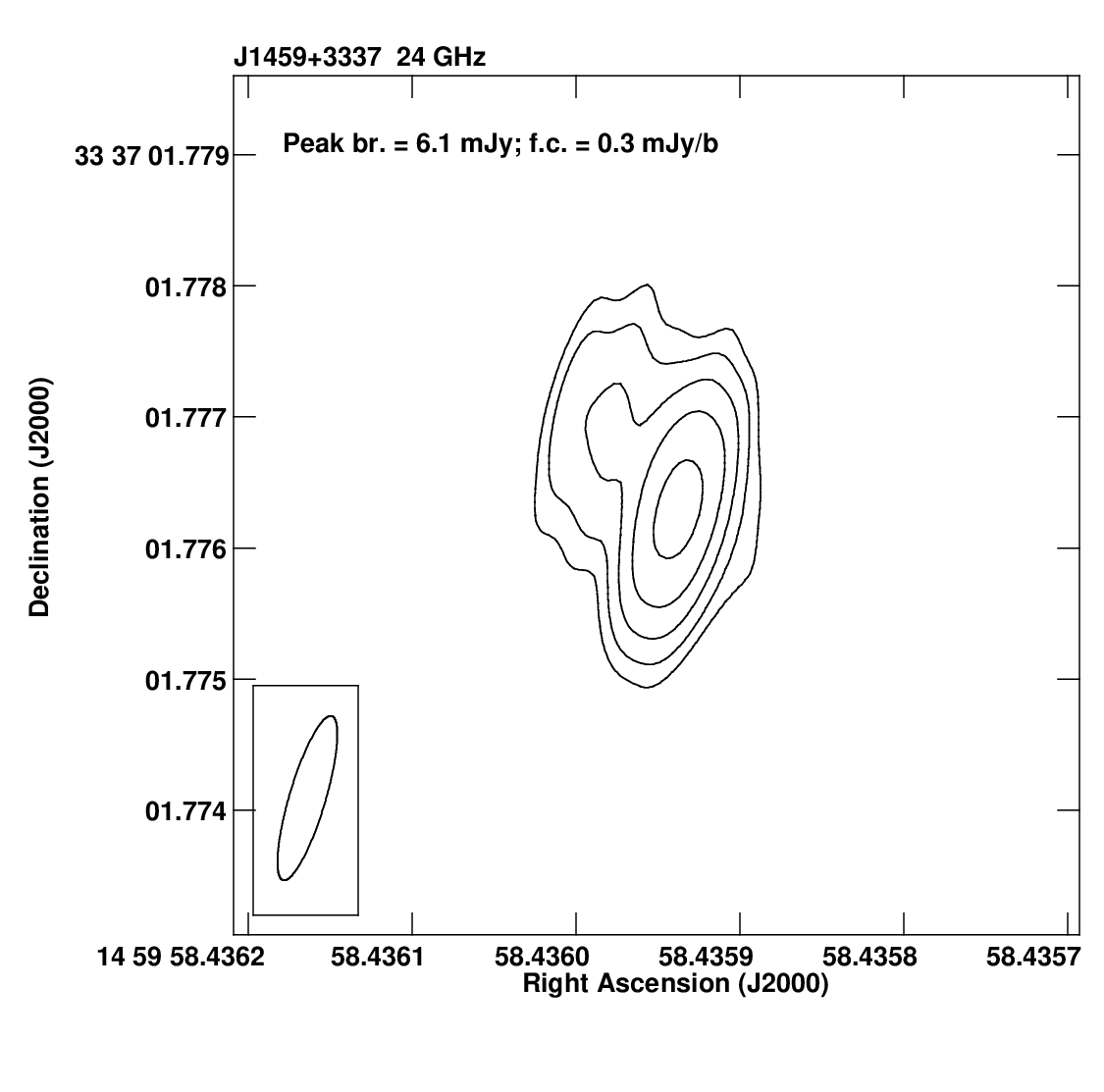}
\vspace{15cm}
\caption{Continued.}
  \end{center}
\end{figure*}

\section{Results}

The large band width of the VLA observations allowed us to determine
the spectral shape roughly continuously from 1 to 22 GHz, with only a
few gaps among X, Ku and K bands. All the 35 sources are unresolved in
the VLA images and we measured the flux densities using the task
\texttt{JMFIT} in \texttt{AIPS} which performs a Gaussian fit on the
image plane. In Fig. \ref{radio_spectra} we plot
the flux densities for each spectral window, together with the
measurements from earlier epochs as it is described in the caption. \\
VLBA observations could detect all the target sources at 15 and
24 GHz. Among the 12 sources observed with the VLBA, 7 sources
($\sim$58 per cent) are resolved or marginally resolved on pc-scale. 
As in the case of VLA, for the
unresolved sources, or 
unresolved components, we measured the flux density using the task
\texttt{JMFIT}, whereas for sources with complex
structure we estimated the total flux density using \texttt{TVSTAT},
which extracts the flux density on a selected polygonal area on the
image plane. 
Errors on the VLA
and VLBA flux densities are estimated by $\sigma = \sqrt{\sigma_{\rm cal}^{2} + \sigma_{\rm
    rms}^{2}}$, where $\sigma_{\rm cal}$ is the uncertainty on the
amplitude calibration (see Section 2), and $\sigma_{\rm rms}$ is the
1-$\sigma$ noise level measured on the image plane. The former
contribution is generally much larger than the latter.\\
Errors on the spectral index are computed assuming the error
propagation theory:\\

$\sigma_{\rm sp} = \sqrt{ \left(\frac{\sigma_{S1}}{S_1} \right)^2 +
\left( \frac{\sigma_{S2}}{S_2} \right)^2} \frac{1}{{\rm ln(\nu_2) -
    ln(\nu_1)}}$\\

\noindent where $S_{i}$ and $\sigma_{Si}$ are the flux density and the flux
density error, respectively, at the frequency $i$ ($\nu_i$). \\

Sloan Digital Sky Survey (SDSS) information from the data release 12
\citep[DR12,][]{alam15} have been used to identify the 
 host and its redshift (either spectroscopic or photometric) of the
 sources still lacking an optical counterpart in 
 previous studies. The updated information is reported in Table
 \ref{vla-flux}.   \\
 
\subsection{Radio spectrum}
\label{sec-spectral}

One of the main characteristics of young radio sources is the convex
radio spectrum that turns over at a frequency related to the
source size/age. In general, as the source deploys in the interstellar
medium (ISM) of the host galaxy, the peak frequency progressively
moves to lower and lower frequencies. The anti-correlation
found between the peak frequency and the linear size \citep{odea97}
implies
that the smaller and younger sources should have the spectral peak above a
few GHz. \\
Following \citet{mo10} and  with the goal of estimating the peak
frequency and how it changes with time,
we fitted the simultaneous multi-frequency radio spectrum for each
epoch
with the pure analytic function:\\

\begin{equation}
{\rm Log(}S{\rm )} = a + {\rm Log (}\nu{\rm )} \times (b + c \,{\rm
  Log (}\nu{)})
\label{eq_spectrum}
\end{equation}

\noindent where $S$ is the flux density, $\nu$ the frequency, and
$a$, $b$, and $c$ numeric parameters. Fits to the radio spectra are
presented in Fig. \ref{radio_spectra}.\\
For two sources, J1052$+$3355 and J1512$+$2219 the fit did not
converge. In particular, the spectrum of J1512$+$2219 is
highly convex, whereas in J1052$+$3355 the lack of data points at low
frequencies prevents the fit in the optically-thick part of the
spectrum.\\ 
Following \citet{mo10} we compute the spectral index below
($\alpha_{\rm b}$) and above ($\alpha_{\rm a}$) the peak frequency. For
some sources we could not estimate either $\alpha_{\rm b}$
or $\alpha_{\rm a}$, due to the lack of data points above or below 
the peak frequency.\\
In 6 sources, J1008$+$2535, J1137$+$3441, J1240$+$2323, J1528$+$3816,
J1530$+$5137, and J1616$+$4632, the spectral shape changes from convex in the first epoch
to flat in 2012, with $\alpha_{\rm a} < 0.4$. 
In addition, the source
J1054$+$5058 shows an inverted spectrum up to 24/32 GHz in all epochs.
The remaining sources keep a convex spectrum, but with some amount of 
flux density variation. \\
In 19 out of the 35 observed sources 
(54 per cent; 14 quasars, 4 galaxies, and 1
object still missing the optical counterpart) we observe a decrease of the flux
density in 
the optically-thin part of the spectrum, and a flux density increase
below the actual peak frequency, which may be consistent with a source in adiabatic
expansion. Although the variability
probed by the VLA monitoring for J0955$+$3335 is consistent with a
source in adiabatic 
expansion, the flux density measured by VLBA images in an intermediate
epoch (in 2010) revealed a temporary increase of the flux density
at 15 and 22 GHz, which is hard to reconcile with an expanding source.
For this reason \citet{mo12} labelled this source a blazar candidate,
and we do not consider this source as a CSO candidate
anymore. The same reasoning applies to the source J1052$+$3355
  (see Section 3.5).\\
The remaining 16 sources (46 per cent;
9 quasars, 6 galaxies, and 1 object still missing an optical counterpart) 
show random flux density increase or
decrease, as it is expected in blazars.\\
In Table \ref{vla-variability} we provide the peak frequency for each
observing epoch.
In 13 sources the peak frequency remains roughly constant, within the
errors, while in 11 sources it moves to lower frequencies with
time. 
Among the latter group, the source J1218$+$2828 represents an outlier: 
between 1.4 and 15 GHz the spectral shape is convex
with the peak frequency shifted towards lower frequencies with respect
to that estimated in earlier epochs. However, above 15 GHz the flux
density increases again, suggesting the presence of an additional
compact component with a highly inverted spectrum. The radio spectrum
becomes clearly inverted within the 2-GHz band width sampled by the
present observations in K-band.
In 4 sources the peak moves to higher
frequencies 
with time, while in 5 sources it moves up and down without any
trend. Remarkably, in all the 18 sources with spectral variability
consistent with adiabatic expansion, the peak frequency is constant or
decreases with time.\\

\subsection{Variability}
\label{sec_variability}

\begin{figure}
\begin{center}
\includegraphics{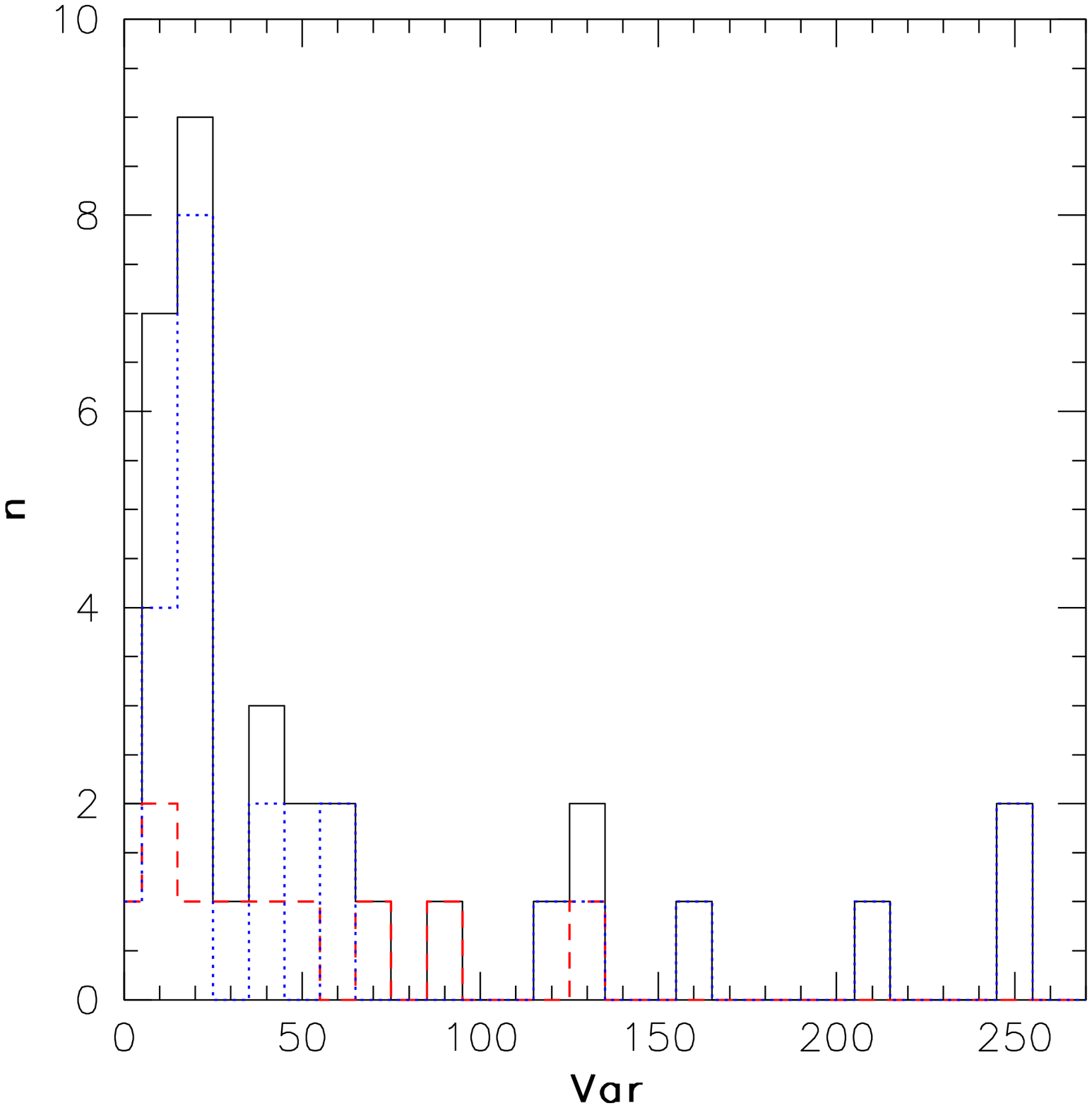}
\vspace{7.5cm}
\caption{Distribution of the variability index computed over the whole
  period $V_{\rm tot}$, for galaxies (red dashed line), quasars
  (blue dotted
  line), and all sources (black solid line).}
\label{histo-var}
\end{center}
\end{figure}

\begin{figure}
\begin{center}
\includegraphics{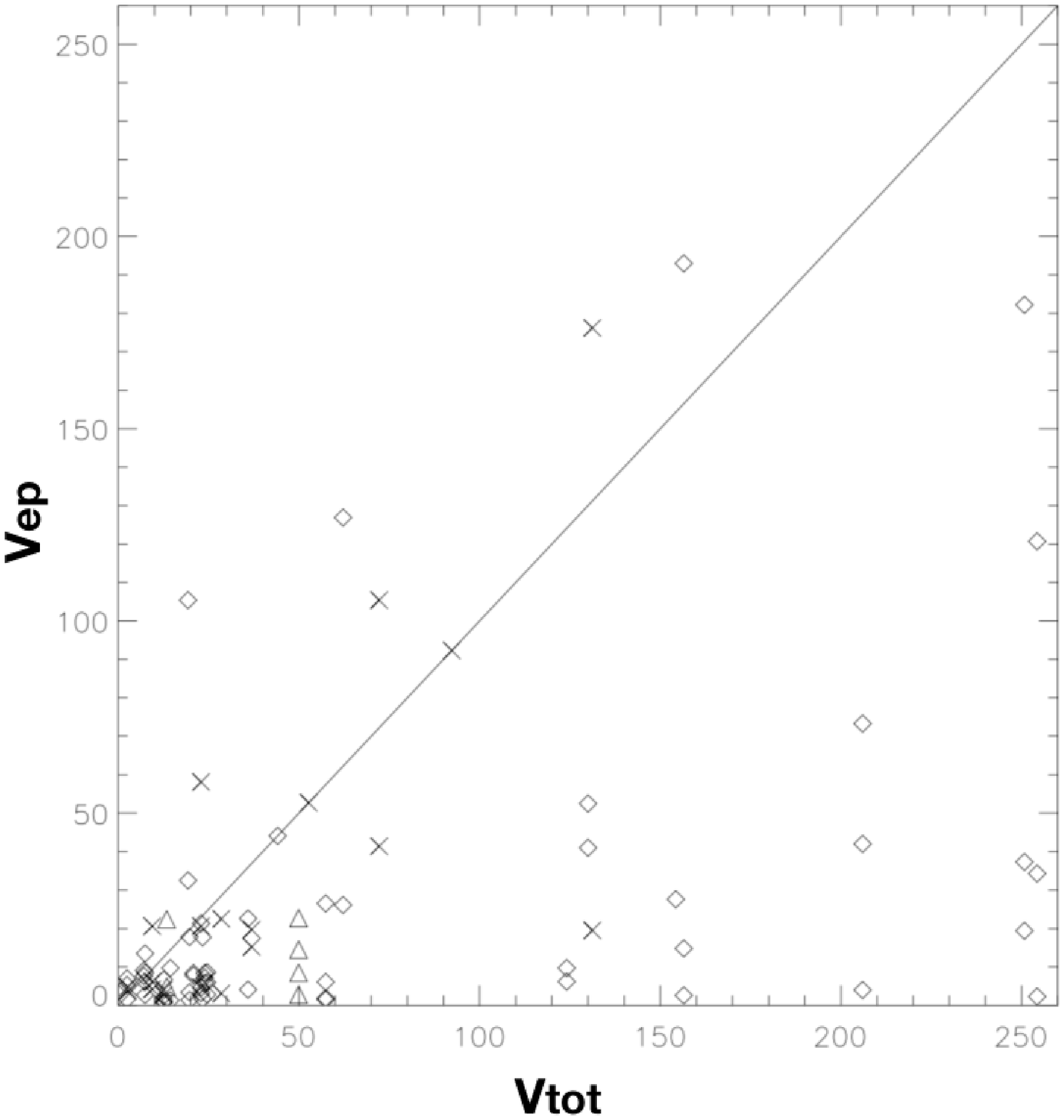}
\vspace{7.5cm}
\caption{Variability index computed for two consecutive epochs $V_{\rm
  ep}$ as a function of the variability index computed over the whole
  period $V_{\rm tot}$ for each source. Crosses, diamonds and
  triangles refer to galaxies, quasars, and sources with unidentified
  optical counterpart, respectively. The line indicates $V_{\rm ep} =
  V_{\rm tot}$.}
\label{fig-var}
\end{center}
\end{figure}

Samples selected at high radio frequencies ($>$ 5 GHz) have proved to
be highly contaminated by blazars \citep{tinti05,
  torniainen05,sadler06,sadler08,hovatta07,mo07,mingaliev12}. 
Flux density and
spectral variability are common properties of blazars, whereas
CSOs are among the least variable sources
with variability of about 10 per cent at most over one year
\citep{odea98}, while some ``secular'' variations are known for a few
objects \citep[e.g. OQ\,208,][]{cstan97}.\\
In order to identify and remove highly variable sources
we performed multi-epoch VLA observations covering
quasi-simultaneously the
frequencies between $\sim$1 GHz and 22 GHz. 
Following \citet{tinti05} we estimate the
variability by means of the parameter $V$ defined as:

\begin{equation}
V = \frac{1}{m} \sum_{i=1}^{m} \frac{(S_{\rm A}(i) - S_{\rm
    B}(i))^{2}}{\sigma_{i}^{2}}
\label{eq-var}
\end{equation}

\noindent where $S_{\rm A}(i)$ and $S_{\rm B}(i)$ are the flux density
at the i-th frequency of two consecutive epochs, $\sigma_{i}$ is the
error on $S_{\rm A}(i) - S_{\rm B}(i)$, and $m$ is the number of
sampled frequencies. We compute the variability index on
consecutive epochs $V_{\rm ep}$
in order to detect a flaring state followed by a period of
quiescence. 
In addition we compute $V_{\rm tot}$ between the
first epoch from \citet{cstan09} and last epoch (data presented in this paper)
in order to determine the long-term variability spanning a decade of
observations (between 1998-1999 and 2012). Results are reported in
Table \ref{vla-variability}. In Fig. \ref{histo-var} we plot the
distribution of $V_{\rm tot}$ for galaxies, quasars, and all
sources. Values range between about 3 and 255. Only 1 source
optically associated with a galaxy (J1624$+$2748) has $V_{\rm
  tot} <$4, and is marked as non variable, NV, in Table
\ref{vla-variability} and in Fig. \ref{radio_spectra}. 
The threshold $V = 4$ indicates variability of 10 per cent
between two epochs. The majority of the objects, $\sim$47 per cent
(16 sources; 13 quasars, 2 galaxies, and 1 object with no optical counterpart) 
have V$\leq$25, indicating that some variability, up to about 30 per
cent, is common on long time scales. These sources are marked as
slightly variable SV in Table \ref{vla-variability} and in
Fig. \ref{radio_spectra}. Sources with a variability index above 25 are
marked V. If the spectrum turns out to be flat, the source is also
labelled F (see Sect. \ref{sec-spectral}).\\ 
In Fig. \ref{fig-var} we plot for each source the variability indices
computed for consecutive epochs $V_{\rm ep}$ 
as a function of the variability index
computed over the whole period $V_{\rm tot}$. Usually the variation
between two consecutive epochs is smaller than that estimated over the
whole period. This is consistent with changes produced by a source
in adiabatic expansion. 10 sources (5 quasars, 4 galaxies,
and 1 object with unidentified optical counterpart) have $V_{\rm ep} >
V_{\rm tot}$, indicating that the variation observed between two
consecutive epochs is larger than the variability derived between
1998$-$1999 and 2012. This behaviour is typical of blazars which
randomly interchange low-activity and high-activity states. \\

\subsection{Dynamical ages}

The radio spectrum of a homogeneous synchrotron source with
the magnetic field frozen in the plasma that is in adiabatic expansion
shows a flux density increase with time in the optically-thick part of
the
spectrum:

\begin{equation}
S_{1} = S_{0} \left( \frac{t_{0} + \Delta t}{t_{0}}\right)^{3}
\label{eq_thick}
\end{equation}

\noindent where $S_{0}$ and $S_{1}$ are the flux densities at the time
$t_{0}$ and $t_{0} + \Delta t$, i.e. the source lifetime at the time
of the last observing epoch. Moreover, the peak frequency moves
towards low frequencies:

\begin{equation}
\nu_{p,1} = \nu_{p,0} \left( \frac{t_{0}}{t_{0} + \Delta t}
\right)^{4}
\label{eq_peak}
\end{equation}

\noindent where $\nu_{p,0}$ and $\nu_{p,1}$ are the flux density at
  the time
$t_{0}$ and $t_{0} + \Delta t$. 
Among the 35 sources observed in 2012, 18 (14 quasars, 3 galaxies, and
1 object with no optical counterpart) show a variability that is
consistent with the expectation of adiabatic expansion. We estimate
the dynamic age of the radio source, $t_{0} + \Delta t$, using
Eq. \ref{eq_thick} and comparing the flux density at 1.4/1.7
GHz during the first and last observing epoch.
The dynamical age estimated by
Eq. \ref{eq_peak} is highly uncertain due to the large uncertainty on
the peak frequency. The estimated dynamical age for the 11 sources
with
significant variability consistent with adiabatic expansion is
reported in Table \ref{tab-lifetime}. Dynamical ages range
between 40 and 270 yr, supporting the idea that these sources are in a
very early phase of their evolution. For the remaining 7 sources the
variation in the optically-thick part of the spectrum is consistent
within the errors, preventing us to set any constraints on their
age. \\

\begin{table}
\caption{Estimated dynamical ages for faint HFP sources whose
  variability is
  consistent with relativistic plasma in adiabatic expansion. Column
  1: source name; column 2: the time (in days) elapsed between the
  first and
  last VLA observations; column 3: source dynamical age (in yr)
  at the
  epoch of last
VLA observations in 2012; Column 4: linear size (pc); column 5:
estimated source expansion velocity in units of the speed of light.}
\begin{center}
\begin{tabular}{ccccc}
\hline
  Source name & $\Delta t$ & $t_{\rm age}$ & $r$ & $v$\\
              & days       & yr            & pc& $c$\\
\hline
J0754$+$3033 & 4713 & 130 & 12 & 0.5\\ 
J0819$+$3823 & 4932 & 240 & -  &\\
J0955$+$3335 & 4709 & 240 & - &\\
J1044$+$2959 & 4576 & 75  & $<$4& $<$0.7\\  
J1052$+$3355 & 4709 & 100 & - &\\
J1107$+$3421 & 4709 & 270 & - &\\
J1309$+$4047 & 4688 & 255 & 6.8& 0.4\\ 
J1420$+$2704 & 4596 & 130 & - &\\
J1459$+$3337 & 4699 & 40  & 7 & 0.9\\ 
J1613$+$4223 & 4720 & 225 & - &\\
J1719$+$4804 & 4720 & 160 & $<$2.5 &$<$0.1\\ 
\hline
\end{tabular}
\end{center}
\label{tab-lifetime}
\end{table}

\subsection{Parsec-scale structure}

A sub-sample of 12 sources from the faint HFP sample were observed
with the VLBA in 2019 January. These sources 
were selected on the basis of their peak frequency below 7 GHz in
order to have VLBA observations at 15 and 24 GHz
in the optically-thin part of the spectrum. 5 sources (about 42 per cent)
are resolved into two components, while 2 sources are marginally resolved at
24 GHz only (Table \ref{phase-vlba}). 
Among the 12 newly observed sources, 6 were target of earlier VLBA
observations at 8.4 and 15 GHz
\citep{mo08,mo12}. The source J1107$+$3421 is resolved in VLBA images
at both epochs, whereas J1002$+$5701 (not observed by VLA in 2012) 
that turned to be slightly resolved
in \citet{mo12} is now unresolved. Remarkable cases are J0819$+$3823
and J1459$+$3337 which show a double structure in the last epoch of
observations, while they were unresolved in earlier
observations. J1420$+$2704 and J1613$+$4223 are unresolved in both
observing epochs. In general, the flux density at 15 GHz in our VLBA
observations in 2019 is
significantly smaller than the flux density measured at the same
frequency in earlier
VLBA observations, in agreement with the trend of decreasing flux density
observed in the VLA monitoring campaign. 
Usually 
S$_{\rm VLBA}/$S$_{\rm VLA}$ at 15 GHz is comparable or higher
than at 24 GHz. In three sources S$_{\rm VLBA}/$S$_{\rm VLA}$ is
higher at 24 GHz. Although the central frequencies of VLA and VLBA
observations are slightly different (14.5 and 21.5 GHz at the VLA and 15
and 24 GHz at VLBA), the flux densities can still be compared. In fact,
even in the case of the source with the steepest spectral index
(J1613$+$4223, $\alpha_{\rm VLA} \sim$1.6), the
difference between the flux density at 14.5 (21.5) GHz and the one
extrapolated at 15 (24) GHz is within the error. 
The spectral index of the sources
has been computed considering the full-resolution images at
both frequencies, since they allow the lowest rms noise levels in
the images. This may cause an artificial steepening of
the spectral index if some extended emission is present. In general
the spectral index ranges between -0.5 and 2.0 (Table
\ref{vlba-flux}), 
indicating the presence
of compact and flat components, like core and mini-hotspots, and
steep-spectrum emission from jets or mini-lobes.  \\

\subsection{Notes on individual sources}

Here we provide a brief description of the sources deserving
  some discussion.
Sources which are considered CSO candidates are marked in boldface.
An asterisk indicates that the VLBA observations of the source are
presented in this paper.\\

\indent {\bf J0754+3033*}: The VLA radio spectrum shows moderate
variability on long-time scales ($V_{\rm tot} \sim 57$), but shows low
variability when consecutive epochs separated by few years or less
($V_{\rm ep} < 2$ between 2003 and 2004, and between 2004 and
  2006; Table \ref{vla-variability}) are considered.
The turnover frequency slightly moves to
lower frequency, from 8.9$\pm$0.6 GHz in 1999 to 6.5$\pm$1.0 GHz in
2012. The flux density in the optically-thin part of the spectrum
decreases with time, whereas it increases in the optically-thick part. 
The dynamical age inferred from its variability is about 130 yr
  (Table \ref{tab-lifetime}).
On pc-scale the source shows a double
structure (Fig. \ref{vlba-images}) with the components separated by
about 1.6 mas (i.e. about 12 pc at the redshift of the source). The
spectral index of the eastern component is 0.6, indicating the
presence of freshly injected/accelerated relativistic particles,
  as one would expect if the component hosts the core and/or a mini-hotspot,
whereas it is steeper ($\alpha \sim$1.3) in the western component
(Table \ref{vlba-flux}). These
characteristics suggest that this source is a CSO candidate.\\  

\indent {\bf J0819$+$3823*}: The VLA radio spectrum is marginally variable,
with $2.7 < V_{\rm ep} < 14$, and $V_{\rm tot} \sim 7$. The peak
frequency is roughly constant within the errors ($\nu_{\rm p} \sim$ 6
GHz), while the flux 
density increases in the optically-thick part of the spectrum and 
decreases in the optically-thin part. The dynamical age computed
  from its variability is about 240 yr (Table \ref{tab-lifetime}).
In the VLBA image at 15 GHz there is diffuse emission on the East of the
compact component that dominates the flux density, and the source size
is about 1 mas. At 24 GHz the compact
component is slightly resolved into two components separated by about
0.4 mas (i.e. the distance between the peak of the two
components). The spectral index integrated over the whole VLBA structure is
about 2.0, while the VLA spectral index between 15 and 22 GHz is about
1.0, suggesting the presence of extended steep-spectrum
emission (Table \ref{vlba-flux}). 
The larger S$_{\rm VLBA}/$S$_{\rm
  VLA}$ at 15 GHz than at 24 GHz supports this interpretation. 
The source was 
unresolved in earlier VLBA observations presented in
\citet{mo12}. Although the complicated pc-scale radio structure
  prevents us from unambiguously classify this source as a CSO, the
  VLA variability is in agreement with what expected for a source in
  adiabatic expansion. For this reason we still consider this object
  as a CSO candidate.\\  

\indent {\bf J0951$+$3451}: The VLA radio spectrum is convex in all the
three observing epochs and the peak frequency is roughly constant
within the errors ($\nu_{\rm p} = 5.6\pm1.0$). The source shows low variability
($V_{\rm ep}$ and $V_{\rm
  tot} \sim$12) and the flux density increases in the
optically thick part of the spectrum, while it slightly decreases in the
optically-thin part. The source is resolved into three
components by VLBA observations, with the central region showing a
flat spectral index \citep{mo12}. These characteristics confirm the
source as a CSO.\\

\indent J1008$+$2533: The VLA radio spectrum had a convex shape only
during the first epoch. In the subsequent epochs the spectrum shows a
complex shape, with an inverted part below 5 GHz and a flattening at
higher frequencies. However, the variability index has relatively
small values $1<V_{\rm ep} <18$, and $V_{\rm tot} \sim$20. At
pc-scale it shows a core-jet structure with a compact component with
an inverted spectrum dominating the radio emission \citep{mo12}. These
characteristics confirm the blazar nature of this source.\\

\indent J1025$+$2541*: The VLA radio spectrum has a convex shape in
all the three observing epochs. The source has some moderate variable
$V_{\rm ep} \sim 15-20$, and $V_{\rm tot} \sim$37. The peak frequency
determined for each epoch is consistent within the errors, with a hint
of decrease from $\sim$4.2 GHz to $\sim$3.3 GHz. At 15 GHz the pc-scale
structure is resolved into two amorphous components whose peaks are separated by
about 1.1 mas (i.e. about 6.5 pc), whereas at 24 GHz it shows
a single component roughly coincident with the brightest part of the
source visible at 15 GHz. The VLBA flux density at 24 GHz is higher
than the flux density observed by the VLA in 2012, suggesting a
blazar-like variability.\\

\indent {\bf J1107+3421*}: The VLA radio spectrum shows moderate 
variability with $2 < V_{\rm ep} < 23$ and $V_{\rm tot} \sim 50$. 
The turnover frequency is roughly constant within the
errors, whereas the flux density in the optically-thick part of the
spectrum increases with time. The dynamical age computed from its
variability is about 270 yr (Table \ref{tab-lifetime}).
The pc-scale radio source is
characterized by two components (Fig. \ref{vlba-images})
separated by 1 mas. The source position angle slightly changes from
-75$^{\circ}$ to -80$^{\circ}$ at 15 and 24 GHz.
Although the double structure was already pointed out by
\citet{mo12}, the flux density ratio at 15 GHz between the components changed
from S$_{\rm E}$/S$_{\rm W}$ $\sim$2 to 1.3 in 2010 and in 2019,
respectively. 
The spectral index of the eastern component is about
0.6, while in the western component $\alpha =$-0.5$\pm$0.4, indicating
an inverted spectrum. This component may be either the core or a very
compact self-absorbed hotspot, like in the case of the HFP sources
J1335$+$5844 and J1735$+$5049 \citep{mo14}. The fractional flux
density S$_{\rm VLBA}/$S$_{\rm VLA}$ is higher at 24 GHz than at 15
GHz (about 80 per cent and 40 per cent at 24 and 15 GHz,
respectively). On the basis of the VLA variability and the
  pc-scale properties, we still consider this source as a CSO candidate.\\

\indent{\bf J1309$+$4047}: The VLA radio spectrum is roughly constant with
$1 < V_{\rm ep} < 7$ and $V_{\rm tot} \sim 2$. The peak frequency is
constant within the errors ($\sim$4.5 GHz). The dynamical age computed
from the variability is about 255 yr (Table \ref{tab-lifetime}).
On pc-scale, the source
shows a double structure whose components are separated by about 0.8
mas, i.e. 6.8 pc at the redshift of the source \citep{mo12}. The steep
spectral index derived from VLBA data make us consider this object a
CSO candidate. \\

\indent J1459+3337*: This radio source was first identified as an
HFP object by \citet{edge96} with a turnover frequency of about 30
GHz. In the two decades thereafter, the peak progressively moved to lower
and lower frequencies, at about 21 GHz and 15 GHz in 1999 and 2003,
respectively, and our new observations in 2012 set the turnover at about
3 GHz. The flux density in the optically-thin part of the spectrum
progressively decreases with time, while in the optically-thick part of
the spectrum it steadily increases (the flux density at 1.4 GHz
progressively increases from $\sim$8
mJy in 1993 to $\sim$50 mJy in 2012). 
The source displays one of the highest
variability index $V_{\rm tot} \sim$250. The dynamical age computed
from its variability is about 40 yr (Table \ref{tab-lifetime}).
We observed a change in the radio morphology of this source: it was
unresolved in VLBA observations in 2005 \citep{mo08}, while in our new
observations it shows a double structure (Fig. \ref{vlba-images}) with the
two components separated 
by about 1.1 mas and 0.9 mas at 15 and 24 GHz, respectively 
(i.e. about 7 pc at the distance of the source). 
The
flux density ratio is $S_{\rm W}/S_{\rm E} \sim$ 3.8 and 3.4 at 15 and
24 GHz, respectively. The spectral index
is relatively steep, with $\alpha \sim$1.0 and $\sim$1.2 in the
eastern and western component, respectively. This source shows one of
the largest discrepancy between VLA and VLBA flux density, 
S$_{\rm VLBA}/$S$_{\rm VLA} \sim$ 20 per cent at both
frequencies, indicating a huge flux density decrement between 2012 and
2019 with no significant variation of the spectral shape between these
frequencies (Table \ref{vlba-flux}). Spectral and morphological
changes may be explained
in terms of either a CSO or a
knot that is moving downstream along the
approaching jet. In the first scenario the two components may be two
asymmetric mini-lobes that are moving away from each other. However,
if we consider that the source
grows from
$<$0.3 mas ($<$2 pc) in 2005 to 1 mas (about 7 pc) in 2019,
we infer a source expansion of about $c$, favouring the interpretation
of the propagation of a knot along the approaching jet that was
produced by a huge flare a few decades ago in a moderately
beamed radio source. Although the
variability associated with a single event hardly lasts longer
than a few years \citep{hovatta08},
there are some cases in which the ejected component can
be followed for longer time. An example is 3C\,84, which underwent a
huge increase of flux density in the 1960's, followed by the ejection
and
expansion of the southern jet \citet{walker94}, whereas not much can
be
said for the northern counterpart due to severe free-free absorption
that prevents its detection below 22 GHz. A similar situation may have
happened in the case of J1459$+$3337, where the component emerging
from the main compact region may be the approaching jet. This
interpretation may be supported by the
slightly different position of the eastern component in our images at
15 and 24 GHz with respect to the brightest one.
For these reasons, we
consider J1459$+$3337 a blazar-like candidate.\\

\indent J1530$+$2705: The VLA radio spectrum shows moderate variability
$4 \leq V_{\rm ep} < 60$ and $V_{\rm tot}          
\sim 23$. Changes of the peak frequency and of flux density in the
optically-thick and in the optically-thin part of the spectrum do not
follow any trend with time. This radio source is hosted in a
nearby galaxy which is part of a group \citep{mo10}. No information on
the pc-scale structure is available, but the erratic variability
suggests that this source is a blazar.\\

\indent{\bf J1613$+$4223*}: The VLA radio spectrum is highly convex with
$\alpha_{\rm b}=-1.9$ and $\alpha_{\rm a} = 1.6$, and
shows a modest variability $0.8 \leq V_{\rm ep} < 16$ and $V_{\rm tot}
\sim 13$. The dynamical age estimated by the variability is
  about 225 yr (Table \ref{tab-lifetime}).
The peak frequency is constant within the errors ($\sim$4.5
GHz). On pc-scale the source is unresolved at both 15 and 24 GHz, with
an upper limit on its angular size of 0.4 mas. The spectral index of
the whole source is very steep, suggesting the presence of
steep-spectrum low-surface brightness emission that may have been
missed at 24 GHz as supported by the much larger S$_{\rm VLBA}/$S$_{\rm
  VLA}$ observed at 15 GHz than at 24 GHz (Table
\ref{vlba-flux}). On
the basis of the VLA variability that is consistent with a source in
adiabatic expansion, we still consider this source as a CSO candidate.\\

\indent{\bf J1624$+$2748}: The VLA radio spectrum is convex and displays
one of the lowest variability estimated for this sample with $V_{\rm
  ep} \sim4$ and $V_{\rm tot} \sim3$. The peak frequency is roughly
constant within the errors. No information on pc-scale structure is
available, but the variability properties make this object a 
very promising CSO candidate.\\

\indent{\bf J1719$+$4804*}: The radio spectrum is convex and displays
high variability with $V_{\rm tot}          
\sim 254$. The peak frequency shifts from 10.8 to 2.8 GHz from 1999 to
2012. The dynamical age computed from the variability is about
  160 yr (Table \ref{tab-lifetime}).
The pc-scale structure is unresolved in our VLBA images, giving
an upper limit on the angular size 
of 0.3 mas, which corresponds to a linear size of 2.5 pc at the
redshift of the source. The
VLBA flux density at 15 and 24 GHz is about 4 mJy and 3.5 mJy,
respectively, indicating
a further decrease with respect to that observed in 2012. The VLBA spectral
index is rather flat $\alpha = 0.4\pm0.2$, suggesting that the radio
emission is dominated by the core or a very compact self-absorbed hotspot, like in the case of the HFP sources
J1335$+$5844 and J1735$+$5049 \citep{mo14}. This source shows one of
the largest discrepancy between VLBA and VLA flux densities with S$_{\rm VLBA}/$S$_{\rm VLA} \sim$ 20 and 30 per cent at 15 and
24 GHz, respectively, indicating a huge flux density decrement between 2012 and
2019 with slight variation of the spectral shape between these
frequencies (Table \ref{vlba-flux}). The VLA variability consistent
with what is expected in case of adiabatic expansion, and the lack of
unambiguous blazar characteristics, make us still consider this
source as a CSO candidate.\\

\section{Discussion}

Ideally, unbeamed young radio sources are characterized by a low
  level of flux 
density variability, low fractional polarization, and 
a double/triple structures dominated by
lobe/hotspot components when studied with sub-arcsecond
resolution. The location of the source core at the centre of a
two-sided radio structure would be the hallmark of a CSO.
In contrast, 
blazars show significant variability,
have pc-scale core-jet structures, and high and variable fractional
polarization. It is therefore clear that the study of variability,
morphology, and polarization is the key to disentangle the nature of a
radio source. \\
HFP sources are all unresolved by arc-second scale VLA observations
and higher resolution observations are necessary for investigating
their structure. With the aim of determining the source structures,
the optically-thin emission of 23
sources from the faint HFP sample have been observed with the VLBA, in
2010 and 2019, and results are reported in \citet{mo12} and in this
paper. Despite the pc-scale resolution, 
14 out of the 23 observed sources with VLBA
observations are unresolved or marginally resolved. The optically-thin
spectral index derived from VLA data (Table \ref{vla-flux} and \citet{mo10}) 
points out that 3 of these
sources (J1002$+$5701, J1436$+$4820, and J1613$+$4223) have
$\alpha_{\rm a} >1.0$, suggesting that the 
emission is dominated by steep-spectrum emission, likely from
mini-lobes. Among the
sources with resolved structure, 5 sources (J0754$+$3033,
J0943$+$5113, J0951$+$3451, J1107$+$3421, and  J1135$+$3624) 
have a double/triple
structure that is consistent with those of CSOs, 
whereas for the other sources the double
morphology may be interpreted in terms of either mini-lobes/hotspots
or core-jet structure. In general, the detection of only two
components makes the classification of these sources rather tentative,
\citep[e.g.][]{snellen00b, deller14}.
J0951$+$3451 shows slight variability between
the first and last epoch, whereas J0754$+$3033 and J1107$+$3421 are
highly variable. The other two sources were found non variable by
\citet{mo10}, and the lack of VLA observations in 2012 prevents us
from the
study of their long-term variability.\\    

Monitoring campaigns of high frequency peaking objects show that
moderate variability is a common characteristics of these
objects.
Earlier studies of the sources from the faint HFP samples pointed out
that about 40 per cent of the target sources are non variable
\citep{mo10}. This percentage drastically decreases to 1 object out
of 35 (3 per cent) when we consider a longer time separation between the
observing epochs. Contrary to what is found for other samples of high
frequency peakers, among the sources showing a random variability
typical of beamed objects, there is a predominance of radio
sources associated with galaxies ($\sim$ 60 per cent) rather than quasars ($\sim$ 43 per cent).\\

\subsection{Variability in young radio sources}

Flux density and spectral variability are not common features of the
class of CSOs, but they are the characteristics of
blazars. Samples selected at high frequencies (in the GHz regime) are more
contaminated by beamed objects than samples selected at lower
frequencies or with different criteria \citep[e.g.][]{coppejans16}.
Variability studies are thus used to discriminate between
CSOs and blazars. 
However, significant variability may be observed also among the
  youngest CSOs 
in which freshly produced bubbles of magnetized plasma
are expanding in a rather inhomogeneous ambient medium, implying an
irregular expansion rate. Moreover, in CSOs 
the time
elapsed between the first and last epoch of the monitoring campaign
corresponds to a significant fraction of the lifetime of the radio
source. It is then quite likely to observe spectral and flux density
changes. 
Dynamical ages estimated on the basis of the variation of the
  optically-thick flux density range
between 40 and 270 yr, supporting the idea that these sources are in a
very early phase of their evolution.
The values derived in this way should be representative of the
order of magnitude of the dynamical ages, owing to the strong
assumption of a single homogeneous component that is
expanding with no effects from the ambient medium. Moreover, the core
activity might not be continuous, and its likely erratic on-off cycle
could perturb the predicted flux density variability.\\
If these sources are actually in a very early stage of their
evolution, their large-scale counterpart should be sought among
low-power radio sources observed in the MHz regime.
In fact, at least for those sources for which
it is possible to ``see'' the epoch-by-epoch evolution, from
Eq. \ref{eq_peak} we expect that the peak frequency would lower by a
factor of about 16 as $\Delta t$ approaches t$_{0}$, falling in the
MHz regime. In parallel the optically-thin flux density decreases as:

\begin{equation}                                                               
S_{1} = S_{0} \left( \frac{t_{0} + \Delta t}{t_{0}}
\right)^{-2\delta}         
\label{eq_thin}                                                                
\end{equation}                                                                 
                                                                               
\noindent where $S_{0}$ and $S_{1}$ are the flux densities at the time         
$t_{0}$ and $t_{0} + \Delta t$, i.e. the source lifetime at the time           
of the last observing epoch, and $\delta$ is the spectral index of             
the electron energy distribution of the relativistic particles                 
(N($E$) $\propto E^{-\delta}$, $\delta = 2\alpha + 1$). 
If in Eq. \ref{eq_thin}
we consider typical values for $\delta$
between 2 and 3, we find that the flux density
would decrease by a factor of about 16$-$60, becoming of (sub-)mJy
level. Only the flux
density below the peak frequency becomes more prominent with
time. Part of these sources may be progenitors of
the population of low-power radio sources \citep[see
  e.g.][]{tingay15,baldi15,baldi19,mingo19} 
and/or MHz peaked spectrum radio sources
\citep[]{coppejans15}, or they may represent short-lived episode of radio
emission from an AGN \citep{czerny09}. The Square Kilometre
  Array, with its huge improvement in sensitivity in the MHz regime
  would enable systematic studies of population of faint 
radio sources, providing a fundamental step in our understanding of
their evolution.\\

For the sources with information on their parsec scale structure and
redshift we
estimate the expansion speed $v$ by: 

\begin{equation}
v = \frac{\theta \; {\rm D_{\rm L}}}{(1+z)^2} \frac{1+z}{\rm t_{\rm age}}
\label{eq-valocity}
\end{equation}

\noindent where $\theta$ is the angular size measured from VLBA
images, D$_{\rm L}$ is the luminosity distance of the source, $z$ is the
redshift, and $t_{\rm age}$ the estimated dynamical age.\\

The expansion speed ranges between 0.1$c$ and 0.7$c$, in agreement with
values estimated for the population of CSOs \citep[see
  e.g.][]{polatidis03,antao12}, with the only exception of the quasar
J1459$+$3337, which turned out to be likely a blazar on the basis of
the pc-scale structure.
Owing to the uncertainty on the source age, the
expansion speed should be considered as an upper limit. 
For all the sources our VLBA observations could not
identify the core region preventing us from investigating if both jets
are expanding at the same velocity or if the ambient medium plays a
role in their growth. Jet-cloud interaction seems to be
common during the first evolutionary phase when the jet is piercing
its way through the dense and inhomogeneous gas of the narrow line
region \citep{dd13}. The presence of an inhomogeneous ambient medium
has been found in some CSOs from the bright HFP sample
\citep{dd00}, where free-free absorption is observed towards only one
of the two 
jets/mini-lobes \citep[e.g. J0428$+$3259 and J1511$+$0518,][]{mo08b} 
and highly asymmetric structures
mark the presence of clouds slowing down the expansion on one side,
preventing its adiabatic expansion and enhancing synchrotron
losses \citep[e.g. J1335$+$5844,][]{mo14}. However, the sources studied
so far are relatively powerful (L$_{\rm 1.4 GHz}$ $>$ 10$^{26}$ W/Hz), 
and there are only a few studies on
the effects the ambient medium has on the expansion of low-power
(L$_{\rm 1.4 GHz}$
$<$ 10$^{26}$ W/Hz) jets
\citep[e.g.][]{kunert10}. Deep and systematic VLBI observations are
necessary for investigating the role of the ambient medium
during the first phase of the source expansion in faint objects.\\

\subsection{Steep spectral shape}

About 20 per cent (7 out of 35) of the sources discussed here have a
rather steep optically-thin VLA spectrum ($\alpha_{\rm a} >$ 1.0), which is
quite uncommon in radio galaxies with active regions. These small and
compact radio sources have somehow different characteristics with
respect to common extended radio galaxies. The equipartition magnetic
fields increases as we consider smaller and smaller sources: from a
few $\mu$G in the lobes of the classical FR-I/II sources
\citep{croston05} to a few mG in compact
steep-spectrum sources \citep{fanti95}, and up to a few tens/hundreds
mG in young HFP
objects \citep{mo08b}. This implies that for a given observing
frequency, the Lorentz factor of the radiating particle is smaller
than in larger sources with weaker fields. Moreover, the radiative
losses are higher, shortening the radiative lifetime of the
relativistic electrons and shifting the frequency break at lower and
lower frequency. In the small sources the
outer components are usually dominated by bright and compact regions,
like mini-hotspots, while no significant emission from the lobes is
observed \citep{tingay03,gugliucci05,mo08b,mo14}, 
supporting the severe losses hypothesis. \\
Alternatively, the steep spectra might be caused by an injection
spectral index that is steeper than what is usually found 
\citep{harwood16,harwood17}. Deep VLBI observations are
necessary to unveil the presence of low-luminosity extended structures
and disentangle between the two scenarios.\\

\section{Summary}

In this paper, we presented results on a multi-epoch multi-frequency
VLA monitoring campaign, and
  pc-scale VLBA observations of CSO candidates 
from the faint HFP sample. The conclusions that we can
  draw from this investigation are:

\begin{itemize}

\item 5 out of the 12 sources (42 per cent) observed with milli-arcsecond
  resolution turned out to be compact doubles. Taking into account
  earlier observations of additional 11 objects, we end up with a
  total of 11 sources showing a morphology consistent with what is
  expected for young radio sources. However, the radio structure and
  the spectral index information are not conclusive and deeper
  pc-scale observations are necessary to probe the nature of the
    sources and identify the locations of the center of activity.

\item 34 out of the 35 sources (97 per cent) observed with VLA have moderate
  to strong long-term variability. Only one source, J1624$+$2748, has
  neither spectral nor flux density variability.   

\item 18 radio sources possess spectral and flux
  density variability that is consistent with a cloud
  of homogeneous magnetized relativistic plasma in adiabatic
  expansion. For the sources with known redshift we estimate the
  dynamical ages which range between a few tens to a few hundreds years.
  The corresponding expansion velocity is mainly between
  0.1$c$ and 0.7$c$, similar to values found in young radio
  sources. However, among these sources, one object shows pc-scale
  properties and an estimated velocity of about $c$, suggesting a
  blazar nature.

\item In 17 sources
  the flux density changes randomly as it is expected in blazars, and
  in 6 sources the spectrum becomes flat in the last observing
  epoch, confirming that samples selected in the GHz regime are highly
  contaminated by beamed objects. 

\item No significant dichotomy is
  found in the flux density variability between galaxies and quasars, with a slightly larger fraction
  of galaxies showing erratic variability typical of beamed
  sources.

\end{itemize}

The fast evolution that we observe in some CSO
candidates suggests that they hardly represent the progenitors of
classical Fanaroff-Riley radio galaxies. Thanks to the huge
improvement in sensitivity and the wide frequency coverage up to 100
GHz, the Next Generation Very Large Array\footnote{ngvla.nrao.edu} will be the optimal instrument for shedding a light on
the nature and fate of these objects.\\

\section*{Acknowledgments}
We thank the anonymous referee for reading the manuscript carefully
and making valuable suggestions.
The VLA and VLBA are operated by the US 
National Radio Astronomy Observatory which is a facility of the National
Science Foundation operated under cooperative agreement by Associated
Universities, Inc. This work has made use of the NASA/IPAC
Extragalactic Database NED which is operated by the JPL, Californian
Institute of Technology, under contract with the National Aeronautics
and Space Administration. AIPS is produced and maintained by the
  National Radio Astronomy
Observatory, a facility of the National Science Foundation
operated under cooperative agreement by Associated Universities, Inc.
Funding for SDSS-III has been provided by
the Alfred P. Sloan Foundation, the Participating Institutions, the
National Science Foundation, and the U.S. Department of Energy Office
of Science. The SDSS-III web site is http://www.sdss3.org/.
SDSS-III is managed by the Astrophysical Research Consortium for the
Participating Institutions of the SDSS-III Collaboration including the
University of Arizona, the Brazilian Participation Group, Brookhaven
National Laboratory, Carnegie Mellon University, University of
Florida, the French Participation Group, the German Participation
Group, Harvard University, the Instituto de Astrofisica de Canarias,
the Michigan State/Notre Dame/JINA Participation Group, Johns Hopkins
University, Lawrence Berkeley National Laboratory, Max Planck
Institute for Astrophysics, Max Planck Institute for Extraterrestrial
Physics, New Mexico State University, New York University, Ohio State
University, Pennsylvania State University, University of Portsmouth,
Princeton University, the Spanish Participation Group, University of
Tokyo, University of Utah, Vanderbilt University, University of
Virginia, University of Washington, and Yale University. 

\section*{Data Availability}
The data uderlying this article are available in the NRAO Data Archive 
(https://science.nrao.edu/observing/data-archive) with the
  project codes AO281 and BO057.  Calibrated
  data are available on request.

\end{document}